\documentclass[english]{iopart}

\usepackage{graphicx}
\usepackage{url}
\usepackage{iopams}

\expandafter\let\csname equation*\endcsname\relax
\expandafter\let\csname endequation*\endcsname\relax

\usepackage{amssymb}
\usepackage[latin1]{inputenc}
\usepackage{amsmath}
\usepackage{amsfonts}                           
\usepackage{epsfig}
\newcounter{fig}
\begin{document}

\title{The lambda extensions of the Ising correlation functions $C(M,N)$}
      
\vskip .1cm 

\author{S. Boukraa$^\dag$,
J.-M. Maillard$^\ddag$, }
\address{\dag LSA, IAESB,
  Universit\'e de Blida, Algeria}
\address{\ddag\ LPTMC, Sorbonne Universit\'e,  Tour 24, 5\`eme \'etage, case 121, \\
 4 Place Jussieu, 75252 Paris Cedex 05, France} 
\ead{maillard@lptmc.jussieu.fr, jean-marie.maillard@sorbonne-universite.fr, 
bkrsalah@yahoo.com}

\vskip .1cm 

\begin{abstract}

We revisit, with a pedagogical heuristic motivation, the lambda extension 
of the low-temperature row correlation functions $ \, C(M,N)$ of
the two-dimensional Ising model. In particular, using these one-parameter series to understand
the deformation theory around selected values of $\ \lambda$, namely $\, \lambda = \, \mathrm{\cos}(\pi \, m/n)$
with $\, m$ and $\, n$ integers, we show that these series yield  perturbation
coefficients, generalizing form factors, that are D-finite functions. As a by-product these exact results
provide an infinite number
of highly non-trivial identities on the complete elliptic integrals of the first and second kind.
These results underline the fundamental role of Jacobi theta functions and Jacobi forms,
the previous D-finite functions being (relatively simple) rational functions of 
Jacobi theta functions, when rewritten in terms of the nome of elliptic functions.

\end{abstract}

\vskip .1cm


\noindent {\bf AMS Classification scheme numbers}: 34M55, 
47E05, 81Qxx, 32G34, 34Lxx, 34Mxx, 14Kxx 

\vskip .2cm


\vskip .2cm

% {\em My name is F2021/BARRY$\_$2021/ Heuristic$\_$JPA.tex}

\vskip .2cm

\today

\vskip .2cm

\section{Introduction}
\label{Introduction}

We revisit, with a pedagogical heuristic motivation, the lambda extension~\cite{Holonomy} 
of the two-point  correlation functions $ \, C(M,N)$ of the two-dimensional
Ising model. For simplicity we will examine in detail the lambda extension
of a particular low-temperature diagonal correlation function, namely $ \, C(1,1)$,
in order  to make crystal clear some structures and subtleties.
However similar structures and results can be obtained on the 
two-point  correlation functions  $ \, C(M,N)$ for the special case $\, \nu = \, -k$
studied in~\cite{OkamotoJPA} where Okamoto sigma-forms of Painlev\'e VI
equations also emerge.

In 1976 Wu, McCoy, Tracy and Barouch~\cite{wmtb} discovered,  in
the scaling limit $ \, T \rightarrow \, T_c\,$ with $ \, N \cdot \, (T \, -T_c) \, $
fixed, that the isotropic diagonal correlation $\, C(N,N)$ is given
by a Painlev{\'e} III equation.
This was generalized in 1980 by Jimbo and Miwa~\cite{jm}
who defined for $\,T< \,T_c$ 
\begin{equation}
\label{sigmam}
\hspace{-0.4in}
\sigma\, \,= \,\,\,
t \cdot \, (t-1) \cdot \, \frac{d}{dt}\ln C(N,N) \, \, -\frac{t}{4}
\quad \,\, \quad  {\rm with} \,\quad \quad \quad \, t \, = \,\, k^2, 
\end{equation}
and for $\,T \, > \,T_c$
\begin{equation}
  \label{sigmap}
\hspace{-0.4in}
\sigma \,\,= \,\,\,
t \cdot \, (t-1)\cdot  \, \frac{d}{dt}\ln C(N,N)\, \, -\frac{1}{4}
\quad \,\quad \,  {\rm with} \quad \quad \,\,\,\,  t \, = \, \, k^{-2}, 
\end{equation}
and in both cases derived the sigma-form of Painlev\'e VI non-linear ODE
satisfied by $\, \sigma$:
\begin{eqnarray}
\label{jmequation}
\hspace{-0.98in}&& \quad  
\left(t \cdot \, (t-1) \cdot \, \frac{d^2\sigma}{dt^2}\right)^2
\\
\hspace{-0.98in}&&  \quad  \,    \quad  
\, = \, \,\, N^2 \cdot \,
\left((t -1) \cdot  \, \frac{d\sigma}{dt} \, -\sigma\right)^2
\, -4 \cdot \,\frac{d\sigma}{dt} \cdot \,
\left((t -1) \cdot \frac{d\sigma}{dt}\, -\sigma\, -\frac{1}{4}\right)
\cdot \,\left(t\frac{d\sigma}{dt}-\sigma\right).
\nonumber 
\end{eqnarray}
The low-temperature diagonal two-point correlation functions $\, C(N,N)$
are (homogeneous) polynomial expressions~\cite{PainleveFuchs,FuchsPainleve}
in the complete elliptic integral  
of the first and second
kind\footnote[1]{In Maple $\, K$ is $ \, \, 2/\pi \, EllipticK(t^{1/2}) \, \, $
  and  $\, E$ is   $ \, \, 2/\pi \, EllipticE(t^{1/2})$. With that normalization
  one has
  $\, K \, = \, \theta_3(0, \, q)^2$  and
  $\, t^{1/2} \, = \, k\, = \, \theta_2(0, \, q)^2/\theta_3(0, \, q)^2$
  and thus 
  $\, k \cdot \, K \, = \, \theta_2(0, \, q)^2$, where $\, q$ denotes the nome.}:
\begin{eqnarray}
\label{EK}
  \hspace{-0.98in}&& \quad \quad \quad  \quad \quad
  K \, = \, \, _2F_1\Bigl([{{1} \over {2}}, \, {{1} \over {2}}], \, [1], \, t\Bigr),
  \, \, \quad \quad \quad 
   E \, = \, \, _2F_1\Bigl([{{1} \over {2}}, \, -{{1} \over {2}}], \, [1], \, t\Bigr).
\end{eqnarray}
In~\cite{Holonomy} it has been underlined that these
correlation functions $\, C(N,N)$
have lambda extensions {\em which are also solutions of} (\ref{jmequation}),
that can be defined using a ``form factor'' low-temperature
expansion~\cite{Holonomy,Lyberg}
(see (9) in~\cite{Holonomy}):   
\begin{eqnarray}
\label{form}
\hspace{-0.98in}&& \quad \quad  \quad   \quad \quad \quad
C_{-}(N, \, N; \, \lambda) \, \, = \, \, \, \,
(1\, -t)^{1/4} \cdot \,
\Bigl( 1 \, + \, \sum_{n=1}^{\infty} \, \lambda^{2\, n} \cdot \, f_{N, \, N}^{(2\, n)} \Bigr), 
\end{eqnarray}
where the form factors~\cite{Holonomy} $\, f_{N, \, N}^{(2\, n)}$ are also
polynomial expressions~\cite{PainleveFuchs,FuchsPainleve} in the complete elliptic integral
of the first and second kind (\ref{EK}).
For instance for the simplest low-temperature correlation
function this form factor expansion reads
\begin{eqnarray}
\label{form11fact}
\hspace{-0.98in}&& \quad \quad  \quad \quad   \quad \quad
C_{-}(1, \, 1; \, \lambda) \, \, = \, \, \, \,
(1\, -t)^{1/4} \cdot \,
\Bigl( 1 \, + \, \sum_{n=1}^{\infty} \, \lambda^{2\, n} \cdot \, f_{1, \, 1}^{(2\, n)} \Bigr), 
\end{eqnarray}
where the first form factors read:
\begin{eqnarray}
  \label{form11f11}
\hspace{-0.98in}&& \quad \quad \quad \quad 
f_{1, \, 1}^{(2)} \, \, = \, \, \,  {{1} \over {2}} \cdot \,
\Bigl(1 \,\,\, -3 \, E \, K \,\, -(t\, -2) \cdot \, K^2\Bigr), 
  \\
  \hspace{-0.98in}&& \quad \quad \quad \quad
 \label{form11f114}
f_{1, \, 1}^{(4)} \, \, = \, \, \,
{{1} \over {24}} \cdot \,
\Bigl(9 \,\, \, -30\, E \, K \,\, -10 \cdot \, (t\,-2) \cdot \, K^2
 \nonumber  \\
  \hspace{-0.98in}&&  \quad \quad \quad \quad \quad \quad \quad  \quad  \quad 
  \, +(t^2\, -6t \, +6) \cdot \, K^4 \,\, +15 E^2\, K^2 \, \,
     +10\cdot (t\, -2) \cdot \, E\, K^3 \Bigr).
\end{eqnarray}
For $\, \lambda \, = \, 1$ we must recover, from (\ref{form11fact}),
the well-known expression of
the {\em low-temperature} two-point correlation function $\, C(1,1) \, = \, E$:
\begin{eqnarray}
\label{form11E}
\hspace{-0.98in}&& \quad   \, \quad  \quad  
C_{-}(1, \, 1; \, 1) \, \, = \, \, \,  E \, \,  = \, \, \,\,
1 \,\, -{{1} \over {4}} \cdot \, t   \,\,  \, -{{3} \over {64}} \cdot \, t^2 \,\,
- {{5} \over {256}} \cdot \, t^3 \,
  \,  -{{175} \over {16384}} \cdot \, t^4 \, \,  \,  \, + \, \, \, \cdots 
\nonumber   \\
 \hspace{-0.98in}&& \quad \quad \quad \quad  \quad  \quad \quad \quad 
 \, = \, \, \,\,
 (1\, -t)^{1/4} \cdot \,
\Bigl( 1 \, \, + \, \sum_{n=1}^{\infty}  \, f_{1, \, 1}^{(2\, n)} \Bigr), 
\end{eqnarray}
which corresponds to write the ratio $\, E/(1\, -t)^{1/4} \, $ as an infinite sum
of polynomial expressions of $\, E$ and $\, K$.

\vskip .1cm

\section{Simple power series expansions and formal calculations}
\label{undergraduate}

For pedagogical reasons we restrict our analysis to the
low-temperature two-point correlation function
$\, C(1,1)$ and its lambda extension. 
Since all these lambda extensions are power series in $\, t$, we can try
to get, order by order, the  series
expansion of   $\, C_{-}(1, \, 1; \, \lambda) \, $
from the non-linear ODE  (\ref{jmequation}). Recalling~\cite{Holonomy} the
form factor expansion (\ref{form11fact}),
we can either see the series expansion in $\, t$ as a deformation of the simple
algebraic function $\, (1 \, -t)^{1/4}$, or {\em more naturally}, see
the series expansion of the lambda-extension
of the low-temperature two-point correlation function $\, C_{-}(1, \, 1; \, \lambda) \, $
as a deformation of
the exact expression $\, C_{-}(1,1) \, = \, E$ ($M$ denotes here a difference
to $\, \lambda^2= \, 1$, see (\ref{correspondence}) below):
\begin{eqnarray}
\label{morenaturallyM1}
\hspace{-0.98in}&& \, \, \, \quad   \quad   \quad  
C_{-}(1, \, 1; \, \lambda) \,   \, = \, \, \,  C_{M}(1, \, 1; \, M)
\nonumber \\
  \hspace{-0.98in}&& \, \, \,
\quad   \quad   \quad  \quad   \quad   \quad   \quad    \, \, = \, \, \,\,\,
E \,\, \,\,  + M \cdot g_1(t)\, \,\, + M^2 \cdot \, g_2(t) \,
\,\, + M^3 \cdot \, g_3(t) \,  \,  \,  \,\, + \, \, \cdots 
\end{eqnarray}
Using the sigma-form of Painlev\'e VI equation (\ref{jmequation})
one can find that this expansion (\ref{morenaturallyM1})
  reads as a series expansion in the variable $\, t$:
\begin{eqnarray}
\label{morenaturally}
\hspace{-0.98in}&& \, \, \,  \, \, \, 
 C_{M}(1, \, 1; \, M) \, \, = \,\, \, \,\, 
 1 \, \, \,\, -{{1} \over {4}} \cdot \, t \,\, \,  \,
 \, -\Bigl({{3} \over {64}}  \, +{{3} \over {256}}\cdot \, M \Bigr) \cdot \, t^2
 \, \,\, \,
 -\Bigl( {{5} \over {256}} \, + {{9} \over {1024}}  \cdot \, M    \Bigr)   \cdot \, t^3
\nonumber   \\
 \hspace{-0.98in}&& \quad \quad \quad \quad
 \, \, -\Bigl(  {{175} \over {16384}} \,   + {{441} \over {65536}}  \cdot \, M     \Bigr)
 \cdot \, t^4\,  \,  \,  \,
 -\Bigl(  {{441} \over {65536}} \,   +{{1407} \over {262144}} \cdot \,  M   \Bigr)
 \cdot \,t^5
 \nonumber   \\
 \hspace{-0.98in}&& \quad \quad  \quad \quad \quad \quad  \, \,
 -\Bigl({{4851} \over {1048576}}  \,  \,
 +{{9281} \over {  2097152}}    \cdot \, M  \,  -{{5} \over {  16777216 }}  \cdot \, M^2 \Bigr)
 \cdot \, t^6
    \,\, \, \,  \, \, + \, \, \cdots 
\end{eqnarray}
Note that this  low-temperature expansion (\ref{morenaturally}) gives for $\, \sigma$
defined by (\ref{sigmam}): 
\begin{eqnarray}
\label{morenaturallysigma}  
\hspace{-0.98in}&& \, \, \,  \, \, \,   \quad  \quad      \quad  \quad    
\sigma \, \,  \, = \,\,\,  \, 
  t \cdot \, (t-1) \cdot \, \frac{d}{dt}\ln C(1, \, 1; \, M) \, \, \, \,  -\frac{t}{4}
 \, \,  \,  \,  = \,\,\, \,  (M\, -4) \cdot \sigma_M, 
\end{eqnarray}
where:
\begin{eqnarray}
\label{morenaturallysigmaM}
\hspace{-0.99in}&& 
 \sigma_M\, \,= \,\,\,\,
  {{3} \over {128}} \cdot \, t^2 \,\, \,\, + {{3} \over {256}} \cdot \, t^3 \,\, \,\,
 +{{3} \over {32768}} \cdot \, (3\, M \, +74) \cdot \, t^4 \, \,\,
 +{{3} \over {65536}} \cdot \, (9\, M \, +94) \cdot \, t^5
\nonumber \\
 \hspace{-0.98in}&& \quad \quad  \quad \quad \quad 
+{{3} \over {8388608}} \cdot \, (9\, M^2 \, +1270\, M \, +8176) \cdot \, t^6
 \,\,\,  \, +  \,\, \, \cdots 
\end{eqnarray}
Recalling the expansions of $\,\, (1\, -t)^{1/4}$
\begin{eqnarray}
\label{recalling}
\hspace{-0.98in}&& \, \, \,\quad
(1\, -t)^{1/4} \,\, \, = \, \, \,\,
1 \,\, \,\, -{{1} \over {4}}\cdot \, t \, \,\,\,  -{{3} \over {32}} \cdot \, t^2 \,\, \,
   -{{7} \over {128}} \cdot \, t^3 \,\, \, - {{77} \over {2048}}\cdot \, t^4 \, \,  \, 
   \, \, + \, \, \cdots 
\end{eqnarray}
one can see that this series  coincides (as it should) with the
series (\ref{morenaturally}) for $\, M= \, 4$
(i.e. $\, \lambda\, = \, 0$ in (\ref{form11fact})). 
Recalling  the expansions of $\, f_{1, \, 1}^{(2)}\,$ and  $\, f_{1, \, 1}^{(4)}$:
\begin{eqnarray}
\label{recalling}
\hspace{-0.98in}&& \quad 
f_{1, \, 1}^{(2)}     \, \, = \, \, \, \,
{{3} \over {64}} \cdot  \, t^2 \,\,  \, + {{3} \over {64}}  \cdot  \, t^3 \,\,  \,
  + {{705} \over {16384}} \cdot \,  t^4 \,\,  + {{321} \over {8192}}  \cdot \, t^5
\,\,  \, +{{18795} \over {524288}}  \cdot  \, t^6  \, \,\,  \, + \, \, \cdots,
\nonumber \\
\hspace{-0.98in}&& \quad 
f_{1, \, 1}^{(4)}     \, \, = \, \, \,
{{5} \over {1048576 }} \cdot \, t^6 \, \,  +{{15} \over {1048576 }} \cdot \, t^7
\,  +{{7335} \over {268435456}} \cdot \, t^8
\, \, +{{2855} \over {67108864 }} \cdot \, t^9
\nonumber \\
\hspace{-0.98in}&&  \quad \quad
\,  +{{4052025} \over {687194767366 }}\cdot \, t^{10}
\,  +{{5215005 } \over { 68719476736  }}\cdot \, t^{11}
 \, \,  \, \,\, + \, \, \, \cdots 
\end{eqnarray}
the series expansion (\ref{morenaturally}) can be seen to match with the (form factor)
expansion (\ref{form11fact}) with (\ref{form11f11}) and  (\ref{form11f114})
(together with the previous expansions (\ref{recalling})) if one has
the following correspondence:
\begin{eqnarray}
\label{correspondence}
\hspace{-0.98in}&& \, \, \,
 \quad \quad  \quad  \quad \quad  \quad \quad  \quad \quad  \quad \quad
M \,  \, = \, \, \, \, 4  \cdot \, (1 \, - \lambda^2). 
\end{eqnarray}
At the first order in $\, \lambda^2$ one gets from (\ref{morenaturally}): 
\begin{eqnarray}
\label{correspondence2}
\hspace{-0.98in}&& \, \, \,
 (1 \, -t)^{1/4} \cdot \, f_{1, \, 1}^{(2)}    
\nonumber   \\
 \hspace{-0.98in}&& \,\,  \,  \quad  \, = \,\, \, \,
 {{3} \over {64}} \cdot  \, t^2 \, \, + {{9} \over {256}} \cdot  \,  t^3
 \, \, + {{441} \over {16384}} \cdot \,  t^4
 \,\,  + {{1407} \over {65536}} \cdot \,  t^5 \, \,
 + {{2319} \over {131072}} \cdot \, t^6  \,\,\, \,  \,  +   \, \,  \cdots 
\end{eqnarray}
in agreement with the exact expression (\ref{form11f11}).
At the second order 
in $\, \lambda^2$ one gets from (\ref{morenaturally}): 
\begin{eqnarray}
\label{correspondence2}
\hspace{-0.98in}&& \, \, \,\,\,\,
 (1 \, -t)^{1/4} \cdot \, f_{1, \, 1}^{(4)}     \, \, \, \, = \, \, \, \,  \,
 {{5} \over {1048576 }} \cdot \, t^6 \, \, \,  +{{55} \over {41943046 }} \cdot \, t^7
\, \,  +{{6255} \over {2684354566 }} \cdot \, t^8
\nonumber   \\
\hspace{-0.98in}&&  \quad \,  \, 
 \, +{{36625} \over {10737418246 }} \cdot \, t^9
\,  \,  +{{3079025} \over {687194767366 }}\cdot \, t^{10}
\,  \,  +{{15116115} \over {2748779069446 }}\cdot \, t^{11}
 \, \, \,  + \, \, \, \cdots 
\end{eqnarray}         
in agreement with the exact expression (\ref{form11f114}).
At the third order 
in $\, \lambda^2 \,$ one gets from (\ref{morenaturally}): 
\begin{eqnarray}
\label{correspondence3}
\hspace{-0.98in}&& \, 
(1 \, -t)^{1/4} \cdot \, f_{1, \, 1}^{(6)}     \, \, \, \, = \, \, \,  \,
{{7} \over {4398046511104}} \cdot \, t^{12}
\,\, \, +{{161} \over {17592186044416}} \cdot \, t^{13}
\nonumber \\
\hspace{-0.98in}&&  \quad \quad \quad \,
 +{{33789} \over {1125899906842624}} \cdot \, t^{14} \, \,
 +{{332703} \over {4503599627370496 }} \cdot \, t^{15}
 \, \, \,  \,  \, \, + \, \, \,  \cdots 
\end{eqnarray}       
Matching the form-factor expansion (\ref{form11fact})  with
the series expansion (\ref{morenaturallyM1})
one gets the following (infinite ...) identities:
\begin{eqnarray}
\label{correspondence2fg} 
  \hspace{-0.98in}&&  \quad
  (1 \, -t)^{1/4} \, \, = \, \,\, 
  E \,\,\,  + \, \, \sum_{n=1}^{\infty} \, \,  4^n \cdot \, g_n(t), \quad \quad \,\,
  (1 \, -t)^{1/4} \cdot \,  f_{1, \, 1}^{(2)}
  \, \, = \, \,  \, - \, \sum_{n=1}^{\infty} \, \,  n \cdot \, 4^n \cdot \, g_n(t),
 \nonumber   \\
  \hspace{-0.98in}&&  \quad \quad \quad \quad \quad 
  (1 \, -t)^{1/4} \cdot \,  f_{1, \, 1}^{(4)}
  \, \, = \, \,  \,\, \sum_{n=1}^{\infty} \,\,   {{n \cdot \, (n\, -1)} \over {2}} \cdot \, 4^n \cdot \, g_n(t),
  \quad \quad \quad \quad \, \cdots 
\end{eqnarray}         
and conversely:
\begin{eqnarray}
\label{correspondence2gf} 
  \hspace{-0.98in}&& \, \, \, \, \, \, \, 
 E  \, \, = \, \,   \, \,
(1\, -t)^{1/4} \cdot \,
\Bigl( 1 \, \, + \, \sum_{n=1}^{\infty}  \, f_{1, \, 1}^{(2\, n)} \Bigr),
    \quad  \quad
  g_1(t)  \, \, = \, \,  \,
  - \,  {{(1\, -t)^{1/4} } \over { 4}} \cdot \, \sum_{n=1}^{\infty} \, \,  n \cdot \, f_{1, \, 1}^{(2 \, n)},
 \nonumber   \\
  \hspace{-0.98in}&&  \quad \quad \quad \quad \quad
 g_3(t)  \, \, = \, \,  \,\,
 {{(1\, -t)^{1/4} } \over { 32}} \cdot \,
 \sum_{n=1}^{\infty} \, \,  n \cdot \, (n \, -1) \cdot \, f_{1, \, 1}^{(2 \, n)},
 \quad \quad \quad \cdots 
\end{eqnarray}  

\vskip .1cm

\subsection{Algebraic subcases}
\label{Algebraicsub}

\vskip .1cm

It had been noticed~\cite{Holonomy}, for $\, \lambda \, = \, \mathrm{\cos}(\pi \, m/n) \, $
where $\, m$ and $\, n$ are integers, and $\,\lambda^2 \, \ne \, 1$, 
that the lambda extension
(\ref{form11fact}) is not only D-finite\footnote[1]{Like in the $\, \lambda^2= \,1$, $\, M=\, 0$ case.},
but is, in fact, an {\em algebraic function}.

\vskip .1cm

\subsubsection{$\, \lambda \, = \, \mathrm{\cos}(\pi/4)$\\}
\label{lambdaPi4}

For instance for
$\, \lambda \, = \, \mathrm{\cos}(\pi/4) \, = \, 1/\sqrt{2} \, \, $, i.e. for $\, M\, = \, 2$,
one finds that (\ref{morenaturally}) is actually the series expansion of
an algebraic expression
\begin{eqnarray}
\label{actually}
\hspace{-0.98in}&& \, \,
(1-t)^{1/16} \cdot \,
_2F_1\Bigl([-{{3} \over {8}}, \, {{1} \over {8}}], \, [{{1} \over {4}}], \, t\Bigr)
\, \, = \, \, \, \, (1-t)^{1/16} \cdot \, \Bigl( {{ 1 \, +(1-t)^{1/2}} \over {2}} \Bigr)^{3/4}
\, \, \, = \, \, \,
  \\
  \hspace{-0.98in}&& \quad 
  1 \, -{{1} \over {4}} \cdot \, t   \,
  -{{9} \over {128}}  \cdot \, t^2 \, \, -{{19} \over {512}} \cdot \, t^3
  \, \, - {{791} \over {32768 }} \cdot \, t^4 \, \, -{{2289} \over {}131072}  \cdot \, t^5
   \, \, -{{56523} \over {4194304}}  \cdot \, t^6 \,   \, + \, \, \cdots \nonumber 
\end{eqnarray}
in agreement of the exact result given in equation (99) of~\cite{Saga}.

\vskip .1cm

\subsubsection{$\, \lambda \, = \,  \mathrm{\cos}(\pi/6)$ \\}
\label{lambdaPi6}

Another example corresponds to  $\, M \, = \, 1$
(i.e. $\, \lambda \, = \, \sqrt{3}/2 \, = \, \mathrm{\cos}(\pi/6)$). The
series (\ref{morenaturally}) reads:
\begin{eqnarray}
\label{globM1}
  \hspace{-0.98in}&& \,\quad \quad \quad \quad  
 1   \, \,\, -{{1} \over {4}}  \cdot \, t \,\, \,  \, -  {{15} \over {256}} \cdot \, t^2
 \,\, \,\, - {{29} \over {1024}}  \cdot \, t^3\, \,\, - {{1141} \over {65536}} \cdot \, t^4
\, \,\, - {{3171} \over {262144}} \cdot \, t^5
\nonumber \\
\hspace{-0.98in}&& \quad \quad \quad  \quad \quad \quad  \,
\, - {{151859} \over {16777216}} \cdot \, t^6 \,  \,\,  \,
                   - {{477697} \over {67108864}}  \cdot \, t^7
                   \, \,  \,   \, \, + \, \, \, \cdots    
\end{eqnarray}
One   first finds that this series (\ref{globM1})  is D-finite, being
the solution of the order-{\em four} linear differential operator: 
\begin{eqnarray}
\label{actuallyM1}
  \hspace{-0.98in}&& \, \,\quad \quad 
  D_t^4 \,\, \, + {{1} \over {3}}\cdot \,
  {{ 19\, t^3-30\, t^2+36\, t-14} \over {(t-1)\, (t^2 -t+1) \, t}}\cdot \, D_t^3
 \nonumber   \\
  \hspace{-0.98in}&&  \quad \quad\quad  \quad 
  \, + {{1} \over {216}} \cdot \,
     {{ 1625\, t^4-3439\,t^3+5091\,t^2-3628\,t+680 } \over { (t-1)^2\, (t^2-t+1) \, t^2}}
     \cdot \, D_t^2
 \nonumber   \\
  \hspace{-0.98in}&&  \quad \quad\quad  \quad 
  \, + {{1} \over {11664}} \cdot \,
     {{ 10033\,t^5-26608\,t^4+53854\,t^3-55334\,t^2+16160\,t+880} \over {
         (t-1)^3\, (t^2-t+1)\, t^3 }} \cdot \, D_t
 \nonumber   \\
  \hspace{-0.98in}&&  \quad \quad \quad \quad 
  \, +\,  {{1} \over {186624 }} \cdot \,
     {{ 3689\,t^5-6725\,t^4+2573\,t^3+8\, t ^2+5200\,t \, -3520} \over {
         (t-1)^4 \, (t^2-t+1) \, t^3}}.          
\end{eqnarray}
In fact the series (\ref{globM1}) is not only D-finite, it is {\em an algebraic series}. Denoting
$\, S(t)$ the series  (\ref{globM1}), and $\, S_{12} = \, S(t)^{12} \, \, $
its {\em twelfth} power,   one can see that $\, S_{12}$
is actually solution of  the quartic equation 
\begin{eqnarray}
 \label{algM6}
\hspace{-0.98in}&&   \quad \quad \quad  \quad  
3^{36} \, \cdot \, t^8\,  \cdot\, S_{12}^4 \,\,\,\,
                   +  2^{10} \cdot  \,  3^{26} \, \cdot \, t^6\,\cdot \, (t \, -1) \, \cdot \, p_{6}  \, \cdot \,S_{12}^3
                   \nonumber \\
\hspace{-0.98in}&& \, \, \quad  \quad  \quad  \quad  \quad  \,\,
                   + 2^{17} \cdot  \,  3^{15} \,  \, \cdot \, t^4 \,\cdot \, p_{12} \,\cdot \, (t-1)^2 \, \cdot \, S_{12}^2
              \, \, \, \,  \,      + 2^{26}  \,\cdot  \, (t-1)\, \cdot  \, p_{24} \, \cdot  \, S_{12}
\nonumber \\
\hspace{-0.98in}&& \, \, \quad  \quad  \quad \quad \quad  \quad  \quad  \quad  
+ 2^{32}   \,\cdot \, (t-1)^4\,\cdot \, (t^2 -t +1)^{12}
\, \, \,\, = \, \,\, \, \, 0, 
\end{eqnarray}
where:
\begin{eqnarray}
\label{wherealgM6}
\hspace{-0.98in}&&
p_{6} \,  \, = \, \,\, \,  5\, \, t^6 \,\, -15\, \, t^5 \,\,
     +138\, \, t^4 \,\, -251\, \, t^3 \,\, +138\, \, t^2 \,\, -15\, \, t \, \, +5,  
\nonumber \\
\hspace{-0.98in}&&
p_{12} \,  \, = \, \, \,
113\, \, t^{12} \,\, -678\, \, t^{11} \, +5829\, \, t^{10} \, -22930\, \, t^9 \,
 +148410\, \, t^8 \,  -463518\, \, t^7
 \nonumber \\
 \hspace{-0.98in}&& \, \, \quad \quad
 +665661\, \, t^6  \, -463518\, \, t^5  \, +148410\, \, t^4 \, 
 -22930\, \, t^3 \,  +5829\, \, t^2 \,  -678\, \, t \,\,  +113, 
 \nonumber
\end{eqnarray}
\begin{eqnarray}
\label{wherealgM6suite}
\hspace{-0.98in}&& p_{24} \,  \, = \, \, \,\,
64\, \, t^{24}\,-768\, \, t^{23}\, +4965\, \, t^{22}\, -22231\, \, t^{21}\,
+3243192\, \, t^{20} \, -31880523\, \, t^{19}
\nonumber \\
\hspace{-0.98in}&& \, \, \quad \quad
+66263383\, \, t^{18}\,
+309635262\, \, t^{17} \, -1791331236\, \, t^{16} \,
+3209457458\, \, t^{15} \nonumber \\
\hspace{-0.98in}&& \, \, \quad \quad
\, -698769519\, \, t^{14} -6199132605\, \, t^{13}
+10265065180\, \, t^{12} -6199132605\, \, t^{11}
\nonumber \\
\hspace{-0.98in}&& \, \, \quad \quad \, -698769519\, \, t^{10} \,
+3209457458\, \, t^9 \, -1791331236\, \, t^8 \, +309635262\, \, t^7
 \\
\hspace{-0.98in}&& \, \, \quad \quad  \, +66263383\, \, t^6
\, -31880523\, \, t^5 \, +3243192\, \, t^4 \,
-22231\, \, t^3 \, +4965\, \, t^2 \, -768\, \, t \, +64.
\nonumber
\end{eqnarray}

\vskip .1cm

\subsubsection{$\, \lambda \, = \,  \mathrm{\cos}(\pi/3)$\\}
\label{lambdaPi3}

Similarly, for $\, M\, = \, 3$  (i.e. $\, \lambda \, = \, 1/2 \, = \, \mathrm{\cos}(\pi/3)$),
the series  (\ref{morenaturally}) reads
\begin{eqnarray}
\label{globM3}
  \hspace{-0.98in}&& \,\quad \quad \quad \quad  
 1  \, \,\, -{{1} \over {4}}  \cdot \, t \,\, \, \, -  {{21} \over {256}} \cdot \, t^2
 \,\, \,- {{47} \over {1024}}  \cdot \, t^3\,\, \,- {{2023} \over {65536}} \cdot \, t^4
\, \, \, - {{ 5985} \over {262144}} \cdot \, t^5
\nonumber \\
\hspace{-0.98in}&& \quad \quad \quad \quad  \quad \quad \quad  \,
\, - {{300315} \over { 16777216}} \cdot \, t^6
\, \, \, \, - {{979737} \over {67108864}}  \cdot \, t^7
 \,  \,  \,  \, \, + \, \, \, \cdots    
\end{eqnarray}
and can be seen to be solution of an order-{\em four} linear differential operator:
\begin{eqnarray}
\label{actuallyM3}
\hspace{-0.98in}&& \quad \quad \quad \quad \quad \quad 
D_t^4 \,\, \,\,   + {{2} \over {3}}\cdot \,
{{ 11 \, t-7} \over {(t-1)  \, t}} \cdot \, D_t^3
 \,\,\, \, + {{1} \over {54}} \cdot \,
 {{ 587\, t^2 \, -737 \, t\, +170} \over {(t-1)^2  \,  \, t^2}}\cdot \, D_t^2
\nonumber \\
\hspace{-0.98in}&&  \quad \quad \quad \quad \quad \quad \quad \quad 
  \, + {{1} \over {1458}} \cdot \,
{{ 2855 \, t^3 -5223 \, t^2 \,  +2130 \, t\, +110} \over {(t-1)^3  \,  \, t^3}}\cdot \, D_t
\nonumber \\
\hspace{-0.98in}&&  \quad \quad \quad \quad \quad \quad \quad \quad \quad \quad 
  \, + {{1} \over {11664}} \cdot \,
{{ 161 \, t^3 -702 \, t^2 \,  +1785 \, t\, -220} \over {(t-1)^4  \,  \, t^3}}.         
\end{eqnarray}
Again, the series (\ref{globM3}) is not only D-finite, it is also
an algebraic series. Denoting $\, S(t)$ the series 
(\ref{globM3}), and $\, S_{6} = \, S(t)^{6} \, $ its {\em sixth} power,   one can see that $\, S_{6}$
is solution of  the quartic equation 
\begin{eqnarray}
 \label{algM3}
\hspace{-0.98in}&&  \quad  \quad \quad  
3^{27}  \,\cdot \, t^4\, \cdot \, S_{6}^4 \,\,  \, \,
- 2^{10}  \,\cdot \,3^{20}  \cdot \, t^4\, \cdot \, (t\, -1) \, \cdot \, (t \, -2) \,  \cdot \, S_{6}^3
\nonumber   \\
  \hspace{-0.98in}&&  \quad \quad \quad \quad 
  + 2^{9}  \,\cdot \,3^{11}    \, \cdot \, t^2 \, \cdot \, p_{4}\, \cdot \, (t-1)^2\, \cdot \, S_{6}^2
  \,\, \, \,\,  
+ 2^{15}   \, \cdot \, (t \, -2)\, \cdot \, p_{8} \, \cdot  \, (t-1)^2\,  \cdot \, S_{6}
\nonumber   \\
  \hspace{-0.98in}&& \quad  \quad \quad \quad \quad 
- 2^{16} \, \cdot \, (t \, -1)^8 \,\, \, = \,\,  \, \, 0.
\end{eqnarray}
where:
\begin{eqnarray}
\label{algM3where}
\hspace{-0.98in}&&\quad \quad \quad \quad \quad 
p_{8} \, \, = \, \,\,  \, 8192\, \, t^8 \,\,  -38912\, \, t^7 \,\,  +82304\, \, t^6 \,\, 
-93704  \, \, t^5 \,\,  +64151\, \, t^4
\nonumber   \\
\hspace{-0.98in}&& \, \, \quad \quad \quad  \quad  \quad \quad  \quad \quad \quad
-20756\, \, t^3 \,\, +6914\, \, t^2 \,\, \, +4\, \, t \,\, -1,
\nonumber   \\
  \hspace{-0.98in}&& \, \, \quad \quad\quad \quad  \quad 
  p_{4} \, \, = \, \, \,
  3584\, \, t^4 \, +5312\, \, t^3 \, -5307\, \, t^2 \, -10\, \, t  \,\, +5.
\end{eqnarray}
Actually (\ref{morenaturally}) provides~\cite{Holonomy} an
{\em infinite number of algebraic functions for selected
  values of} $\, \lambda$, namely  $\, \lambda \, = \, \mathrm{\cos}(\pi \, m/n)$,
or $\, M \, = \, 4 \cdot \, \mathrm{\sin}^2(\pi \, m/n)$, with $\, m$ and $\; n$ integers.

\subsection{The $\, g_n$'s are, at first sight, DD-finite}
\label{Atfirstsight}

The form factor expansion  (\ref{form11fact}) is well-suited~\cite{Holonomy}
to analyse the deformation of the $\, (1\, -t)^{1/4}$
algebraic solution of the sigma-form of Painlev\'e VI equation (\ref{jmequation}).
We underlined in~\cite{Holonomy}
the fact that all the form factors
$\, f_{1, \, 1}^{(2\, n)}$ are D-finite (polynomials in $\, E$ and $\, K$).

Let us now see the series expansion (\ref{morenaturally}) as a (one-parameter)
deformation (\ref{morenaturallyM1}) of the
$\, C(1,1) \, = \, E \, $ low-temperature exact expression: 
\begin{eqnarray}
\label{morenaturallyM}
\hspace{-0.98in}&& \, \, \,   \, \,     \quad  \quad    
C_{M}(1, \, 1; \, M) \, \, = \, \, \, \,\,
E \, \, \, \,   + M \cdot g_1(t)\,  \, + M^2 \cdot \, g_2(t) \,
\,\,  + M^3 \cdot \, g_3(t) \,\,\, \,  + \, \, \cdots 
\end{eqnarray}
At first sight these $\, g_n(t)$'s {\em have no reason to be D-finite}. The
series expansion of $\, g_1(t)$ reads:
\begin{eqnarray}
\label{g1}
  \hspace{-0.98in}&&  \quad \quad  \quad  \quad    
 g_1(t) \, \, = \, \, \, -{{3 } \over { 256}} \cdot \, t^2 \, \, \,
 -{{ 9} \over {1024 }} \cdot \, t^3 \, \, \,
 - {{441 } \over { 65536}} \cdot \, t^4 \, \,  \, \,
 - {{ 1407} \over { 262144}} \cdot \, t^5
 \nonumber \\
  \hspace{-0.98in}&& \quad   \quad     \quad    \quad \quad  \quad                    
  \, -{{9281 } \over { 2097152}}  \cdot \, t^6 \,  \, \,
  -{{ 31405} \over { 8388608}}\cdot \, t^7
  \, \,\,  - {{13877397 } \over { 4294967296}} \cdot \, t^8
   \,\,  \,  \,   \, \, + \cdots 
\end{eqnarray}
Inserting   (\ref{morenaturallyM}) in the sigma form of Painlev\'e VI  non-linear ODE
(\ref{jmequation}) (with $\, \sigma$ defined by (\ref{sigmam})),
one gets straightforwardly, at the first order in $\, M$, that $\, g_1(t)\, $
is {\em DD-finite}\footnote[1]{A D-finite function
  is a function solution of a linear ODE with polynomial coefficients. A DD-finite function
  is a  function solution of a linear differential equation whose coefficients are
  D-finite functions~\cite{DDFinite}.}~\cite{DDFinite}:
it is solution of an order-three linear differential operator $\, {\cal L}_3\, $
with coefficients {\em that are themselves D-finite}
(they are polynomials of hypergeometric $\, _2F_1 \, $ functions).
This order-three linear differential  operator is of the form
$\,  {\cal L}_3 \, \, = \, \, \,  {\cal L}_1 \cdot \, L_E \, $
where the order-two linear differential operator $\, L_E$ is
the operator annihilating the complete elliptic integral of the
second kind $\, E$, and 
where the order-one DD-finite operator $\,  {\cal L}_1 \, $ reads: 
\begin{eqnarray}
\label{DDfiniteEK}
\hspace{-0.98in}&& \, \, \, \quad \quad 
{\cal L}_1    \, \, = \, \, \, \,  \,
K^3 \cdot \, (t-1)^2 \cdot \, \Bigl( 2 \cdot \, (t -1)  \cdot \, t\,  D_t \, +5\, t -3 \Bigr)
\nonumber \\
\hspace{-0.98in}&& \quad \quad \quad \quad 
-E \, K^2 \cdot \, (t-1) \cdot \,
\Bigl( 4 \cdot  \, (t -1)  \cdot \, (t-2) \cdot \, t \, D_t \, \, \,  +10\, t^2-27\, t+13 \Bigr)
 \nonumber \\
  \hspace{-0.98in}&&\quad  \quad \quad \quad 
  -K \, E^2 \cdot \, (t-1) \cdot \,
  \Bigl( 10 \cdot \,  (t -1) \cdot \, t \, D_t \, \, +26 \, t  -17 \Bigr)
 \nonumber \\
  \hspace{-0.98in}&&\quad  \quad \quad \quad 
  +E^3 \cdot \,
  \Bigl( 2 \cdot  \, (t -1) \cdot \, (t-2) \cdot \, t \,  D_t  \, \,  +3\, t^2 -14\, t +7  \Bigr )
 \nonumber \\
  \hspace{-0.98in}&& \quad \quad \quad  \quad  \quad 
 \, \, = \, \, \, \,  \,
   2\, \cdot \, \Bigl( (t-2) \cdot \, E^3 \, \,  -5 \cdot\, (t-1) \cdot \, K\, E^2 \, \, \, 
\nonumber \\
\hspace{-0.98in}&& \quad \quad \quad \quad \quad \quad 
\, \, \, -2\, \, (t-1) \cdot \, (t-2)  \cdot  \, E\, K^2
\, \, +\, (t-1)^2  \cdot  \, K^3 \Bigr) \cdot \, (t-1) \cdot \, t  \cdot \,  D_t
\nonumber \\
\hspace{-0.98in}&& \quad \quad \quad \quad \quad \quad \quad \quad 
+\, (t-1)^2 \cdot \, (5\, t-3) \cdot \, K^3 \, \,
- \, (t-1) \cdot \, (10\, t^2-27\, t+13) \cdot \, E \, K^2
\nonumber \\
\hspace{-0.98in}&&\quad \quad \quad \quad \quad \quad \quad \quad 
\, \, -\, (t-1) \cdot \, (26\, t-17) \cdot \, K \, E^2
\, \, + \, (3\, t^2-14\, t+7)   \cdot \, E^3.            
\end{eqnarray}
At first sight $\, g_1(t)$ is  DD-finite and one easily verifies
that the series expansion (\ref{g1}) is actually solution of 
the order-three  DD-finite linear differential
operator  $\,\,\,  {\cal L}_3 \, \, = \, \, \,  {\cal L}_1 \cdot \, L_E$. Could
it be possible that $\, g_1(t)$ is, in fact, D-finite ?

\vskip .2cm

\section{The $\, g_n(t)$'s are D-finite}
\label{actuallydfinite}

In order to see that the $\, g_n(t)$'s are  D-finite, let us recall that
there actually exists an {\em exact closed expression}~\cite{Saga} for the lambda extension
$\, C(1,1; \lambda)$. This requires to rewrite everything in terms
of the {\em nome}~\cite{IsingCalabi} variable $\, q\, $ and use extensively
Jacobi theta functions.  This
exact expression has been given in equation (98) of~\cite{Saga}:
\begin{eqnarray}
 \label{Saga}
  \hspace{-0.98in}&& \quad 
 C_{-}(1, \, 1; \, \lambda)  \, \, \, = \, \, 
 {{ -\, \theta_2'(u, \, q)} \over { \sin(u) \cdot \, \theta_2(0, q) \cdot \, \theta_3(0, q)^2 }}
 \quad \quad  \,  \, 
 \hbox{where:} \quad   \,  \,   \quad
 \lambda \, \, = \, \,  \, \mathrm{\cos}(u),                 
\end{eqnarray}
where $\, \,\theta_2'(u, \, q)\,\, $ denotes the partial derivative of  $\, \,\theta_2(u, \, q)\,\, $
with respect to $\, u$.
This exact expression, when rewritten in terms of the $\, t$ variable, is, at first sight, 
a {\em differentially algebraic function}\footnote[2]{A differentially algebraic function~\cite{Tutte} is a function
  $\, f(t)$ solution of a polynomial relation $\, P(t, \, f(t), \, f'(t), \, \cdots \, f^{(n)}(t)) \, = \, \, 0$,
where $\, f^{(n)}(t)$ denotes the $n$-th derivative of  $\, f(t)$ with respect to $\, t$.}.
Let us write (\ref{Saga}) as
\begin{eqnarray}
 \label{Sagaother}
 \hspace{-0.98in}&&
 {{ f(u) } \over { \sin(u) \cdot \, \theta_2(0, q) \cdot \, \theta_3(0, q)^2 }}
 \, \,   \,  \, \, 
 \hbox{where:} \,  \,  \,   \, \, 
 f(u) \, \, = \, \, \, -\, \theta_2'(u, \, q),  \quad
 \sin(u) \, = \, \, \Bigl({{M} \over {4}}\Bigr)^{1/2}, 
\end{eqnarray}
where $\, M$ is defined by (\ref{correspondence}), 
and let us perform the Taylor expansion\footnote[2]{One has, at first sight,
  a Puiseux series in $\, M^{1/2}$ but all the coefficients
  for $\, M^{-1/2}$, $\, M^{1/2}$, $\, M^{3/2}$, ...
  here are equal to zero because all the even derivative
  $\, f^{(2\, n)}(0)$  are equal to zero.}
of $\,\, f(u)/\sin(u) \,\,$ in $\, M$:
\begin{eqnarray}
\label{SagaTaylor}
\hspace{-0.98in}&& \quad 
 {{ f\Bigl( \mathrm{\arcsin}( (M/4)^{1/2} )\Bigr) } \over { (M/4)^{1/2}}}
 \, \, = \,  \, \, \,\,\,   f^{(1)}(0)
 \,\,\,   \,  +  {{1} \over {24}} \cdot \, \Bigl(f^{(3)}(0) \, +  f^{(1)}(0) \Bigr) \cdot \, M 
\nonumber \\
  \hspace{-0.98in}&& \quad \quad 
  \, + {{1} \over {1920}} \cdot \,
  \Bigl(f^{(5)}(0) \, +  10 \, f^{(3)}(0)\, +  9 \, f^{(1)}(0)\Bigr) \cdot \, M^2
 \\
  \hspace{-0.98in}&& \quad \quad \quad 
 \, +  {{1} \over {322560}} \cdot \,
 \Bigl(f^{(7)}(0) \, +  35 \, f^{(5)}(0)\, +  259 \, f^{(3)}(0) \, +  225 \cdot \,  f^{(1)}(0)\Bigr)
 \cdot \, M^3
 \, \, \, \, \, + \, \, \cdots \nonumber
\end{eqnarray}
where $\, f^{(n)}(u)$ denotes the $\, n$-th derivative\footnote[5]{Note, in this Taylor series
(\ref{SagaTaylor}), that the terms  corresponding to even derivatives $\, f(0)$,  $\, f^{(2)}(0)$, ...,
$\, f^{(2\, n)}(0)$, are identically zero, since the odd derivatives of $\, \theta_2(u, \, q)$
with respect to $\, u$ vanish:  $\, \theta_2^{(2\, n\, +1)}(u, \, q) \, ) \, =\, 0$.} 
of $\, f(u)$ (with respect to $\, u$).
From this Taylor expansion (\ref{SagaTaylor}) one gets the following exact expressions
for $\, g_1(t)$, $\, g_2(t)$, etc ... (and even the first term $\, g_0(t) \, = \, E$):
\begin{eqnarray}
 \label{Sagag1}
  \hspace{-0.98in}&&
g_0(t) \, \, = \, \,\, E \, \, = \,\, \, \,
   -\,{{\theta_2^{(2)}(0, q)  } \over {\theta_2(0, q) \cdot \, \theta_3(0, q)^2  }},
\nonumber \\   
\hspace{-0.98in}&&                
g_1(t) \, \, = \, \,
 -{{1} \over {24}} \cdot \, {{ \theta_2^{(4)}(0, \, q) \, +\theta_2^{(2)}(0, \, q)
  } \over {\theta_2(0, q) \cdot \, \theta_3(0, q)^2  }},
   \nonumber \\   
  \hspace{-0.98in}&&
g_2(t) \, \, = \, \, \,
 -{{ 1} \over { 1920}} \cdot \,
  {{ \theta_2^{(6)}(0, \, q) \, +10 \cdot \, \theta_2^{(4)}(0, \, q)
 \, +9 \cdot \,  \theta_2^{(2)}(0, q) } \over { \theta_2(0, q) \cdot \, \theta_3(0, q)^2 }}, 
\\   
\hspace{-0.98in}&& 
g_3(t) \, \, = \, \, -{{ 1} \over { 322560}} \cdot \,
{{ \theta_2^{(8)}(0, \, q) \,  + 35 \cdot \, \theta_2^{(6)}(0, \, q)
\, +259 \cdot \, \theta_2^{(4)}(0, \, q)
 \, +225 \cdot \,  \theta_2^{(2)}(0, q) } \over { \theta_2(0, q) \cdot \, \theta_3(0, q)^2 }}, 
\nonumber \\   
  \hspace{-0.98in}&&
 g_4(t)  = 
-{{ 1} \over {92897280}} \cdot \,
{{ N_4 } \over { \theta_2(0, q) \cdot \, \theta_3(0, q)^2 }},
\, \, \, 
g_5(t)  = 
-{{ 1} \over {40874803200}} \cdot \,
{{ N_5 } \over { \theta_2(0, q) \cdot \, \theta_3(0, q)^2 }},
\nonumber 
\end{eqnarray}
where
\begin{eqnarray}
\label{Sagag1where}
 \hspace{-0.98in}&& \quad \quad
N_4 \, \, = \, \, \,\,
\theta_2^{(10)}(0, \, q) \, \, \,  +84 \cdot \, \theta_2^{(8)}(0, \, q)
\, \, \,  + 1974 \cdot \, \theta_2^{(6)}(0, \, q)
\nonumber \\   
\hspace{-0.98in}&&\quad \quad \quad \quad \quad  \quad \quad \quad \quad  
 \, +12916 \cdot \, \theta_2^{(4)}(0, \, q) \,\,  +11025 \cdot \,  \theta_2^{(2)}(0, q),
 \nonumber \\   
\hspace{-0.98in}&& \quad \quad
 N_5 \, \, = \, \, \,\,
 \theta_2^{(12)}(0, \, q) \,\,  \,  +165 \cdot \, \theta_2^{(10)}(0, \, q)
 \, \, \,  +8778 \cdot \, \theta_2^{(8)}(0, \, q)
\, \,  + 172810 \cdot \, \theta_2^{(6)}(0, \, q)
 \nonumber \\   
 \hspace{-0.98in}&& \quad \quad\quad \quad \quad \quad  \quad \quad \quad
 \, +1057221\cdot \, \theta_2^{(4)}(0, \, q)
\, \, \, +893025 \cdot \,  \theta_2^{(2)}(0, q),           
\end{eqnarray}
and where $\, \, \theta_2^{(2 \, n)}(u, \, q)\, $
denotes the $ \, (2\, n)$-th partial derivative of
$\, \, \theta_2(u, \, q)\, $ with respect to $\, u$.

Let us recall that {\em ratios} of D-finite expressions are {\em not} (generically\footnote[1]{The denominator
must not be an algebraic function.} ...)
D-finite: they are  {\em differentially algebraic}~\cite{Tutte}. Section (\ref{Atfirstsight})
suggests that the $\, g_n(t)$'s are {\em DD-finite} (or DDD-finite, ...): the previous expressions (\ref{Sagag1}) of
the  $\, g_n(t)$'s as {\em ratio} of derivatives of theta functions confirms this prejudice.
On the other hand, all these  $\, g_n(t)$'s are
{\em globally bounded series}~\cite{ChristolDiag} (see (\ref{g1})), 
and we have seen, so many times in physics, and in particular the two-dimensional
Ising model,  the emergence of globally bounded series as
a consequence of the frequent occurrence of
{\em diagonals of rational functions}~\cite{ChristolDiag,2F1,Heun,From} (or $\, n$-fold
integrals~\cite{Heegner,Khi5,ze-bo-ha-ma-05b,bo-gu-ha-je-ma-ni-ze-08,diagsuscept,bo-bo-ha-ma-we-ze-09,IsingCalabi,IsingCalabi2}). This
may suggest, on the contrary, that the $\, g_n(t)$'s could be D-finite. 

\vskip .1cm

\subsection{Expansions of the $\, g_n(t)$'s in the $\, t$ variable}
\label{gnexpansion}
From the previous exact expressions (\ref{Sagag1}) in terms of theta functions, one can obtain
the series expansions of the  $\, g_n(t)$'s in the $\, t$ variable and try to see
if these $\, g_n(t)$'s are  solutions of linear differential operators. 

From these expansions (\ref{Sagag1}), rewritten in $\,t$, one can get large enough series
in $\, t$ to see that $\, g_1(t)$ is in fact solution of an order-six
linear differential operator $\, L_6 \,$ which is actually the direct sum (LCLM) of an order-four
linear differential operator $\, L_4$ and of the order-two linear differential operator $\, L_E$
having $\, E \, = \,  _2F_1([{{1} \over {2}}, \, -{{1} \over {2}}], \, [1], \, t)$
as a solution. Furthermore one finds that this order-four linear differential
operator  $\, L_4$ is homomorphic to the {\em symmetric third power} of this order-two
linear differential operator $\, L_E$, with an intertwiner reading:
\begin{eqnarray}
\label{R1}
\hspace{-0.98in}&&
{{3} \over {8}} \, R_1 \, \, = \, \,\,
 \\
\hspace{-0.98in}&&  \, \,\,    \, \, = \, \,\, 
(t-1) \cdot \, t^3 \cdot \, D_t^3 \, \, \,
+{{3} \over {2}}\cdot \, (t-1) \cdot \, t^2 \cdot \, D_t^2 \,\,
\,  -{{1} \over {4}} \cdot \, (3\, t +1) \cdot \, t \cdot \, D_t
 \, \, \, + {{3} \over {8}} \cdot  \, {{t^2 +1} \over {t-1}}.
 \nonumber 
\end{eqnarray}
One finally finds that the series expansion (\ref{g1}) is exactly the linear combination
of  $\, E$ and the order-three linear differential  operator (\ref{R1})
acting on $\, E^3$:
\begin{eqnarray}
\label{gg1}
\hspace{-0.98in}&&
\quad   \quad \quad \quad\quad  \quad
g_1(t) \, = \, \,\,
{{1} \over {24}} \cdot \, E \,\, + \, {{1} \over {24}} \cdot \,  R_1(E^3)
\nonumber \\   
\hspace{-0.98in}&& \quad \quad \quad \quad \quad\quad \quad \quad
\, = \, \,\, \,  {{1} \over {24}} \cdot \, E \, \,\,\,
- \,   {{1} \over {8}} \cdot \, K \, E^2 \, \,\,
   - \,   {{ t \, -1 } \over {12}} \cdot \,  K^3.
\end{eqnarray}
Similar calculations can be performed for $\, g_2(t)$.   The series $\, g_2(t)$
can also be seen to be D-finite, being
solution of an order-{\em twelve} linear differential operator
which turns out to be the direct-sum (LCLM) of the
previous order-two linear differential operator $\, L_E$, of the previous
order-four $\, L_4$, and of an order-six linear differential
operator homomorphic to the {\em symmetric fifth power}
of $\, L_E$ with the following order-five intertwiner:
\begin{eqnarray}
\label{R2}
\hspace{-0.98in}&&\quad \quad 
-{{5} \over {8}} \, R_2 \,\, \, = \, \, \, \,
{{4} \over {3}} \cdot \, (t-1)  \cdot \, (t-2) \cdot \, t^5 \cdot \, D_t^5 \, \,
\, \, +\, {{5} \over {2}} \cdot  \,  (t-1) \cdot \, (4\,t -9)\cdot \, t^4 \cdot \, D_t^4
\nonumber \\
\hspace{-0.98in}&& \quad \, \, \quad \quad 
\, \, \, +5  \cdot \, (2\,t \, -3)  \cdot \, (t \, -3) \cdot  \, t^3 \cdot \, D_t^3
 \, \,\,  -{{5} \over {24}} \, {{24\, t^3 \, -122\, t^2 \, +59\, t \, +103
 } \over { t-1}} \cdot \,t^2 \cdot \, D_t^2
\nonumber \\
\hspace{-0.98in}&& \quad\, \,  \quad \quad \quad \quad 
\, + {{1} \over {24}}  \cdot \, {{ 90\,t^4 -488\, t^3-7\,t^2 \, +774\, t \, -1
 } \over {(t \, -1)^2}} \cdot \, t \cdot \, D_t
\nonumber \\
\hspace{-0.98in}&&    \, \,  \quad \quad \quad \quad \quad \quad  \quad  \quad 
\, -{{5} \over {96}} \cdot \, {{36\,t^5 \, -205\,t^4-59\,t^3 \, +409\,t^2 \, +23\, t \, -12
} \over {(t \, -1)^3}}.
\end{eqnarray}
One finally finds that the series expansion for $\, g_2(t)$ is exactly the
linear combination of  $\, E$, of the order-three linear differential  operator (\ref{R1})
acting on $\, E^3$, and of
the order-five linear differential  operator (\ref{R2}) acting on $\, E^5$:
\begin{eqnarray}
\label{gg2}
\hspace{-0.98in}&&
\quad   \quad \quad \quad 
g_2(t) \,\, = \, \,\,\,
{{3} \over {640}} \cdot \, E \,\,\, \,
+  {{1} \over {192}} \cdot \,  R_1(E^3)\, \,\,\,
 + {{1} \over {1920}} \cdot \,  R_2(E^5)
\nonumber \\   
\hspace{-0.98in}&& \quad \quad \quad  \quad \quad \quad \quad
\, = \, \,  \,\,
   {{3} \over {640}}  \cdot \, E \,\, \, \, \, \,
   - \, {{1} \over {64}}  \cdot \, E^2 \, K
   \, \, \, \, - \, {{t \, - 1} \over {96}}  \cdot \, K^3 
 \\   
  \hspace{-0.98in}&& \quad \quad \quad  \quad  \quad  \quad \quad \quad \quad
  \,  + {{1} \over {128}}  \cdot \, E^3 \, K^2  \,\,   \,
  +   {{t \, - 1} \over {64}}  \cdot \, E \, K^4 \,\,\,
  +  {{(t \, - 1) \, (t \, -2)} \over {240}}  \cdot \, K^5 .
 \nonumber 
\end{eqnarray}
Similar calculations can be performed for $\, g_3(t)$. They are displayed in \ref{g3}.

\vskip .1cm

{\bf Remark:} All these $\, R_1(E^3)$, $\, R_2(E^5)$,  ...
which are {\em homogeneous} polynomials in the complete elliptic integrals
$\, E$ and $\, K$,
{\em can be directly expressed in terms of simple ratios of theta functions}:
\begin{eqnarray}
\label{Rn}
 \hspace{-0.98in}&& \,  \quad \quad
 R_1(E^3)  \, \, = \, \, \,
 - \, {{ \theta_2^{(4)}(0, \, q) } \over { \theta_2(0, \, q) \cdot \, \theta_3(0, \, q)^2 }},
\quad \quad  \quad 
R_2(E^5)  \, \, = \, \, \,
-  \,{{ \theta_2^{(6)}(0, \, q) } \over { \theta_2(0, \, q) \cdot \, \theta_3(0, \, q)^2 }},
   \nonumber \\   
\hspace{-0.98in}&& \quad \quad\, 
R_3(E^7)  \, \, = \, \, \,
  -  \,{{ \theta_2^{(8)}(0, \, q) } \over { \theta_2(0, \, q) \cdot \, \theta_3(0, \, q)^2}},
\quad \quad  \quad \cdots 
\end{eqnarray}
where $\, \theta_2^{(n)}(u, \, q) \, $ denote the $\, n$-th derivative of
  $\, \theta_2(u, \, q)$ with respect to $\, u$.

\vskip .1cm

One can conjecture the following expression for (\ref{morenaturally}):
\begin{eqnarray}
\label{morenaturally2}
\hspace{-0.98in}&& \,  \quad \quad 
C_{M}(1, \, 1; \, M) \, \, = \, \, \,\,
E \, \, \, + \, M \cdot (c_1^{(1)} \cdot \, E \,  +  c_2^{(1)} \cdot \, R_1(E^3))
\nonumber \\
\hspace{-0.98in}&& \quad \quad \quad  \quad
  \, + \, M^2 \cdot ( c_1^{(2)} \cdot \, E \, +  c_2^{(2)} \cdot \, R_1(E^3)\,
  +   c_3^{(2)} \cdot \, R_2(E^5))
\nonumber \\
\hspace{-0.98in}&& \quad \quad \quad \quad
  \, + \, M^3 \cdot ( c_1^{(3)} \cdot \, E + \,  c_2^{(3)} \cdot \, R_1(E^3)
                   + \,  c_3^{(3)} \cdot \, R_2(E^5)  + \,  c_4^{(3)} \cdot \, R_3(E^7))
                   \, \,  \, \, \, + \, \, \cdots
 \nonumber \\
\hspace{-0.98in}&& \quad \quad \quad
\, \, = \, \, \,\,
(1\, + c_1^{(1)} \cdot \, M \, + c_1^{(2)} \cdot \, M^2
\, + c_1^{(3)} \cdot \, M^3 \, + \, \cdots  ) \cdot \,      E
\nonumber \\
  \hspace{-0.98in}&& \quad \quad \quad \quad
  \,\,  +   ( c_2^{(1)} \cdot \, M \, + c_2^{(2)} \cdot \, M^2
  \, + c_2^{(3)} \cdot \, M^3 \, + \, \cdots ) \cdot \,  R_1(E^3)
   \\
  \hspace{-0.98in}&& \quad \quad \quad \quad 
  \,\,  +  ( c_3^{(2)} \cdot \, M^2 \, + c_3^{(3)} \cdot \, M^3
  \, + \, \cdots ) \cdot \,  R_2(E^5)
  \, \,  \, \, +  ( c_4^{(3)} \cdot \, M^3 \, + \, \cdots) \cdot \,  R_3(E^7)
  \nonumber \\
  \hspace{-0.98in}&& \quad \quad \quad \quad \quad \quad \quad \quad \quad \quad
 \,\,\, + \, \, \, \cdots
\nonumber
\end{eqnarray}
where the $\, c_i^{(j)}$'s are constants obtained
from equations (\ref{Sagag1}) and (\ref{Rn}) (see (\ref{gg1}), (\ref{gg2})). One
can encapsulate these results in the following closed formula, deduced from (\ref{Saga})
and its Taylor expansion (see also (\ref{Sagaother})): 
\begin{eqnarray}
\label{encapsulate}
\hspace{-0.98in}&& \,  \quad \quad \quad 
C_{M}(1, \, 1; \, M) \, \, = \, \, \,\,
    - {{2} \over {\sqrt{M}}} \cdot  \,
  {{ \theta_2'\Bigl(\mathrm{\arcsin} {{\sqrt{M}} \over {2}}, \, q\Bigr)} \over { \theta_2(0, q) \cdot \, \theta_3(0, q)^2 }}
\nonumber \\
 \hspace{-0.98in}&& \,  \quad \quad  \quad \quad 
 \, \, = \, \, \,\,  -\, {{2} \over {\sqrt{M}}} \cdot  \,
 \sum_{p=0}^{\infty} \, \, \Bigl(\mathrm{\arcsin} {{\sqrt{M}} \over {2}}\Bigr)^{(2\, p \, +1)} \cdot \,
  {{ \theta_2^{(2 \, p \, +2)}(0, \, q) } \over {  \theta_2(0, \, q) \cdot \, \theta_3(0, \, q)^2 \cdot \,   (2\, p \, +1)!}}
\nonumber \\
 \hspace{-0.98in}&& \,  \quad \quad  \quad \quad 
 \, \, = \, \, \,\,  {{2} \over {\sqrt{M}}} \cdot  \,
\sum_{p=0}^{\infty} \, \,  \Bigl(\mathrm{\arcsin} {{\sqrt{M}} \over {2}}\Bigr)^{(2\, p \, +1)}
  \cdot \, {{ R_p(E^{(2\, p \, +1)})} \over {(2\, p \, +1)!}}.         
\end{eqnarray}

\vskip .1cm
         
\section{lambda-extensions and globally bounded series}
\label{globallybounded}

Let us consider the series expansion (\ref{morenaturally})
for values of the parameter $\, M \, \ne 0$
not yielding the previous algebraic function series
(i.e. $\, M \, \ne \ 4 \cdot \, \sin^2(\pi m/n)$ where $\, m$ and $\, n$ are integers). These
series are\footnote[2]{They are solutions of a non-linear ODE,
  the sigma-form of Painlev\'e VI.} {\em differentially algebraic}~\cite{Tutte}: is
it possible that such series could be D-finite for selected values of $\, M$? 

Let us change $\, t \, $ into $\,\, 16 \, t \,$ in the series expansion
(\ref{morenaturally}). One gets the following expansion:
\begin{eqnarray}
\label{glob}
\hspace{-0.98in}&& \,  \, \, \, \, 
1 \, \, \, -4\, t \, -(12+3\, M)  \cdot \, t^2 \, \, \, -(80+36\, M) \cdot \, t^3
\,\, \,  -(700+441\, M) \cdot \, t^4
\nonumber \\
  \hspace{-0.98in}&& \quad
\, -(7056+5628\, M) \cdot \, t^5
\, \, \,   -(77616 +74248\, M -5\, M^2) \cdot \, t^6
\nonumber \\
  \hspace{-0.98in}&& \quad\,
  -(906048 +1004960\, M -220\, M^2) \cdot \, t^7 \,
  -(11042460 +13877397\, M-6255\, M^2) \cdot \, t^8
 \nonumber \\
 \hspace{-0.98in}&& \quad \,
 -(139053200 +194712812\, M-146500\, M^2) \cdot \, t^9
\nonumber \\
\hspace{-0.98in}&& \quad 
\, -(1796567344 +2767635832\, M  -3079025\, M^2) \cdot \, t^{10}
\, \, \, \,\, + \, \,\, \cdots 
\end{eqnarray}
One sees immediately that this (generically) {\em differentially  algebraic}
series provides, {\em for any integer} $\, M$,
{\em an infinite number  of series with integer coefficients}.
In fact one can see that the series expansion (\ref{morenaturally}) (or the
series expansion (\ref{glob})) is a  {\em globally bounded series}\footnote[1]{A series
with rational coefficients and non-zero radius of convergence
is a globally bounded series~\cite{ChristolDiag} if it can be recast into a series with integer
coefficients with one rescaling $\,\, t \, \rightarrow \, \, N \, t \,$
where $\, N$ is an integer.} when $\, M$
is {\em any rational number}. 
One thus obtains the quite puzzling result that an {\em infinite number of}
(at first sight ...) {\em differentially algebraic series} can be
{\em globally bounded series}.

\vskip .1cm

Quite often we see the emergence of {\em globally bounded series}~\cite{ChristolDiag} 
as solutions of  D-finite
linear differential operators, and more specifically as
{\em diagonals of rational functions}~\cite{ChristolDiag,2F1,Heun,From}
(this is related to the so-called Christol's conjecture~\cite{Christolconj}).
Along this line it is tempting to imagine that such  {\em globally bounded} situation
could correspond to cases where the globally bounded series are in fact D-finite.
If this is not the case, it will thus be tempting to imagine that such
{\em globally bounded} situation could correspond to particular ratio of D-finite
functions, namely ratio of diagonals of rational functions (or even rational functions
of diagonals).

\vskip .1cm

\subsection{The $\, M \, = \, 5$ case.}
\label{M5}

Let us restrict to simple integer values of $\, M$ and see whether the corresponding
globally bounded series (\ref{morenaturally}) are D-finite.

Let us  consider an integer $\, M$ different from $\, M=\, 0$ (the D-finite solution
$\, C(1,1)$), and different from $\,M=\,  1, \, 2, \, 3, \, 4$,
which correspond to algebraic functions.  For simplicity we will consider
the integer coefficient series (\ref{glob}) for $\, M=\, 5$.
The  $\, M= \, 5\, $ series  (\ref{glob})  reads:
\begin{eqnarray}
\label{globM5}
\hspace{-0.98in}&& \,\quad \quad \,\,\,
1 \,\,\, -4\, t\,\,\, -27\, t^2\,\,\, -260\, t^3\,\, \,-2905\, t^4\,\, \,-35196\, t^5
\,\,\, -448731\, t^6\,\,\, -5925348\, t^7
\nonumber \\
\hspace{-0.98in}&& \quad \quad \quad  \,\,
 -80273070\, t^8\,\, -1108954760\, t^9 \,\,-15557770879\, t^{10}\,\, -220998916404\, t^{11}
\nonumber \\
\hspace{-0.98in}&& \quad \quad \quad  \quad \quad 
\, -3171743667652\, t^{12}\,\, -45915042520880\, t^{13}
\, \, \, \, \, + \, \, \, \cdots    
\end{eqnarray}
One  finds that this series (\ref{globM5}) does not seem to be  D-finite:
one does not find any linear differential operator
even with a thousand coefficients. Let us recall the strategy we have used in~\cite{Tutte}:
we study the series with integer coefficients
modulo small increasing primes $\, p=3, \, 5, \, 7, \, 11, \, 13, \, \cdots$
and seek for the linear differential operator annihilating
these series modulo such a prime.

For the prime $\, p=\, 3 \, $ the series (\ref{globM5}) mod. 3
is solution of an order-one linear differential operator (of degree one in $\, t$):
\begin{eqnarray}
\label{globM5mod3}
\hspace{-0.98in}&& \,\quad \quad \quad \quad \quad \quad \quad \quad \quad \quad
2\, \, t \, \, \, +(t \, +2) \cdot \, t \, \, D_t.
\end{eqnarray}
For the prime $\, p=\, 5\,$ the series (\ref{globM5}) mod. $ 5\,$
is solution of an order-one linear differential operator (of degree two in $\, t$):
\begin{eqnarray}
\label{globM5mod5}
\hspace{-0.98in}&& \,\quad \quad \quad \quad \quad \quad \quad
3 \, \cdot \, t \cdot \,(t \, +1)\,\,  \, +(t^2 \, +2\, t \, +2) \cdot \, t \, \, D_t.
\end{eqnarray}
These series mod. 3 or 5, are not only D-finite, they are
in fact {\em algebraic series} mod. 3 or 5:
\begin{eqnarray}
  \label{algebraic}
  \hspace{-0.98in}&& p=\, 3,  \quad \quad  \quad 
  F^{16} \, \, +2 \cdot \, (t^2 +t +1 ) \cdot \,  F^8 \, \, + (t \, +2) \cdot \,   t^6
  \, \, \, = \, \, \, 0,
\\
 \hspace{-0.98in}&& p= \, 5, \quad \quad  \quad 
(t^2 +4 t  +1)^5 \cdot \,  F^4 \, \,  +4 \cdot \, (t \, +3)^4 \cdot \,  (t \, +4)^4
\, \,  \,= \, \, \, 0.
\end{eqnarray}

For the prime $\, p=\, 7$ the series (\ref{globM5}) mod. 7
is solution of an order-three linear differential operator
(of degree three in $\, t$):
 \begin{eqnarray}
 \label{globM5mod7}
\hspace{-0.98in}&& \,\quad 
 2 \, \cdot \, t \cdot \, (t \, +2) \,\,
+(9\, t^3 \, + 13\, t^2 \, +4\, t \, +9) \cdot \, t \, \, D_t\, \, 
 +(5\, t^3 \, + 16 \, t^2 \, +9 \, t \, +19) \cdot \, t^2 \, \, D_t^2
 \nonumber \\
  \hspace{-0.98in}&& \quad  \quad  \quad
 +(t^3 \, + 4 \, t^2 \, +3 \, t \, +6) \cdot \, t^3 \, \, D_t^3, 
 \end{eqnarray}
This mod. 7 series is also algebraic, but finding the corresponding
characteristic polynomial equation (like (\ref{globM5mod7}) previously)
requires more than one thousand coefficients.

For the next primes we get more and more involved linear differential operators
of increasing orders and degrees of the polynomials in $\, t$.  One
finds for the prime $\, p= \, 11$ an  order 5
and a degree in $\, t$ also equal to five, and one gets for the following primes 
\begin{eqnarray}
\label{prime}
\hspace{-0.98in}&&  \quad \quad
p\, = \, 13, \quad  order \, =  degree \, = \, 6,
\quad \quad \quad \,   \, \, 
p\, = \, 17, \quad  order \, =  degree \, = \, 8,
\nonumber \\
 \hspace{-0.98in}&& \quad \quad
 p\, = \, 19, \quad  order \, =  degree \, = \, 9,
 \quad \quad \quad  \, \, \, 
 p\, = \, 23, \quad  order \, =  degree \, = \, 11,
 \nonumber \\
 \hspace{-0.98in}&& \quad \quad
 p\, = \, 29, \quad  order \, =  degree \, = \, 14,
 \quad \quad \quad 
 p\, = \, 31, \quad  order \, =  degree \, = \, 15,
 \nonumber \\
 \hspace{-0.98in}&& \quad \quad
 p\, = \, 37, \quad  order \, =  degree \, = \, 18,
 \quad \quad \quad 
 p\, = \, 41, \quad  order \, =  degree \, = \, 20, \quad \cdots
 \nonumber 
\end{eqnarray}
An inspection of the corresponding linear differential operators
strongly suggests that the orders and degrees
of the polynomials in $\, t$ of the linear
differential operator grow (linearly) with the prime $\, p\, $
according to the formula:
\begin{eqnarray}
\label{primegener}
\hspace{-0.98in}&& \,\quad \quad \quad \quad \quad \quad \quad \quad \quad \quad
order \, =  \, degree \, = \, \,  \,\, {{ p \, -1} \over {2}}.
\end{eqnarray}
These results have to be compared with the same mod. prime calculations
for the D-finite (possibly algebraic)
series (\ref{glob}) for $\, M= \, 0, \, 1, \, 2, \, 3, \, 4$.
In that case, since there is a linear differential operator
(in characteristic zero), the series modulo a prime is solution of
the mod. prime reduction of that linear differential operator, however
for small primes the series modulo a prime can be solution of
a linear differential operator of smaller order (order one, ...). Therefore
the previous analysis modulo increasing primes
provides linear differential operators of increasing orders, but
very quickly saturating to the order of the  linear differential operator
in characteristic zero. 

These calculations, thus, strongly suggest that the integer coefficient series
(\ref{globM5}) is  {\em not D-finite} but
 is {\em only differentially algebraic}.

Similar calculations can be performed for any
integer $\, M \, \ge 5$ (or any integer $\, M\, \le -1$) with similar results.
Similar calculations can be performed {\em for any rational number}
$\, M$  with similar results ruling out D-finiteness.
Let us display miscellaneous algebraic equation for the series
for various $\, M$ and modulo various primes: 
\begin{eqnarray}
  \hspace{-0.98in}&& \,\quad \quad
  M=6,\,\quad p=3, \quad \quad \quad 
  (t^3 +1) \cdot \, F^{2} \, +2 \cdot \,  (t^2 +t +1 ) \, \, \, = \,\,  \, \, 0,
  \nonumber\\
  \hspace{-0.98in}&& \,\quad \quad
  M=7, \quad\, p=3, \quad \quad \quad  F^4 \, \, +(t \, +2) \, \, \, = \, \, \, \, 0,
  \nonumber\\
  \hspace{-0.98in}&& \,\quad \quad 
  M=7,\,\quad p=7, \quad \quad\quad 
  (t \, +1)^7 \cdot \, (t +3)^7\cdot \,  (t \, +5)^7 \cdot \,  F^6
   \nonumber\\
  \hspace{-0.98in}&& \,
  \quad \quad \quad \quad \quad  \quad \quad \quad \quad \quad  \quad  \quad \quad  \quad  \quad \, \,
  +6\, \cdot \,  (t \, +6)^6 \cdot \, (t^2 \, +2 t \, +5)^6
    \, \, \, = \, \, \, \, 0,
  \nonumber\\
  \hspace{-0.98in}&& \,\quad \quad
  M=11,\,\quad p=3, \quad \quad \quad  F^{16} \, \,
  +2\cdot \,  (t^2 +t +1 ) \cdot \,  F^8 \, \, + (t \, +2)  \cdot \, t^6
  \, \, = \, \,\,  \, 0.
  \nonumber
\end{eqnarray}

All these calculations suggest that the infinite number of
integer coefficient series (\ref{glob}), for any
integer $\, M \, \ge 5$ (or any integer $\, M\, \le -1$), are {\em not} D-finite,
as well as the infinite number of globally bounded series
(\ref{morenaturally}) or (\ref{glob}) 
when $\, M$ is any rational number,  thus providing an
{\em infinite set of globally bounded differentially algebraic series}
(far beyond the D-finite diagonals of rational functions~\cite{ChristolDiag,2F1,Heun,From}
providing so many globally bounded  series, see Christol's
conjecture~\cite{Christolconj}).

The question to see whether these globally bounded series
could be ratio of particular D-finite functions,
namely ratio of diagonals of rational
functions\footnote[2]{Or more generally rational functions of diagonals of rational functions.}
remains open.

\vskip .1cm

{\bf Remark:} Finding that a series is actually the  ratio of particular D-finite functions
can be a difficult task, possibly some tour-de-force, requiring a lot of (guessing) intuition.
Conversely, there are very few papers, in the literature, addressing the question of ruling out the
possibility that a series can be the  ratio of  D-finite functions, or even
ruling out the possibility that a series can be  DD-finite~\cite{DDFinite}. Here
we have a prejudice that the series (\ref{glob}) for integer values $\, M \, \ge \, 5$ are
not ratio of diagonals of rational functions, but we are not able to prove
such a no-go result, even for specific integer vales of $\, M$.

\vskip .1cm
   
\section{Other one-parameter deformations: deformations of algebraic functions}
\label{other}

The ``form factor'' expansion (\ref{form}) 
(see (9) in~\cite{Holonomy}) amounts to seeing the  lambda-extension of the correlation function
$\, C_{-}(N, \, N; \, \lambda) \, $
as a deformation of the algebraic solution $\, (1\, -t)^{1/4}$. With section (\ref{Algebraicsub}) 
we have seen that there are many other (algebraic) values of the parameter $\, \lambda \,$
for which the  lambda-extension 
$\, C_{-}(N, \, N; \, \lambda)$ becomes an {\em algebraic function}~\cite{Holonomy}.  Let us consider
``form factor'' expansions~\cite{Holonomy} similar to (\ref{form11fact}), but corresponding to
seeing the   lambda-extension  as a deformation around these other algebraic functions
(see (\ref{actually}),  (\ref{algM6}),  (\ref{algM3})).  

Recalling the exact expressions of the $\, g_n(t)$'s in terms of theta functions displayed
in (\ref{Sagag1}) and (\ref{Sagag1where}), it is worth noticing that similar expressions
can also be obtained for the form factors $\, f_{1,1}^{(2\, n)}$. One gets respectively
(with\footnote[1]{Note that
$\, (1\, -t)^{1/4} = \, \theta_4(0, q)/\theta_3(0, q)$
 with
$ \,  \, \theta'_1(0, q) =\, \theta_2(0, q)\, \theta_3(0, q)\, \theta_4(0, q)$. } $\,f_{1,1}^{(0)} = \, 1$):
\begin{eqnarray}
\hspace{-0.98in}&& \, \,   \quad  \quad 
(1\, -t)^{1/4} \cdot  f_{1,1}^{(0)}  \,  \,  = \, \,  \, 
  {{ \theta^{(1)}_1(0; q)} \over {\theta_2(0, q) \cdot \, \theta_3(0, q)^2  }}, 
                   \nonumber \\
 \label{f11theta}
  \hspace{-0.98in}&& \, \,  \quad  \quad 
 (1\, -t)^{1/4} \cdot  f_{1,1}^{(2)}  \,  \,  = \, \,  \, 
 {{1} \over {2}} \cdot \,
 {{ \theta^{(3)}_1(0, q) \, + \theta^{(1)}_1(0, q) } \over {
        \theta_2(0, q) \cdot \, \theta_3(0, q)^2 }}, 
\end{eqnarray}
\begin{eqnarray}
\label{f11thetasuite}
\hspace{-0.98in}&& \, \,\quad    \quad 
(1\, -t)^{1/4} \cdot  f_{1,1}^{(4)}  \,  \,  = \, \,  \, 
 {{1} \over {24}} \cdot \,  
 {{ \theta^{(5)}_1(0, q) \, + 10 \cdot \, \theta^{(3)}_1(0, q)
\, + 9 \cdot \, \theta^{(1)}_1(0, q) } \over {  \theta_2(0, q) \cdot \, \theta_3(0, q)^2  }}, 
  \nonumber \\
\hspace{-0.98in}&& \, \,  \quad   \quad 
(1\, -t)^{1/4} \cdot  f_{1,1}^{(6)}  \, \,  = \, \,   \,
 \nonumber \\
  \hspace{-0.98in}&& \, \,  \quad     \quad   \quad      \quad   
         \, \,  = \, \,   \,   {{1} \over {720}} \cdot \, 
{{\theta^{(7)}_1(0, q) \,   + 35 \cdot \, \theta^{(5)}_1(0, q) \, + 259 \cdot \, \theta^{(3)}_1(0, q) \,
   + 225 \cdot \, \theta^{(1)}_1(0, q)  } \over {\theta_2(0, q) \cdot \, \theta_3(0, q)^2  }},
                     \nonumber
\end{eqnarray}
\begin{eqnarray}
\label{f11thetasuitesuite}                    
  \hspace{-0.98in}&& \, \,  \quad \quad 
 (1\, -t)^{1/4} \cdot  f_{1,1}^{(8)}  \, \,  = \, \,   \,
        {{1} \over {40320}} \cdot \, 
           {{ {\cal N}_9  } \over {\theta_2(0, q) \cdot \, \theta_3(0, q)^2  }},
 \nonumber \\
  \hspace{-0.98in}&& \, \,  \quad \quad 
(1\, -t)^{1/4} \cdot  f_{1,1}^{(10)}  \, \,  = \, \,   \,
 {{1} \over {3628800}} \cdot \, 
   {{ {\cal N}_{11}  } \over {\theta_2(0, q) \cdot \, \theta_3(0, q)^2  }},
 \nonumber \\
  \hspace{-0.98in}&& \, \,  \quad \quad
 (1\, -t)^{1/4} \cdot  f_{1,1}^{(12)}  \, \,  = \, \,   \,
   {{1} \over {479001600}} \cdot \, 
   {{ {\cal N}_{13}  } \over {\theta_2(0, q) \cdot \, \theta_3(0, q)^2  }},
 \quad \quad  \quad \cdots 
 \nonumber     
\end{eqnarray}
where
\begin{eqnarray}
\label{f11thetasuitesuitewhere}                    
  \hspace{-0.98in}&& \, \,  \quad \quad 
{\cal N}_9 \, \, = \, \, \,
 \theta^{(9)}_1(0, q) \,  + 84 \cdot \,\theta^{(7)}_1(0, q) \,   + 1974 \cdot \, \theta^{(5)}_1(0, q)
  \nonumber \\
  \hspace{-0.98in}&& \, \, \quad \quad \quad  \quad  \quad \quad \quad \quad \quad \quad \quad 
 \, + 12916 \cdot \, \theta^{(3)}_1(0, q) \, + 11025 \cdot \, \theta^{(1)}_1(0, q),
\nonumber \\
  \hspace{-0.98in}&& \, \, \quad \quad 
{\cal N}_{11}  \, \, = \, \, \,   \theta^{(11)}_1(0, q) \, 
   +165 \cdot  \, \theta^{(9)}_1(0, q) \,  + 8778 \cdot \,\theta^{(7)}_1(0, q) \,   + 172810 \cdot \, \theta^{(5)}_1(0, q)
\nonumber \\
  \hspace{-0.98in}&& \, \, \quad  \quad \quad  \quad  \quad \quad \quad \quad \quad \quad \quad 
                     \, + 1057221 \cdot \, \theta^{(3)}_1(0, q) \, + 893025 \cdot \, \theta^{(1)}_1(0, q),
\nonumber  \\
  \hspace{-0.98in}&& \, \, \quad \quad 
{\cal N}_{13}  \, \, = \, \, \,   \theta^{(13)}_1(0, q) \,  + 286 \cdot \, \theta^{(11)}_1(0, q) \, 
   + 28743 \cdot  \, \theta^{(9)}_1(0, q) \,  + 1234948 \cdot \,\theta^{(7)}_1(0, q)
\nonumber \\
  \hspace{-0.98in}&& \, \, \quad  \quad \quad  \quad \quad 
\,   + 21967231  \cdot \, \theta^{(5)}_1(0, q)   \, + 128816766  \cdot \, \theta^{(3)}_1(0, q)
\, + 108056025 \cdot \, \theta^{(1)}_1(0, q),
 \nonumber 
\end{eqnarray}
and where $\, \theta^{(2\, n\, +1)}_1(u, q)\, $ denotes the $\, (2\, n\, +1)$-th partial derivative of
the Jacobi theta function $\, \theta_1(u, q)$ with respect to $\, u$. 
Let us remark that these terms can be obtained
similarly to  (\ref{Sagag1}) and (\ref{Sagag1where}),  using now the expansion of
$\, f(\mathrm{\arccos}(\lambda))/\sqrt{1 \, -\lambda^2} \, \, $
around $\, \lambda \, = \, \, 0$,  which corresponds to $\, u \, = \, \pi/2$, and, then, use
$\,\, \theta_2^{odd}(\pi/2, \, q) \, = \, \, -\, \theta_1^{odd}(0, \, q) \, $
and $\, \,\theta_2^{even}(\pi/2, \, q) \, = \, \, 0$.

\vskip .1cm
\vskip .1cm

{\bf Remark 1:} Similarly to (\ref{encapsulate}) one can encapsulate
the previous results in the following closed formula, deduced from (\ref{Saga})
and its Taylor expansion:
\begin{eqnarray}
\label{encapsulate2}
\hspace{-0.98in}&& \,  \,  \,   \quad \quad \quad
C_{-}(1, \, 1; \, \lambda) \, \, = \, \, \,\,
-  {{ \theta_2'\Bigl(\mathrm{\arccos} \, \lambda, \, q\Bigr)} \over {
    \sqrt{1 \, -\lambda^2} \cdot \, \theta_2(0, q) \cdot \, \theta_3(0, q)^2 }}
   \\
 \hspace{-0.98in}&& \, \quad  \quad \quad \quad
 \, \, = \, \, \,\,  \, {{1} \over {\sqrt{1 \, -\lambda^2}}} \cdot  \,
 \sum_{p=0}^{\infty} \, \, \Bigl(\mathrm{\arcsin}  \, \lambda\Bigr)^{(2\, p)} \cdot \,
     {{ \theta_1^{(2 \, p \, +1)}(0, \, q) } \over {
         \theta_2(0, \, q) \cdot \, \theta_3(0, \, q)^2 \cdot \,   (2\, p)!}}.
\nonumber
\end{eqnarray}

\vskip .1cm
{\bf Remark 2:} Introducing ratios of theta functions $S^{(2\, n \, +1)}$ by:
\begin{eqnarray}
  \label{Sp}
  \hspace{-0.98in}&& \, \, \, \,   \quad \quad  \quad \quad  \quad \quad \quad
  S^{(2\, n \, +1)}   \, \,  = \, \,   \, \,
 {  { \theta_1^{(2\, n+1)}(0, q) } \over  { \theta_1^{(1)}(0, q) } },
\end{eqnarray}
and the quantities $\,  \kappa^{(2\, n \, +1)}$'s related to the form factors
$ f_{1,1}^{(2 \, n)}$'s introduced in (\ref{form11fact}):
\begin{eqnarray}
  \label{kappa}
  \hspace{-0.98in}&& \, \, \, \,   \quad  \quad   \quad \quad  \quad \quad \quad
  f_{1,1}^{(2 \, n)}  \, \,  = \, \,   \,  ( 2\, n \, +1) \cdot \kappa^{(2\, n \, +1)},
\end{eqnarray}
one can deduce, from the previous relations (\ref{f11theta}),
the expression of the  $\,  S^{(2\, n \, +1)}   $'s in terms of these  $\,  \kappa^{(2\, n \, +1)}$'s:
\begin{eqnarray}
  \label{kappadeduce}
 \hspace{-0.98in}&&   \, \,  \quad  \quad
S^{(1)}     \, \,   = \, \,   \, \,\kappa^{(1)},
 \nonumber \\
  \hspace{-0.98in}&& \, \,  \quad  \quad
 {{S^{(3)}}\over{3!}}     \, \,   = \, \,   \, \,
 \kappa^{(3)} \,  - {{1} \over { 6 }} \cdot \,  \kappa^{(1)},
 \nonumber \\
  \hspace{-0.98in}&& \, \,   \quad   \quad
{{S^{(5)}}\over{5!}}     \, \,   = \, \,   \, \,
\kappa^{(5)} \,  -{{1} \over {2 }} \cdot \,  \kappa^{(3)} \, +{{1} \over {120 }} \cdot \,  \kappa^{(1)},
  \\
  \hspace{-0.98in}&& \, \,  \quad   \quad
{{S^{(7)}}\over{7!}}     \, \,   = \, \,   \, \,
 \kappa^{(7)} \,  -{{5} \over {6}} \cdot \,  \kappa^{(5)}
          \,   +{{13} \over {120 }}  \cdot \,  \kappa^{(3)} \, -{{1} \over {5040 }} \cdot \,  \kappa^{(1)},
 \nonumber \\
  \hspace{-0.98in}&& \, \,   \quad   \quad
{{S^{(9)}}\over{9!}}     \, \,  = \, \,   \, \,
 \kappa^{(9)} \,  -{{7} \over {6}} \cdot \,  \kappa^{(7)} \,   +{{23} \over {72 }}  \cdot \,  \kappa^{(5)}
    \, -{{41} \over {3024 }} \cdot \,  \kappa^{(3)}\, +{{1} \over { 362880 }} \cdot \,  \kappa^{(1)},
  \quad   \quad  \cdots \nonumber
\end{eqnarray}
The coefficients in these linear combinations (\ref{kappadeduce}) correspond exactly to the linear
combinations we had to introduce for
the ($n$-fold integrals) $ \, \tilde{\chi}^{(2\, n \, +1)}$'s in the  analysis of the susceptibility
of the square Ising model, see for instance
equation (8) in~\cite{Khi6}, {\em but in the high temperature regime}:
\begin{eqnarray}
\label{chi5}
  \hspace{-0.98in}&& \, \, \quad  \quad \quad  \quad \quad  \quad \quad  \quad
  \Phi^{(5)} \, \, \,   = \, \,   \, \,
    \tilde{\chi}^{(5)} \, \,  -{{1} \over {2 }} \cdot \, \tilde{\chi}^{(3)}  \, \, +{{1} \over {120 }} \cdot \,  \tilde{\chi}^{(1)}.                     
\end{eqnarray}
Along these lines we give, in \ref{Appendixchin}, a Taylor expansion similar to (\ref{encapsulate2})
but for the lambda extension of $\, C(0,0, \, \lambda)$, instead of $\, C(1,1, \, \lambda)$ in  (\ref{encapsulate2}).
From these expansions one deduces linear combinations (\ref{encapsulate72}) (similar to (\ref{kappadeduce})),
corresponding exactly to the linear combinations we had to introduce for
the  ($n$-fold integrals) $ \, \tilde{\chi}^{(2\, n)}$'s in the analysis of
the susceptibility of the square Ising model, see for instance
equation (26) in~\cite{Khi6}, {\em in the low temperature regime}:
\begin{eqnarray}
\label{chi6}
  \hspace{-0.98in}&& \, \, \quad  \quad \quad  \quad \quad  \quad \quad  \quad
  \Phi^{(6)} \, \, \,   = \, \,   \, \,
    \tilde{\chi}^{(6)} \,  -  {{ 2  } \over {3}} \cdot \, \tilde{\chi}^{(4)} \, +{{ 2 } \over {45}} \cdot \, \tilde{\chi}^{(2)}.                
\end{eqnarray}

\vskip .1cm

\subsection{Other one-parameter deformations: deformation of $\, M \, = \, 2$ (i.e. $u \, = \pi/4$).}
\label{otherM2sub}

Recalling that
 one finds that (\ref{morenaturally}) is actually, for $\, M\, = \, 2$, the series expansion
 of an algebraic function (\ref{actually}), one can try to write
 the series (\ref{morenaturally}) as a deformation of this
 $\, M\, = \, 2 \, $ algebraic function (\ref{actually}):
\begin{eqnarray}
\label{actually2}
\hspace{-0.98in}&& \, \, \quad  \quad  \quad   \quad  \quad   \quad 
C_{\rho}(1, 1; \rho) \,  \, = \, \,\,  \, \, G_0(t)
\, \, \, \, +   \rho \cdot \, G_1(t) \,\,\, \, +  \rho^2 \cdot \, G_2(t)
\, \,\, \,\,  + \, \,\, \,  \cdots
\end{eqnarray}
where
\begin{eqnarray}
\label{actually2G0}
\hspace{-0.98in}&& \, \,  \quad   \quad \quad  
G_0(t) \,  = \, \,  \,  \,
 (1-t)^{1/16} \cdot \, \Bigl( {{1 \, \, +(1-t)^{1/2}} \over {2}} \Bigr)^{3/4}
\\
\hspace{-0.98in}&&\quad   \quad \quad  \quad \quad 
\,  = \, \,  \,  \, \, 1 \, \,\,  \, - {{1 } \over {4}} \cdot \, t
\, \,  \, \, -{{9 } \over {128}} \cdot \, t^2 \, \, \, \, 
- {{19} \over {512}} \cdot \, t^3    \,  \,\, \,  - {{791 } \over {32768}} \cdot \,  t^4 \,\, \,
-{{2289 } \over {131072}} \cdot \, t^5
\nonumber \\
\hspace{-0.98in}&&\quad   \quad \quad \quad \quad \quad \quad \quad \quad
 \label{actually2G0bis}
\,  - {{ 56523 } \over {4194304}} \cdot \, t^6 \, 
\,\,  \,  - {{182193 } \over {16777216}} \cdot \, t^7
\, \, \, \, + \, \, \, \cdots 
\end{eqnarray}
and where $\, \rho \, = \, M \, -2$.

Let us introduce
\begin{eqnarray}
\label{actually2G0u}
\hspace{-0.98in}&& \, \,  \quad   \quad \quad \quad  \quad \quad \quad  \quad
G_0(t) \,  = \, \,  \,
-\, \sqrt{2} \cdot \,
{{ \theta^{(1)}_2(\pi/4, q)} \over {\theta_2(0, q) \cdot \, \theta_3(0, q)^2  }},        
\end{eqnarray}
which actually coincides with the algebraic expression (\ref{actually2G0}).
Let us also introduce the $\, S_n$'s defined as
\begin{eqnarray}
\label{Si}
\hspace{-0.98in}&& \, \, \quad \quad \quad \quad \quad \quad\quad \quad \quad  \, 
  S_n       \, \, = \, \, \, \,
  {{ \theta_2^{(n)}(\pi/4, \, q) } \over {\theta'_2(\pi/4, \, q) }},  
\end{eqnarray}
where $\, \theta_2^{(n)}(u, \, q)$ denotes the $\, n$-th partial
derivative with respect to $\, u$ of  $\, \theta_2(u, \, q)$.
Similarly to (\ref{encapsulate}) one can write (\ref{actually2}) as
\begin{eqnarray}
\label{encapsulate3}
\hspace{-0.98in}&& \,  \,  \,   \, \,  \,   \, 
 C_{\rho}(1, \, 1; \, \rho)
   \, \, = \, \, \,\,  \, {{\sqrt{2} \cdot \, G_0(t)} \over {\sqrt{\rho\, +2}  }}   \cdot  \,
 \sum_{p=0}^{\infty} \, \, 
 \Bigl( \mathrm{\arcsin} \Bigl( {{\sqrt{\rho\, +2}} \over {2}} \Bigr) \, -{{\pi} \over {4}} \Bigr)^{(p \, -1)} \cdot \,
    {{ S_p } \over {    (p \, -1)! }}.
\end{eqnarray}
Again one can ask whether
the $\, G_n(t)$'s in (\ref{actually2}) are D-finite, and, again, polynomials
in the complete elliptic integrals $\, E$ and $\, K$. One can find that
(\ref{actually2}), or (\ref{encapsulate3}), can be written as
\begin{eqnarray}
\label{actually2rho}
\hspace{-0.98in}&& \quad  
{{C_{\rho}(1, 1; \rho)} \over {G_0(t) }}  \,  \, \,  = \, \,\,  \,   1
\, \, \, \, +  \rho \cdot \, \Bigl({{1} \over {4}} \cdot \, S_2  \, \,  -{{1} \over {4}} \Bigr)
\,\,\,\,  + \rho^2 \cdot \, 
\Bigl( {{1} \over {32}} \cdot \, S_3
\,  \,- \, {{1} \over {16}} \cdot \, S_2   \,\, + {{3} \over {32}}   \Bigr)
\nonumber \\
 \hspace{-0.98in}&&  \quad \quad   \quad    \quad  
 +  \rho^3 \cdot \,
 \Bigl(   {{1} \over {384}} \cdot \, S_4  \, \, -  {{1} \over {128}} \cdot  \, S_3 \, \,
   +  {{13} \over {384}} \cdot  \, S_2  \,  \, - {{5} \over {128}}  \Bigr)
\nonumber \\
\hspace{-0.98in}&& \quad  \quad \quad     \quad  
 +   \rho^4 \cdot \, 
 \Bigl( {{1} \over {6144}} \cdot \, S_5  \, \, - {{1} \over {1536}} \cdot \, S_4 \, \,
 + {{17} \over {3072}} \cdot \, S_3 \, \, - {{19} \over {1536}} \cdot \, S_2 \,
 \, + {{35} \over {2048}} \Bigr) \, \,\, \,
\nonumber \\ 
\hspace{-0.98in}&&  \quad \quad  \quad     \quad  
 +    \rho^5 \cdot \,  \Bigl( {{1} \over {122880}} \cdot \, S_6  \, \,
 -{{1} \over {24576}} \cdot \, S_5  \, \, +{{7} \over { 12288}} \cdot \, S_4 \, \,
 - {{23} \over { 12288}} \cdot \, S_3 \, 
 \nonumber \\
 \hspace{-0.98in}&&  \quad   \quad  \quad
 \quad     \quad \quad    \quad     \quad      \quad
 \, +{{263} \over {40960}} \cdot \, S_2 \, \, - {{63} \over {8192}}   \Bigr)
\ \, \,\, \, \, \,\, \, + \, \,\, \,        \cdots
\end{eqnarray}
where the $\, S_n$'s are defined by (\ref{Si}).
It is crucial
to note that {\em all these ratio} (\ref{Si})
{\em are actually  polynomial expressions} in the complete elliptic integrals
$\, E$ and $\, K$. The first $\, S_n$'s  read:
\begin{eqnarray}
\label{actually2rho}
\hspace{-0.98in}&& 
S_2 \, \, = \, \, \, {{ 2} \over { t }} \cdot \,
\Bigl(1  \,\, -(1\, -t)^{1/2}\Bigr) \cdot \, E
\, \, \,  \,  - {{ 1} \over { 2\, t }} \cdot \,
\Bigl( (t\, -4) \cdot \, (1\, -t)^{1/2} \,  \, -(3\, t \, -4) \Bigr) \cdot \, K,
\nonumber \\  
\hspace{-0.98in}&&
S_3 \, \, = \, \, \, {{1} \over { 4}} \cdot \,
\Bigl( 6 \cdot \, (1\, -t)^{1/2} \, \,  -(t \, -2) \Bigr) \cdot \, K^2 \, \,  \, \, -3 \, E\, K, 
\nonumber
\end{eqnarray}
\begin{eqnarray}
\hspace{-0.98in}&&
S_4 \, \, = \, \, \,
{{3} \over { t}} \cdot \, \Bigl( (t\, -4) \cdot \, (1\, -t)^{1/2} \,\,  \, -(3\, t \, -4) \Bigr)
\cdot \, E \, K^2 \,\, \,
-{{ 6} \over { t }} \cdot \, (1 \, \, -(1\, -t)^{1/2}) \cdot \, E^2 \, K
\nonumber \\  
\hspace{-0.98in}&& \quad \quad  \quad \quad 
\, + \, {{1} \over { 8 \, t}} \cdot \, \Bigl( (t^2 \, -28\,t \, +48) \cdot \, (1\, -t)^{1/2}
\,  \, -(21\, t^2 \, -68\, t \, +48) \Bigr) \cdot \, K^3,  \,
\nonumber
\end{eqnarray}
\begin{eqnarray}
\hspace{-0.98in}&&
S_5 \, \, = \, \, \, 15 \, E^2 \, K^2 \,  \, \, \, \, - \, {{5} \over { 2 }} \cdot \,
\Bigl( 6 \cdot \, (1\, -t)^{1/2} \,  \, -(t \, -2) \Bigr) \cdot \, E \, K^3
\nonumber \\  
\hspace{-0.98in}&& \quad  \quad  \quad  \quad  \, 
- \, {{1} \over { 16 }} \cdot \,
\Bigl( 60 \cdot \,(t \, -2) \cdot \,  (1\, -t)^{1/2}\,  \, \,  -(t^2  \,+24\, t \,  -24) \Bigr)
\cdot \, K^4,  
\nonumber
\end{eqnarray}
\begin{eqnarray} 
\hspace{-0.98in}&&
S_6 \, \, = \, \, \, -\, {{1} \over {32 \, t}} \cdot \,
\Bigl( (t^3 \, -168\, t^2 \, +944\, t \, -960) \cdot \,  (1\, -t)^{1/2} \,
\nonumber \\  
\hspace{-0.98in}&& \quad  \quad  \quad  \quad  \quad \quad  \quad  \quad 
-(183\, t^3 \, -1160\, t^2 \, +1936\, t \, -960) \Bigr) \cdot \, K^5
\nonumber \\  
\hspace{-0.98in}&&\quad \quad    \quad  \quad 
 - {{ 15} \over { 8 \, t }} \cdot \,  \Bigl( (t^2 \, -28\,t \, +48) \cdot \, (1\, -t)^{1/2}
\, \,\,  \, -(21\, t^2 \, -68\, t \, +48) \Bigr)  \cdot \, E \, K^4
\nonumber \\  
\hspace{-0.98in}&&\quad  \quad  \quad  \quad 
- {{ 45} \over { 2\, t }} \cdot \,
\Bigl( (t\, -4) \cdot \, (1\, -t)^{1/2} \,\,\,  -(3\, t \, -4) \Bigr)    \cdot \, E^2 \, K^3
\nonumber \\  
\hspace{-0.98in}&&\quad  \quad  \quad  \quad  \quad  \quad \quad \quad  \quad 
 + {{ 30} \over { t }} \cdot \, \Bigl(1 \, \,  \, -(1\, -t)^{1/2}\Bigr) \cdot \, E^3 \, K^2.
\end{eqnarray}
Let us note that these selected ratio of theta functions (\ref{Si})
are not only polynomials in $\, E$ and $\, K$,
but {\em homogeneous polynomials} in $\, E$ and $\, K$.
 The $\, G_n(t)$'s will be D-finite, and
 again polynomials in $\, E$ and $\, K$, as a consequence of the fact that the $\, S_n$'s are
 polynomial expressions of $\, E$ and $\, K$.

The expansion of  $\, G_1(t)$ reads:
\begin{eqnarray}
\label{G1}
\hspace{-0.98in}&& \,  \,   \quad  \,
G_1(t) \, \, = \, \, \,  \, 
-{{3} \over {256}}  \,t^2 \, \, \,  -{{9} \over {1024}}  \,t^3 \, \, 
\, -{{441} \over {65536}}  \,t^4 \, \, \, 
-{{1407 } \over {262144}} \,t^5 \, \,\,   -{{18557 } \over {4194304}}  \,t^6
\nonumber \\
 \hspace{-0.98in}&& \quad  \quad \quad \quad \quad 
\, \, \, \, -{{62755 } \over {16777216}} \,t^7 \,\, \, \, 
-{{13852377 } \over {4294967296}} \,t^8 \, \, 
\, \,\, \,  -{{48531703} \over {17179869184}}  \,t^9
\, \,  \,  \,  \, \,  \, + \, \, \, \cdots 
\end{eqnarray}
The first $\, G_1(t)\, $ reads 
\begin{eqnarray}
\label{G1exact}
\hspace{-0.98in}&& \quad 
G_1(t) \, \, = \, \, \,   G_0(t) \cdot \, \tilde{G}_1(t)
\quad \quad \quad \quad \quad \quad \quad \quad \quad \hbox{where :}
\nonumber \\
 \hspace{-0.98in}&& 
  \quad
\tilde{G}_1(t) \, \, = \, \, \,\,
{{1} \over {4}}\cdot \, S_2  \, \,\,  -{{1} \over {4}} 
 \\
\hspace{-0.98in}&& \quad  \quad   \quad
\, \, = \, \, \, \,  - {{1} \over {4}} \, \,  \,\,\,
+ \Bigl({{ 1 \, -(1-t)^{1/2}} \over { 2 \, \, t}}\Bigr) \cdot \, E \, \, \,\,\,
- {{  (t\, -4) \cdot \, (1-t)^{1/2} \, -(3\, t \, -4)  } \over { 8 \, \, t }} \cdot \, K
\nonumber
\end{eqnarray}
\begin{eqnarray}
\label{G1suite}
 \hspace{-0.98in}&& \quad  \quad   \quad                  
 \, \, = \, \, \,\, 
 - {{1} \over {4}} \, \, \,\,  \,   + {{E } \over { 2 \, \, t}} \, \,\,
 +{{ (3\, t \, -4) } \over { 8 \, \, t}}  \cdot \,  K\, \, \, \,\,
 - (1-t)^{1/2} \cdot \, \Bigl({{E } \over { 2 \, \, t}}
 \, + {{(t -4) } \over { 8 \, \, t  }} \cdot \, K  \Bigr)
  \nonumber \\
  \hspace{-0.98in}&& \quad  \quad   \quad
  \, \, = \, \, \, \,
  -{{ 3} \over { 256}} \cdot \, t^2 \,\, \,  - {{ 3} \over { 256}} \cdot \, t^3 \,\, 
 - {{687 } \over { 65536}} \cdot \, t^4 \, \, - {{303 } \over { 32768}}   \cdot \, t^5 \,\, 
- {{34355 } \over {4194304 }} \cdot \, t^6
\nonumber \\
\hspace{-0.98in}&& \quad  \quad  \quad   \quad  \quad \quad\,
 -{{ 30681} \over { 4194304}} \cdot \, t^7
\,\,  - {{28298151 } \over { 4294967296}}  \cdot \, t^8
 \,\,  - {{6422951 } \over { 1073741824}}  \cdot \, t^9
 \, \,  \,  \,  \, + \, \,  \, \cdots                   
\end{eqnarray}
and the next two read
\begin{eqnarray}
\label{G2G3exact}
\hspace{-0.98in}&& \quad 
G_2(t) \, \, = \, \, \,   G_0(t) \cdot \, \tilde{G}_2(t)
  \nonumber \\
  \hspace{-0.98in}&& \quad  \quad  \quad 
  \, \, = \, \, \,\,   {{ 5} \over { 16777216}} \cdot \, t^6
  \,\,   + {{ 55} \over { 67108864}}  \cdot \, t^7
 \, \,  +{{ 6255} \over { 4294967296}} \cdot \, t^8
 \, \, + {{36625 } \over {17179869184 }} \cdot \, t^9
\nonumber \\
 \hspace{-0.98in}&& \quad  \quad  \quad \quad \quad  \quad \quad
 \,  + {{3079025 } \over { 1099511627776}} \cdot \, t^{10}
\, \, \, + {{ 15116115} \over { 4398046511104}} \cdot \,  t^{11} \,
 \, \, \,  \, \, + \, \, \, \cdots 
\end{eqnarray}
where
\begin{eqnarray}
\label{G2tildeEXACT}
\hspace{-0.98in}&& \quad 
\tilde{G}_2(t) \, \, = \, \, \, \,
      {{1} \over {32}}\cdot \, S_3  \, \, \,
      - \, {{1} \over {16}} \cdot \, S_2 \,  \,\, + {{3} \over {32}}  
\nonumber \\
 \hspace{-0.98in}&&  \quad  \quad \quad  \quad  \, \, = \, \, \,\,
{{3} \over {32}} \,\, \,
  - {{ 1 \, -(1-t)^{1/2}} \over { 8 \, \, t}} \cdot \, E \, \, \,\,\,
 - {{  (t\, -4) \cdot \, (1-t)^{1/2} \, -(3\, t \, -4)  } \over { 32 \, \, t }} \cdot \, K
\nonumber \\
 \hspace{-0.98in}&&  \quad  \quad \quad  \quad \,
 - {{3} \over {32}} \cdot \, E \, K  \,\, \, \,
 + \, {{ 6\cdot \, (1-t)^{1/2} \, -(t \, -2)  } \over { 128 \, \, t }} \cdot \, K^2
\nonumber \\
\hspace{-0.98in}&&  \quad    \quad \,                  
 \, \, = \, \, \, \,   {{5} \over { 67108864}}  \cdot \, t^6 \,\, \,
 + {{ 15} \over { 67108864}}  \cdot \, t^7 \,\, \,
 + {{7305 } \over {17179869184 }}   \cdot \, t^8 \, \, \, +{{ 2825} \over { 4294967296}}  \cdot \,t^9
\nonumber  \\
 \hspace{-0.98in}&& \quad  \quad  
\,  + {{3978105 } \over {4398046511104 }} \cdot \,  t^{10} \,\,
+ {{ 5075805} \over { 4398046511104}} \cdot \,  t^{11} \,\,
+ {{ 1575278229} \over { 1125899906842624}}   \cdot \, t^{12}
\, \, \,+ \, \, \, \cdots
\nonumber 
\end{eqnarray}
and 
\begin{eqnarray}
\label{G2G3exactbis}
\hspace{-0.98in}&& \quad \quad 
G_3(t) \, \, = \, \, \,   G_0(t) \cdot \,  \tilde{G}_3(t)
 \nonumber \\
  \hspace{-0.98in}&& \quad \quad  \quad   \quad  \, \, = \, \, \,\, 
- {{7} \over {281474976710656 }}  \cdot \,  t^{12} \, \,\, 
  - {{ 161 } \over { 1125899906842624}}  \cdot \,  t^{13}
 \nonumber \\
  \hspace{-0.98in}&& \quad \quad  \quad  \quad  
   \,  - {{ 33789} \over { 72057594037927936}}  \cdot \, t^{14} \, \,\, 
   - {{ 332703} \over { 288230376151711744}}  \cdot \, t^{15}
\\
  \hspace{-0.98in}&& \quad  \quad \quad  \quad    \,
- {{4379312\, 7 } \over {18446744073709551616 }}  \cdot \, t^{16} \, \, \,\, 
 - {{318184713 } \over {73786976294838206464 }}  \cdot \, t^{17} \,\, 
   \, \, \, + \, \, \, \cdots
 \nonumber         
\end{eqnarray}
where:
\begin{eqnarray}
\label{G3exact}
\hspace{-0.98in}&&  
 \tilde{G}_3(t) \, \, = \, \, \, \,\, 
 {{1} \over {384}} \cdot \, S_4 \, \,\,  \, -  {{1} \over {128}} \cdot  \, S_3 \,\, \,\, 
   +  {{13} \over {384}} \cdot  \, S_2  \, \,\,  \, - {{5} \over {128}} 
  \\
  \hspace{-0.98in}&&  
 \, \, = \, \, \,\,
 - {{5} \over {128}}  \, \,  \,\,  \,\,
 + {{13} \over {192}}  \cdot \,
 {{ 1 \, -(1 \, -t)^{1/2}} \over { t }} \cdot \, E  
   \nonumber \\
  \hspace{-0.98in}&&  \quad  \quad \quad  \quad \,\,\, \,\,
 -{{13} \over {768}}  \cdot \, {{ (t-4) \cdot \, (1 \, -t)^{1/2} \,
     -(3 \, t \, -4) } \over { t }} \cdot \, K
   \nonumber \\
  \hspace{-0.98in}&&  \quad \quad  \quad \quad  \quad \,
  + {{3 } \over {128}} \cdot \, E \, K \, \,\,
  -{{ 6 \cdot \, (1 \, -t)^{1/2} \ -(t-2) } \over { 512  }} \cdot \, K^2 
\nonumber \\
 \hspace{-0.98in}&&  \quad \quad  \quad \quad  \quad \,
 -{{ 1 \, - (1 \, -t)^{1/2} } \over { 64 \, t }} \cdot \, E^2 \, K  \,\,  \,  
+{{ (t-4) \cdot  \, (1 \, -t)^{1/2} \,  -(3 \, t \, -4) } \over {128 \, t }} \cdot \, E \, K^2
  \nonumber \\
 \hspace{-0.98in}&&  \quad \quad   \quad   \quad \quad  \quad \,                  
 + {{ (1-t)^{1/2} \cdot \, (t^2 -28\,t +48)  \,  \,
     -(21 \, t^2 -68\,t +48) } \over { 3072\, t }} \cdot \, K^3
  \nonumber \\
 \hspace{-0.98in}&& \quad  \quad \,            
 \, \, = \, \, \, \,    -{{ 7} \over {281474976710656 }}  \cdot \, t^{12}
 \, \, \, - {{ 21} \over { 140737488355328}}   \cdot \, t^{13}
 \\
\hspace{-0.98in}&& \quad \quad  \quad  \quad  \quad  \quad    \, \,
 - {{ 36603} \over { 72057594037927936}}  \cdot \, t^{14}  \, \,\, 
 -  {{93149 } \over { 72057594037927936}} \cdot \, t^{15}
 \, \, \,\,  \, + \, \, \, \cdots \nonumber        
\end{eqnarray}                

We have obtained similar results for the next  $\, G_n(t)$'s, namely polynomial
expressions in $\, E$ and $\, K$ with algebraic function coefficients. 

\vskip .3cm

Similar results can be obtained for the other values
$\, \lambda = \, \mathrm{\cos}(\pi \, m/n)$ ($m$ and $\, n$ integers)
yielding algebraic functions for the lambda-extension $\, C(1, 1; \lambda)$. Again,
the (form-factor-like) expansion (\ref{actually2})
around each of these algebraic functions can be written
in a similar way as (\ref{actually2rho})
in terms of the corresponding ratio of theta functions
\begin{eqnarray}
\label{newSi}
  \hspace{-0.98in}&& \, \,\quad \quad \quad \quad \quad \quad \quad \quad \quad  \, 
 S_n  \, \, = \, \, \, \,
   {{ \theta_2^{(n)}(\pi m/n, \, q) } \over {\theta'_2(\pi m/n, \, q) }},  
\end{eqnarray}
where $\, \theta_2^{(n)}(u, \, q)$ denotes the $\, n$-th partial derivative  with respect to $\, u$
of  $\, \theta_2(u, \, q)$. 
It becomes much more difficult to see whether these new $\, S_n$'s are actually polynomial expressions
in $\, E$ and $\, K$ with more and more involved
algebraic coefficients.  One finds that these  new $\, S_n$'s are D-finite, but the reduction to 
polynomial expressions in $\, E$ and $\, K$ becomes a difficult task, in general. Let us display
a few examples.

\vskip .1cm

\subsection{Other one-parameter deformations: deformation of $\,  u \, = \pi/6$.}
\label{otherpisur6}

For $\, u \, = \, \pi/6 \, $ we find that the corresponding $\, S_2$
\begin{eqnarray}
\label{newSipi6}
  \hspace{-0.98in}&& \, \, \quad \quad \quad\quad \quad \quad  \, 
     {{1} \over {\sqrt{3}}} \cdot \,  S_2
         \, \, = \, \, \, \,
  {{1} \over {\sqrt{3}}} \cdot \,
  {{ \theta_2^{(2)}(\pi/6, \, q) } \over {\theta'_2(\pi/6, \, q) }}
  \nonumber \\
\hspace{-0.98in}&&  \quad\quad \quad \quad \quad \quad \quad \quad
  \, \, = \, \, \, \,\,  1 \,\, \, - {{3} \over {128}} \cdot \, t^2 \, \,
  \, \, - {{3} \over {128}} \cdot \, t^3 \, \, \,
  - {{339} \over {16384}} \cdot \, t^4  \,\, \,\, + \,\, \cdots 
\end{eqnarray}
is solution of 
an order-{\em eight} linear differential operator
which is the LCLM (direct-sum) of two order-four
linear differential operators $\, L_4$ and $\, M_4$. The first order-four
linear differential operator  $\, L_4\, $ is the {\em symmetric product}\footnote[2]{This paper belonging
  to the symbolic computation literature and not
  pure mathematics, we use the standard Maple (DEtools) terminology of symmetric powers and symmetric
  products of linear differential operators~\cite{Weil}. Note that "symmetric product" is not a proper mathematical name
  for this construction on the solution space; it is a homomorphic image of the tensor product.
  The (Maple/DEtools) reason for choosing the name symmetric$\_$product is the resemblance with the function symmetric$\_$power. }
of the two order-two linear differential operators
\begin{eqnarray}
\label{twoOper}
\hspace{-0.98in}&& \, \quad \quad \quad\quad  \quad
D_t^2 \,\,\,
+ {{1} \over {3}} \cdot \, {{10\, t^3-15\, t^2+9 \, t -2} \over {
    (t^2-t+1)\, t\, (t-1)}}  \cdot\, D_t
\nonumber \\
\hspace{-0.98in}&&  \quad  \quad \quad  \quad\quad\quad  \quad \quad \quad \,
+ {{1} \over {12}} \cdot \, {{11\, t^6 -33\, t^5 +47\, t^4 -39\, t^3 +3\, t^2 +11\,t -5
  } \over {t^2 \, (t-1)^2\, (t^2 -t +1)^2}},
\nonumber \\
\hspace{-0.98in}&&  \quad\quad \quad\quad  \quad \,
D_t^2 \,\,\,
+ {{1} \over {4}} \cdot \, {{t^6-3\, t^5 +15\, t^4 -25\, t^3 +15\, t^2 -3\, t +1
  } \over {t^2 \, (t-1)^2\, (t^2-t+1)^2}}, 
\end{eqnarray}
having, respectively, the two hypergeometric solutions:
\begin{eqnarray}
\label{twoOpersol}
\hspace{-0.98in}&& \, \,\quad \quad \quad
t^{5/6} \,\cdot \, (1\, -t)^{5/6} \,\cdot \, (t^2-t+1)^{-1/2} \cdot \,
_2F_1\Bigl([{{7} \over {6}}, \, {{5} \over {2}}], \, [{{7} \over {3}}], \, \,  t \Bigr),
 \\
\hspace{-0.98in}&&  \quad \,
\label{twoOpersol2}
t^{1/2} \, \cdot \, (1\, -t)^{1/2} \,\cdot \, (t^2-t+1)^{-1/4} \cdot \,
_2F_1\Bigl([ -{{1} \over {12}}, \, {{7} \over {12}}], \, [1], \, \,
{{27} \over {4}}\,{{ t^2 \cdot \, (1\, -t)^2 } \over { (1 \, -t \, +t^2)^{3} }} \Bigr).
\end{eqnarray}
Let us first note that the first hypergeometric function
$ H \, = \, \, \,_2F_1([7/6,5/2],[7/3], t)$ is actually an
{\em algebraic function}. It is solution of the polynomial equation:
\begin{eqnarray}
\label{Alg}
\hspace{-0.98in}&& \, \,\, \,  \quad \quad
3^{21}\, \cdot  \, t^8\, \, (t \, -1)^8 \, \cdot  \, H^8
\, \, \, \, 
\, \,  +2^{17} \cdot \, 3^{11} \,\cdot \, t^4\,\cdot \, (t^2-t+1)
\, \cdot\, (t-1)^4 \, \cdot  \, H^4
\nonumber \\
\hspace{-0.98in}&&  \quad  \quad  \quad  \quad\,
\, \,
+2^{26} \,\cdot  \, (t-2)\, \cdot\, (2\, \, t-1)\, \cdot\,
(t+1)\,\cdot \, (32\, \, t^6-96\, \, t^5\, +219\, \, t^4 \, -278\, \, t^3
\nonumber \\
\hspace{-0.98in}&&  \quad  \quad \quad  \quad  \quad \quad
\, +219\, \, t^2-96\, \, t+32) \, \cdot  \, H^2
\, \, \, \,  \,  - 2^{32} \,\cdot  \, (t^2-t+1)^2
\,\, \, \, = \, \,  \, \,\, \, 0.
\end{eqnarray}
For the second solution (\ref{twoOpersol2}), we use the identities 
\begin{eqnarray}
\label{identities}
\hspace{-0.98in}&& \, \,\, \,  \quad \quad
_2F_1\Bigl([ -{{1} \over {12}}, \, {{7} \over {12}}], \, [1], \, \,
       {{27} \over {4}}\,{{ t^2 \cdot  \, (1\, -t)^2 } \over { (1 \, -t \, +t^2)^{3} }} \Bigr) 
     \nonumber \\
\hspace{-0.98in}&&  \quad  \quad  \, \,  \quad \quad
\, \,= \, \,\,
-6 \cdot \, {{t  \, \cdot \, (t-1)} \over {(t^2-t+1)^{1/2}}}
\, \cdot  \,  {{ d H_2} \over {dt}} 
\,\,\,\,
+ {{1} \over {2}} \, \, {{(2\, \, t-1)\, \cdot \, (t^2-t-2)} \over {(t^2-t+1)^{3/2}}}
\, \cdot  \,  H_2, 
\end{eqnarray}
where the pullbacked hypergeometric function $\, H_2$ reads:
\begin{eqnarray}
\label{identities}
\hspace{-0.98in}&& \, \,\, \,  \quad \quad \quad \quad\quad
H_2 \, \,= \, \,  \, _2F_1\Bigl([ {{1} \over {12}}, \, {{5} \over {12}}], \, [1], \, \,
{{27} \over {4}}\,{{ t^2 \cdot \, (1\, -t)^2 } \over { (1 \, -t \, +t^2)^{3} }} \Bigr)
  \nonumber \\
\hspace{-0.98in}&&  \quad  \quad  \quad \quad \quad \quad \quad \quad \quad \quad
  \, \,= \, \,\, \, (t^2-t+1)^{1/4} \, \cdot \,
  _2F_1\Bigl([ {{1} \over {2}}, \, {{1} \over {2}}], \, [1], \, \,t \Bigr), 
\end{eqnarray}
Consequently, the relevant solution of the order-four linear differential operator $\, L_4$
will be  a linear combination
$\, \, \, \alpha(t) \cdot \, E \, + \,  \beta(t) \cdot \, K\,\,  $
of the two complete
elliptic integrals $\,  E$, $\, \, K$,   $\, \alpha(t)$ and  $\,  \beta(t)$
being (quite) involved algebraic functions. 

The other order-four linear differential operator $\, M_4 \, $
is, at first sight, slightly more difficult to analyze.
In fact we are in the typical situation of an
{\em absolute factorization}\footnote[1]{A linear differential operator
  $\, L \, \in \, \mathbb{C}(x)[d/dx]$
  is called absolutely reducible~\cite{CompointWeil} if it admits a factorization
over an {\em algebraic extension} of $\, \mathbb{C}(x)$.} of this order-four
linear differential operator, and this can be seen performing
the {\em exterior square} of that  order-four linear
differential operator~\cite{CompointWeil}.
Some calculations are displayed in \ref{AppendixA}. These calculations  strongly suggest
 that the relevant solution of the order-four linear differential operator $\, M_4$
will also be of the form  $\, \, \, \alpha(t) \cdot \, E \, + \,  \beta(t) \cdot \, K$, the functions  
$\, \alpha(t)$ and  $\,  \beta(t)$  being (very) involved algebraic functions of $\, t$.

Fortunately, one can get that result in a much more straight way, if one remarks that
the two order-four linear differential operators $\, L_4$ and $\, M_4$
are actually (non-trivially) homomorphic. Introducing
$\, \rho \, = \, \, t^{2/3} \cdot \, (1\, -t)^{2/3}$, one finds that a conjugate of $\, M_4$
is  actually  homomorphic to the first order-four linear differential operator $\, L_4$:
\begin{eqnarray}
\label{LCLML4M4}
\hspace{-0.98in}&& \, \,\, \,  \quad \quad\quad \quad \quad \quad \quad \quad
L_4 \cdot \,  I_3  \, \, \, = \, \, \,   \,
J_3 \cdot \, \Bigl( {{1} \over {\rho}} \cdot \, M_4  \cdot \,  \rho\Bigr),
\end{eqnarray}
where $\, I_3$ and $\, J_3$ are (slightly involved) order-three intertwiners.

\vskip .1cm

Therefore we have shown that the relevant solution of the order-eight linear differential operator
will be of the form  $\, \, \,  \alpha(t) \cdot \, E \, + \,  \beta(t) \cdot \, K$,
$\, \alpha(t)$ and  $\,  \beta(t) \, $    being (quite) involved algebraic functions of  $\, t$.

Again, one finds that $\, S_2$ is D-finite,
but the reduction to polynomials in the complete elliptic integrals $\, E\, $
and $\, K \, $ is far from obvious.

\vskip .1cm

\subsubsection{Deformation of $\,  u \, = \pi/6$: the $\, S_3$ term \\}
\label{otherpisur6S3}

The next $\, S_n$, namely
\begin{eqnarray}
\label{newSipi6bis}
  \hspace{-0.98in}&& \, \,\quad \quad \quad \quad\quad \quad\quad \quad \quad  \, 
  S_3       \, \, = \, \, \, \,
  {{ \theta_2^{(3)}(\pi/6, \, q) } \over {\theta'_2(\pi/6, \, q) }},  
\end{eqnarray}
is solution of a linear differential operator of order {\em twelve}
with coefficient polynomials in $\, t$ of
degree $\, 67$.  This is a quite large  order (twelve) linear differential
operator, that we will not give here.
This order-twelve  linear differential operator is actually the direct sum
of an order-three operator and an order-nine operator $\, L_9$.
The order-three linear differential operator $\, L_3$ reads
\begin{eqnarray}
\label{L3pi6}
\hspace{-0.98in}&& \, \, \quad 
 L_3 \, \, = \, \, \,\, D_t^3 \,\,\,
 +6 \cdot \, {{ q_{12}} \over {  q_{6}  \cdot \, (t\, -1) \cdot \,
     (t\, +1) \cdot \,  (t \, -2) \cdot  \, (2\, t \, -1)
     \cdot \, (t^2 \, -t \, +1) \cdot \,t}} \cdot \, D_t^2
 \nonumber \\
 \hspace{-0.98in}&& \, \,   \quad      \quad    \quad      \quad       \,
 + {{ r_{12}} \over {  q_{6}  \cdot \, (t\, -1)^2 \cdot \,
     (t\, +1) \cdot \,  (t \, -2) \cdot \, (t^2 \, -t \, +1) \cdot \,t^2}} 
                    \cdot \, D_t
 \nonumber \\
 \hspace{-0.98in}&& \, \,      \quad   \quad     \quad  \quad      \quad      \quad      \quad           \,
 \, \, + {{3} \over {2}} \cdot \, {{ r_6} \over {  q_{6}  \cdot \, (t -1)
     \cdot \, (t\, +1) \cdot \,  (t \, -2) \cdot  \, (2\, t \, -1)   \cdot \, t}}, 
\end{eqnarray}
where:
\begin{eqnarray}
\label{L3pi6where}
\hspace{-0.98in}&& \, \, \quad      \quad    
q_{12}  \, \, = \, \, \,
t^{12} \, -6\, t^{11} +2536\, t^{10}-12625\, t^9 \, +18414\, t^8 \, +2028\, t^7 \, -31302\, t^6
\nonumber \\
  \hspace{-0.98in}&& \, \,   \quad   \quad      \quad     \quad      \quad   \quad
         \, +33849\, t^5 \, -16458\, t^4 \, +4084\, t^3 \, -528\, t^2 \, +7\, t \, \, \, -1,
\nonumber \\
  \hspace{-0.98in}&& \, \,   \quad      \quad
  q_{6}  \, \, = \, \, \,
  t^6 \, \, -3\, t^5 \, +1518\, t^4 \, -3031\, t^3 \, +1518\, t^2 \, -3\, t \, \, \, +1,
 \nonumber \\
  \hspace{-0.98in}&& \, \,   \quad      \quad
  r_{12}   \, \, = \, \, \,
  t^{12} \, -6\, t^{11} +4881\, t^{10} -24350\, t^9 \, +24459\, t^8 \, +48198\, t^7 \, -120498\, t^6
  \nonumber \\
  \hspace{-0.98in}&& \, \,  \quad     \quad      \quad    \quad      \quad   \quad \,
  +90597\, t^5 -20496\, t^4 -5105\, t^3 +2304\, t^2 +15\, t  \, -2,
\nonumber \\
  \hspace{-0.98in}&& \, \,   \quad      \quad
  r_6  \, \, = \, \, \,
  59\, t^6 \, -177\, t^5 \, +4512\, t^4 \, -8729\, t^3 \, +4512\, t^2 \, -177\, t \, \, +59.     
\end{eqnarray}
Let us denote  $\, L_K$ the order-two linear differential operator
annihilating the complete elliptic integral
of the first kind $\, K \, = \, _2F_1([1/2,1/2],[1],\, t)$:
\begin{eqnarray}
\label{recall}
  \hspace{-0.98in}&& \, \, \quad \quad \quad \quad \quad \quad 
  L_K \, \, = \, \, \,  \, \,  D_t^2 \, \, \,\,
  + {{2\, t \, -1} \over {t \cdot \, (t-1)}} \cdot \, D_t \, \, \,
         +{{ 1} \over { 4 \, t \cdot \, (t \, -1)}},              
\end{eqnarray}
This order-three linear differential operator (\ref{L3pi6})
is  actually homomorphic to the symmetric square of operator $\, L_K$, 
with order-two intertwiners. Consequently the solutions of $\, L_3$
are (quadratic) homogeneous polynomials in $\, E$ and $\, K$.
Actually one finds that the solution of $\, L_3$ given by (\ref{L3pi6}) reads:
\begin{eqnarray}
\label{L3pi6sol}
  \hspace{-0.98in}&& \, \,      \quad \quad 
  \mathrm{Sol}(L_3) \,  \, = \, \, \, \,
  {{(t \, -2)^3}\over{(t^2\, -t+\, 1)}} \cdot \, K^2 \,  \,\, +9 \cdot \, E \, K
 \nonumber \\
  \hspace{-0.98in}&& \, \, \,  \,\, \,  \quad \quad 
\,  \, = \, \, \,  \,  1 \, \,  \,
+{{177 } \over {32}} \, t^2 \, \, 
+ {{177 } \over {32}} \,  t^3 \,\, 
+{{1095} \over {8192}} \, t^4 \,\,  
-{{21561} \over {4096}} \, t^5 \, \, 
-{{1384095} \over {262144}} \, t^6 \, \, 
+{{22467} \over {262144}} \, t^7
\nonumber \\
\hspace{-0.98in}&& \, \,     \quad \quad
\quad \quad  \quad \quad \quad \quad   \quad  \,
 +{{2927958291} \over {536870912}} \, t^8 \,\,  \, 
+{{730823955} \over {134217728}} \, t^9  \, \,   \, \,\, \,  + \, \, \, \cdots 
\end{eqnarray}

The order-nine linear differential operator  $\, L_9$ can be seen
to be the {\em symmetric product} 
of an order-three linear differential operator $\, A_3$,
and of the  order-three linear differential operator, which is the
{\em symmetric square} of the order-two  linear differential operator
$\, L_K$ annihilating $\, K \, = \, \, _2F_1([1/2,1/2],[1],t)$
\begin{eqnarray}
\label{L9first}
\hspace{-0.98in}&& \, \, \quad \quad  \quad    \quad \quad    \quad     \quad \quad    \quad   
  L_9 \,\, = \, \, \, \mathrm{SymProd}\Bigl( \mathrm{Sym}^2(L_K),  \,  A_3  \Bigr). 
\end{eqnarray}
The order-three linear differential operator $\, A_3$ reads
\begin{eqnarray}
\label{A3}
  \hspace{-0.98in}&& \, \, \quad \quad   \quad \quad     
 A_3 \,\, \, = \, \, \, \,  D_t^3 \, \, \, \,
 + {{r_8 \cdot \,  (2\,t -1) } \over {
     q_6 \cdot t \cdot \, (t\, -1) \cdot  \, (\, t^2 \, - \, t \, +1) }} \cdot \, D_t^2
  \nonumber \\
\hspace{-0.98in}&& \, \, \quad   \quad    \quad   \quad  \quad  \quad  
\, - {{5}\over{9}} \cdot {{r_6 \cdot  \, (\, t^2 \, - \, t \, +1)  } \over {
    q_6 \cdot t^2 \cdot (t\, -1)^2 }} \cdot \, D_t
\,\,\,\,\,  +{{5}\over{18}} \cdot  {{ r'_6 \cdot (2\, t\, -1)  } \over {
    q_6 \cdot t^2 \cdot (t\, -1)^2   }},
\end{eqnarray}
where:
\begin{eqnarray}
\label{L3pi3where}
\hspace{-0.98in}&& \, \, \quad      \quad    
r_{6}  \, \, = \, \, 52-156 t-3009\,t^2 \, +6278 t^3-3009 t^4-156 t^5 \, +52 t^6,
\nonumber \\
  \hspace{-0.98in}&& \, \,   \quad      \quad
  r'_{6}  \, \, = \, \, \,
  r_6\, \,-2106 \, \cdot  \, t\cdot  \, (t-1) \cdot  \,(t-2)\cdot  \, (t+1),
 \nonumber \\
  \hspace{-0.98in}&& \, \,   \quad      \quad
 q_{6}   \, \, = \, \, \,  5\, r_6 \, \, +16038 \cdot  \, t^2 \cdot  \,(t-1)^2,
  \nonumber \\
  \hspace{-0.98in}&& \, \,   \quad      \quad
  r_8   \, \, = \, \, \,  5\, r_6\cdot  \,  (t^2-t+1) \,\,
  +17172 \cdot  \, t^2\cdot  \,  (t-1)^2 \,\, +15471\cdot  \,  t^3 \cdot  \,  (t-1)^3.     
\end{eqnarray}
The solutions of this order-three linear differential operator $\, A_3 \,$
are actually {\em algebraic functions} satisfying
\begin{eqnarray}
\label{satisfying}
  \hspace{-0.98in}&& \, \, \quad \quad 
 432 \cdot \, (t^2-t+1)^4 \cdot \, F^4 \,  \,\, \, 
 -72  \cdot \, P_6 \cdot \, (t^2-t+1)^2 \cdot \, F^2
 \nonumber \\
  \hspace{-0.98in}&& \, \, \quad  \quad \quad
 -16 \cdot \, (t-2) \cdot \, (2\,t-1) \cdot \, (t +1) \cdot \, (t^2 -t +1) \cdot \,
     \Bigl(P_6 \,\, +972 \cdot \, t^2  \cdot \,(t -1)^2\Bigr)\cdot \, F
  \nonumber \\
  \hspace{-0.98in}&& \, \, \quad \quad \quad \quad \quad
 +6480 \cdot \, t^2 \cdot \, (t-1)^2 \cdot \, (t^2-t+1)^3 \,  \,\, -P_6^2
  \,\,  \, = \,  \,\,\, 0,                     
\end{eqnarray}
where the polynomial $\, P_6$ reads:
\begin{eqnarray}
\label{satisfyingP}
\hspace{-0.98in}&& \, \, \quad \quad \quad \quad \quad \quad 
 P_6 \, \,  = \, \,  \,   \,
  4 \cdot  \, (t^2 \, -t \, +1)^3  \,  \, -243 \cdot \, t^2 \cdot \, (1\, -t)^2.
\end{eqnarray}
The well-suited solution of the order-three linear differential operator $\, A_3$ reads:
\begin{eqnarray}
\label{SolA3}
  \hspace{-0.99in}&&  \quad   
Sol(A_3) \, \, = \, \, \, 
1 \, \,  \,
-{{1} \over {2}} \, t \,\,  \, \, 
- {{165} \over {64}} \, t^2 \, \, \, 
- {{165} \over {128}} \, t^3 \,\,  \,
+ {{26655} \over {16384}} \, t^4 \,  \,
+ {{101085} \over {32768}} \, t^5 \,  \,
+  {{6546741} \over {4194304}} \, t^6
  \nonumber \\
\hspace{-0.98in}&& \, \,    \quad   \quad    \quad   \quad   \quad   \,
 - {{12198135} \over {8388608}}  \, t^7 \,  \, 
 -{{3182706057} \over {1073741824}} \, t^8 \, \,
 - {{3159215679} \over {2147483648}}\, t^9 \,  \,\, \, +\, \, \, \cdots 
\end{eqnarray}
The solution of the order-nine linear differential operator $\, L_9$ reads:
\begin{eqnarray}
\label{solL9pisur6}
  \hspace{-0.98in}&& \, \, \, \,  \, \,  
  \mathrm{Sol}(L_9) \, \, = \, \, \,   \,
  \mathrm{Sol}(A_3) \cdot \,  K^2 \, \, \,   = \, \, \, \, \,
  1   \, \,  \, \,
- {{159} \over {64 }} \, t^2   \, \, 
- {{159} \over {64 }} \, t^3 \,   \, 
+ {{2973} \over {16384 }}  \, t^4 \,  \,  
+  {{23325} \over {8192 }} \, t^5
  \nonumber \\
\hspace{-0.98in}&& \, \,    \quad  \quad   \quad   \,
+{{11858901} \over {4194304 }} \, t^6 \,  \,
+{{510591} \over {4194304}} \, t^7 \,  \,
- {{2771276211} \over {1073741824 }}\, t^8 \,
- {{695778099} \over {268435456}}\, t^9
\,   \, \, + \, \, \, \cdots         
\end{eqnarray}
The series expansion of (\ref{newSipi6bis}) reads: 
\begin{eqnarray}
\label{newSipi6bisexp}
  \hspace{-0.99in}&&  \quad  \quad 
  -S_3       \, \, = \, \, \, \,
 -{{ \theta_2^{(3)}(\pi/6, \, q) } \over {\theta'_2(\pi/6, \, q) }}
   \, \, \,\,  = \, \, \,\, 
1 \, \,  \, 
+ {{3 } \over { 16 }} \, t^2 \, \, 
+ {{3 } \over { 16 }} \, t^3 \, \, 
+ {{339 } \over { 2048 }} \, t^4 \, \, 
+{{147 } \over { 1024 }} \, t^5 \, \, 
 \nonumber \\
\hspace{-0.98in}&& \, \,    \quad  \quad   \quad  \quad 
\,+{{ 262047} \over {2097152 }} \, t^6  \,  \, 
+{{ 230109} \over {2097152 }} \, t^7 \, \, 
+{{1632105} \over {16777216}} \, t^8   \, \, 
+{{365061} \over {4194304}} \, t^9 \,  \, \,  \,  \,  \, 
+ \, \, \, \cdots  
\end{eqnarray}
Recalling the series expansions (\ref{L3pi6sol}) and  (\ref{solL9pisur6}),
one actually finds that this series (\ref{newSipi6bisexp}) is exactly:
\begin{eqnarray}
\label{newSipi6bisactually}
\hspace{-0.98in}&& \, \, \quad  \quad  \quad
- S_3       \, \, = \, \, \, \,
 - {{ \theta_2^{(3)}(\pi/6, \, q) } \over {\theta'_2(\pi/6, \, q) }}
   \, \, \, = \, \, \, \,\,   {{1} \over {3 }} \cdot \, \mathrm{Sol}(L_3)
 \,  \, \, + \,  {{2} \over {3 }} \cdot \, \mathrm{Sol}(L_9)
 \nonumber \\
\hspace{-0.98in}&& \, \,   \,  \,   \quad  \quad  \quad  \quad   \quad  
  \, \, \, = \, \, \,
     {{1} \over {3 }} \cdot \,
     \Bigl( {{(t \, -2)^3}\over{(t^2\, -t+\, 1)}} \cdot \, K^2 \,
 +9 \cdot \, E \, K \Bigr)
   \,  \, \, \, +  {{2} \over {3 }} \cdot \, \mathrm{Sol}(A_3) \cdot \,  K^2.
\end{eqnarray}

\vskip .1cm

{\bf Remark:} More generally, for $\, u \, = \, \pi/6$, one has
\begin{eqnarray}
\label{encapsulate4}
\hspace{-0.98in}&& \,  \,  \, \quad \quad   \quad  \quad   \, \,  \,   \, 
 C_{\rho}(1, \, 1; \, \rho)
 \, \, = \, \, \, 
 - 2 \cdot \, {{ \theta_2'\Bigl({{\pi} \over {6}}, q\Bigr)  } \over {\sqrt{\rho\, +1} \cdot \,  \theta_2(0, q) \cdot \, \theta_3(0, q)^2  }} 
\nonumber   \\
 \hspace{-0.98in}&& \,  \,  \, \quad \quad \,   \,  \quad \quad  \quad   \quad \quad  \quad   \quad  \quad  \quad    \, \,  \,   \,
                     \times \, \sum_{p=0}^{\infty} \, \, 
 \Bigl( \mathrm{\arcsin} \Bigl( {{ \sqrt{\rho\, +1}} \over {2}} \Bigr) \, -{{\pi} \over {6}} \Bigr)^{p} \cdot \,
    {{ S_{(p\, +1) } \over {    (p)! }}},
\end{eqnarray}
where:
\begin{eqnarray}
\label{encapsulate4Sn}
  \hspace{-0.98in}&& \,  \,  \,   \, \,  \,   \, \quad  \quad  \quad  \quad  \quad  \quad  \quad   \quad
  S_n \, \, = \, \, \,  {{ \theta_2^{(n)}(\pi/6, \, q) } \over {\theta'_2(\pi/6, \, q) }}.  
\end{eqnarray}

\vskip .1cm

\subsection{Other one-parameter deformations: deformation of $\,  u \, = \pi/3$.}
\label{otherpisur3}

{\bf Note:} To avoid any confusion with the linear differential operators
introduced in the $\, u \, = \, \pi/3$ case (see subsection \ref{otherpisur6} and \ref{AppendixAM4})
we will add an extra $\, (3)$ subscript for the linear differential operators
of this  $\,  u \, = \pi/3$ case. 

\vskip .1cm

For $\, u \, = \, \pi/3\, $ we also find that 
\begin{eqnarray}
\label{newSipi3}
  \hspace{-0.98in}&& \, \, \quad  \quad\quad \quad \quad  \, 
  \sqrt{3} \cdot \, S_2       \, \, = \, \, \, \,
  \sqrt{3} \cdot \,  {{ \theta_2^{(2)}(\pi/3, \, q) } \over {\theta'_2(\pi/3, \, q) }}
  \nonumber \\
  \hspace{-0.98in}&& \, \,    \quad \quad \quad  \quad \quad  \quad  \quad \quad
  \, \, = \, \, \, \, \, 
  1 \,\,\,\, -{{9} \over {128 }} \cdot \, t^2 \,\, \, -{{9} \over {128 }} \cdot \, t^3 \,
  \,\, -{{261} \over {4096 }} \cdot \, t^4 \,  \, \,  \,  \,\, + \, \, \, \cdots 
\end{eqnarray}
is actually D-finite: it is solution of a (slightly involved)
{\em order-eight linear differential operator} $\, L_8^{(3)}$.
In fact, revisiting the calculations performed in
section \ref{Atfirstsight}, but this time with a perturbation
around an algebraic solution $\, A(t)$ (see (\ref{algM3})),  one easily finds, using
the sigma-form of Painlev\'e VI non-linear differential equation (\ref{jmequation}), 
that the first correction term $\, G_1(t)$ is solution of an
order-three linear differential operator, with
very involved algebraic coefficients depending on the algebraic
solution $\, A(t)$ and its derivatives. 
This provides lower order linear differential operators, but
with a price to pay, namely very involved algebraic
coefficients. In fact one can study directly the previous order-eight
linear differential operator. 

\vskip .1cm

If one conjugates this  order-eight linear differential operator $\, L_8^{(3)} \, $ by $\, t^{4/3}$, changing
$\, L_8^{(3)}$ into  ${\tilde L}_8^{(3)} \, = \, \,  t^{-4/3}  \cdot \,  L_8^{(3)} \cdot \,  t^{4/3}$, 
one can easily see that this new order-eight  linear differential operator ${\tilde L}_8^{(3)}$
is actually the direct-sum (LCLM)  of two order-four  linear differential operators:
$\,  {\tilde L}_8^{(3)} \, = \, \, \mathrm{LCLM}(L_4^{(3)}, \, M_4^{(3)}) \, = \, \, L_4^{(3)} \oplus \, M_4^{(3)}$. Furthermore,
one finds that these two order-four  linear differential operators are
non-trivially homomorphic, after performing a conjugation
of one of the two  linear differential operator by
$\, \rho \,\,  = \, \, t^{1/3} \cdot \, (1\, -t)^{1/3}$ 
\begin{eqnarray}
\label{newSipi3Homo}
\hspace{-0.98in}&& \, \,\quad \quad  \quad \quad \quad  \quad \quad  \quad
M_4^{(3)} \cdot \, I_3 \, \, \, = \, \, \, \,
J_3   \cdot \, \Bigl({{1} \over {\rho}} \cdot \, L_4^{(3)} \cdot \, \rho\Bigr),         
\end{eqnarray}
where $\, I_3$ and $\, J_3$ are order-three intertwiners.
Let us focus on the simplest  order-four linear differential operator, namely $\, L_4^{(3)}$:
\begin{eqnarray}
\label{newSipi3L4}
\hspace{-0.98in}&& \, \,
 L_4^{(3)} \, \, = \, \, \, \, \,
 D_t^4  \, \, \, \,
 + {{4} \over {3}} \cdot \, {{ 9\, t-5 } \over { (t-1) \cdot \, t}} \cdot \, D_t^3  \,\, \, \, 
 + {{1} \over {9}}  \cdot \, {{ 337\, t^2-373\, t+73 } \over { (t-1)^2 \cdot \, t^2}} \cdot  \, D_t^2
 \\
\hspace{-0.98in}&& \, \, \quad     \, \, 
+ {{1} \over {54}}  \cdot \, {{1590\, t^3-2627\, t^2+1085\, t-42 } \over {
    (t-1)^3 \cdot \, t^3}} \cdot \, D_t \, \, \, 
+ {{1} \over {162}}   \cdot \,{{ 350\, t^3-769\, t^2+485\, t-84 } \over {(t-1)^4 \cdot \, t^3 }}.
\nonumber 
\end{eqnarray}
We have a prejudice that this order-four linear differential operator
could correspond to an absolute factorisation~\cite{CompointWeil}, and could be
written\footnote[1]{This prejudice comes from subsection (\ref{otherpisur6}),
  see (\ref{twoOper}).} 
as a {\em symmetric product} of two order-two
linear differential operators (see also \ref{AppendixA}). In order to
check this scenario, let us calculate the {\em exterior square}
of that order-four linear differential operator. One finds that
it is actually the direct-sum (LCLM) of two order-three linear differential operators
\begin{eqnarray}
\label{newSipi3Algebrdirect}
  \hspace{-0.98in}&& \, \,  \quad \quad  \quad \quad \quad  \quad \quad 
  \mathrm{Ext}^2\Bigl( L_4^{(3)}  \Bigr)   \, \, = \, \, \,
  \mathrm{LCLM}(A_3^{(3)}, \, B_3^{(3)}) \, \, = \, \, \, A_3^{(3)} \, \oplus \, B_3^{(3)}, 
\end{eqnarray}
where the second order-three linear differential operator $\, B_3^{(3)}$ is exactly
the {\em symmetric square}
of an order-two linear differential operator $\, A_2^{(3)}$
\begin{eqnarray}
\label{newSipi3Algebr}
\hspace{-0.98in}&& \, \, \quad \quad \quad 
A_2^{(3)} \, \, = \, \, \,\,
D_t^2 \, \, \, +{{2} \over {3}} \cdot \, {{7\, t \, -4} \over {t \cdot \, (t-1)}} \cdot \, D_t \, \,\, 
        +{{1} \over { 36}} \cdot \, {{117 \, t^2-133\, t \, +21} \over { t^2 \cdot \, (t \, -1)^2}},     
\end{eqnarray}
which has the two {\em algebraic function} solutions:
\begin{eqnarray}
\label{newSipi3Algebrsol}
  \hspace{-0.98in}&& \, \,\,
  t^{-1/2} \cdot \, (1\, -t)^{-1/6} \cdot \,
  _2F_1\Bigl([{{ 5} \over { 6}},  \, {{ 3} \over { 2}}], \, [{{ 5} \over {3 }}],  \, t    \Bigr),
  \, \quad
    t^{-7/6} \cdot \, (1\, -t)^{-1/6} \cdot \,
   _2F_1\Bigl([{{ 1} \over { 6}},  \, {{ 5} \over { 6}}], \, [{{ 1} \over {3 }}],  \, t    \Bigr).
 \nonumber 
\end{eqnarray}
Recalling (\ref{recall}) the order-two linear differential operator $\, L_K$
annihilating the complete elliptic integral
of the first kind $\, K \, = \, _2F_1([1/2,1/2],[1],\, t)$, let us consider
the {\em symmetric product} of the  order-two linear differential operator $\, A_2$
and of  $\, L_K$. One finds that this symmetric product is non-trivially
homomorphic to some conjugate of $\, L_4$
\begin{eqnarray}
\label{recall}
  \hspace{-0.98in}&& \, \,  \quad \quad \quad \quad  \quad  \quad 
 \mathrm{SymProd}(A_2^{(3)}, \, L_K) \cdot \, I_2 \, \, = \, \, \,\, 
 J_2 \cdot \,   \Bigl( {{1} \over {\rho}} \cdot \, L_4^{(3)} \, \cdot \, \rho    \Bigr),         
\end{eqnarray}
where $\, \rho \, = \, t^{1/6} \cdot \, (1\, -t)^{1/6}$, and where $\, I_2$ and $\, J_2$
are order-two intertwiners.
This shows that the solution of $\, L_4^{(3)}$ (and thus $\, M_4^{(3)}$),
and therefore the solution of
the order-eight linear differential operator $\, L_8^{(3)}$,
are actually of the form $\,\, \alpha(t) \cdot \, E \, + \, \beta(t) \cdot \, K\,$
where $\, \alpha(t)$ and $\, \beta(t)\,$ are  {\em algebraic functions.} 

\vskip .2cm

{\bf Remark:} Note, eventually, that these two order-four linear differential
operators  $\, L_4^{(3)}$ and $\, M_4^{(3)}$ can, in fact, be seen to be (non-trivially) homomorphic to
some well-suited conjugates of the two order-four operators  $\, L_4$ and  $\, M_4$
emerging for $\, u \, = \, \pi/6 \, $
in the previous subsection (\ref{otherpisur6}).

\vskip .1cm
\vskip .1cm

\subsubsection{Deformation of $\,  u \, = \pi/3$: the $\, S_3$ term \\}
\label{otherpisur3S3}

 The next $\, S_n$, namely
\begin{eqnarray}
\label{newSipi3bis}
  \hspace{-0.98in}&& \, \,\quad \quad \quad \quad\quad \quad\quad \quad \quad  \, 
  S_3       \, \, = \, \, \, \,
  {{ \theta_2^{(3)}(\pi/3, \, q) } \over {\theta'_2(\pi/3, \, q) }},  
\end{eqnarray}
is solution of a linear differential operator of order {\em twelve} with
coefficient polynomials in $\, t$ of degree $\, 52$.  This is a quite large
order twelve linear differential operator, that we will not give here.
This order twelve  linear differential operator is actually the direct sum
of an order-three operator and an order-nine linear differential operator $\, L_9$.
The order-three linear differential operator $\, L_3$ reads:
\begin{eqnarray}
\label{L3pisur3}
\hspace{-0.98in}&& \, \, \quad \quad \quad 
 L_3^{(3)} \, \, = \, \, \, D_t^3 \,\,
 +{{ 6 \cdot \, (64 \, t^{4}-170\, t^{3}+40\, t^{2}+3\, t \, -1) } \over {
     (128 \,  t^2 \, + \, t \, - 1) \cdot \, (t\, -1)
     \cdot \,   (t \, -2) \cdot \,t}} \cdot \, D_t^2
 \nonumber \\
 \hspace{-0.98in}&& \, \,   \quad    \quad \quad   \quad    \quad      \quad       \,
 + {{ (128\,  t^{5} -410\, t^{4} -55\, t^{3} +218\, t^2 \, -11\, t \, + 2) } \over {
     (128 \,  t^2 \, + \, t \, - 1)   \cdot \, (t\, -1)^2 \cdot \,  (t \, -2) \cdot \,t^2}} 
 \cdot \, D_t
 \nonumber \\
 \hspace{-0.98in}&& \, \,      \quad  \quad \quad  \quad
 \quad  \quad      \quad      \quad      \quad           \,
 \, \, - {{ 3\cdot \,  (32 \,  t^2 \, + 5\, t \, - 5) } \over {
 2 \cdot \,  (128 \,  t^2 \, + \, t \, - 1)
     \cdot \, (t -1)^2 \cdot \,  (t \, -2)   \cdot \, t}}. 
\end{eqnarray}
This order-three linear differential operator (\ref{L3pisur3})
is  actually homomorphic to the symmetric square of the order-two linear
differential operator $\, L_K$, annihilating $\, K \, = \, _2F_1([1/2,1/2],[1],\, t)$,
with order-two intertwiners. Consequently the solutions of $\, L_3$
are (quadratic) homogeneous polynomials in $\, E$ and $\, K$.
Actually one finds that the solution of $\, L_3^{(3)}$ given by (\ref{L3pisur3}) reads:
\begin{eqnarray}
\label{L3pi3sol}
  \hspace{-0.98in}&& \, \,     \quad \quad \quad 
  \mathrm{Sol}(L_3^{(3)}) \,  \, = \, \, \, \,\,
  4 \cdot \,  (t \, -2) \cdot \, K^2 \,\, +9 \cdot \, E \, K
 \nonumber \\
  \hspace{-0.98in}&& \, \,   \, \,   \quad   \quad \quad \quad \quad 
\,  \, = \, \, \,   1 \,\, -{{15 } \over {32}} \, t^2 \, \,- {{15 } \over {32}} \,  t^3 \,\,
-{{3513} \over {8192}} \, t^4 \,\, -{{1593} \over {4096}} \, t^5 \,\,
-{{92895} \over {262144}} \, t^6 \,\, -{{85245} \over {262144}} \, t^7
\nonumber \\
\hspace{-0.98in}&& \, \,   \quad   \quad \quad
\quad \quad  \quad \quad \quad \quad   \quad  \,
  -{{161330925} \over {536870912}} \, t^8 \, \,\,
  -{{37507821} \over {134217728}} \, t^9
 \,  \, \,\,  + \, \, \, \cdots 
\end{eqnarray}

The order-nine linear differential operator  $\, L_9^{(3)}\,$ can be seen
to be the {\em symmetric product}
of an order-three linear differential operator $\, A_3^{(3)}$
and of the  order-three linear differential operator which is the
symmetric square of the order-two  linear differential operator
$\, L_K$ annihilating $\, K \, = \, \, _2F_1([1/2,1/2],[1],t)$:
\begin{eqnarray}
\label{L9first}
  \hspace{-0.98in}&& \, \, \quad  \quad  \quad  \quad    \quad    \quad    \quad    \quad   
    L_9^{(3)} \,\, = \, \, \, \mathrm{SymProd}\Bigl( \mathrm{Sym}^2(L_K),  \,  A_3^{(3)}  \Bigr). 
\end{eqnarray}
The order-three linear differential operator $\, A_3$ reads:
\begin{eqnarray}
\label{A3}
  \hspace{-0.98in}&& \, \,    
 A_3^{(3)} \,\, \, = \, \, \, \, \,  D_t^3 \, \,\, 
 + {{ 16\,t^3 \,-94\,t^2 \,+165\, t \, -55} \over {
     t \cdot \, (t\, -1) \cdot  \, (16\, t^2 \, -55 \, t \, +55) }} \cdot \, D_t^2
 \\
\hspace{-0.98in}&& \, \,     
\, + {{ 32\, t^4 \, -130\, t^3 \, +75\, t^2 \, +110\, t \, -55 } \over {
    9 \cdot \, t^2 \cdot \, (t\, -1)^2 \cdot  \, (16\, t^2 \, -55 \, t \, +55) }} \cdot \, D_t
\, - {{64\,t^3 \, -240\, t^2 \, +165\, t \, +55 } \over {
    18 \cdot \, t^2 \cdot \, (t\, -1)^2 \cdot  \, (16\, t^2 \, -55 \, t \, +55)}}.
 \nonumber
\end{eqnarray}
The solutions of this order-three linear differential operator $\, A_3^{(3)}\, $
are actually {\em algebraic functions}
satisfying the algebraic equation:
\begin{eqnarray}
\label{algeqfirst}
\hspace{-.98in}&& \quad \quad \quad   \, \,  
27 \cdot \, F^4 \,\,  \,   \,  
-18 \cdot \,  (16 \,  t^2 \, - \, t \, + 1) \cdot \,   F^2 \,\,  \,  \,  
-4 \, (t \, -2) \cdot \, (128 \,  t^2 \, + \, t \, - 1) \cdot \,  F
\nonumber \\
\hspace{-.98in}&& \quad \quad \quad \quad \quad \quad \quad 
- (256 \,  t^4 \, -\,   752 \,  t^3 \, +\,   753 \,  t^2 \, -2\, t \, + 1)
\,\, \,\,    = \,\,\,\,\,    0. 
\end{eqnarray}
The well-suited solution of the order-three linear differential
operator $\, A_3$ reads:
\begin{eqnarray}
\label{SolA3}
  \hspace{-0.98in}&& \, \, \quad   
  \mathrm{Sol}(A_3^{(3)}) \, \, = \, \,\, \, \,\,   1 \, \, \,\,   -{{1} \over {2}} \, t \,\,\,
  + {{9} \over {64}} \, t^2 \, \,\, + {{9} \over {128}} \, t^3 \,\, \,
  + {{747} \over {16384}} \, t^4 \,\,
  + {{1089} \over {32768}} \, t^5 \,\, +  {{108603} \over {4194304}} \, t^6
  \nonumber \\
\hspace{-0.98in}&& \, \,    \quad   \quad    \quad   \quad   \quad   \,
 + {{176679} \over {8388608}}  \, t^7 \,  \, +{{18959247} \over {1073741824}} \, t^8 \, \,
 + {{32508009} \over {2147483648}}\, t^9
 \, \,\,   \,\, +\, \, \, \cdots 
\end{eqnarray}
The solution of the order-nine linear differential operator reads:
\begin{eqnarray}
\label{solL9pisur3}
  \hspace{-0.98in}&& \, \, \, \,  \, \,  
 \mathrm{Sol}(L_9^{(3)}) \, \, = \, \, \,   \,  \mathrm{Sol}(A_3^{(3)}) \cdot \,  K^2 \, \, \,   = \, \, \,\,
  1   \, \,\, +{{15} \over {64 }} \, t^2 \,  \, \,
  + {{15} \over {64 }} \, t^3 \, \,  \, + {{3513} \over {16384 }}  \, t^4
  \,  \,\,  +  {{1593} \over {8192 }} \, t^5
  \nonumber \\
\hspace{-0.98in}&&   \quad  \quad   \quad   
+{{743115} \over {4194304 }} \, t^6 \,\, +{{681825} \over {4194304}} \, t^7
\, +{{161265045} \over {1073741824 }}\, t^8 \,
 + {{37482261} \over {268435456}}\, t^9
 \,   \, \,\,\,   + \, \, \, \cdots         
\end{eqnarray}
The series expansion of (\ref{newSipi3bis}) reads: 
\begin{eqnarray}
\label{newSipi3bisexp}
  \hspace{-0.98in}&& \, \, \quad  \quad 
  - S_3       \, \, = \, \, \, \,
 - {{ \theta_2^{(3)}(\pi/3, \, q) } \over {\theta'_2(\pi/63 \, q) }}
   \, \, \, = \, \, \,
     1 \, \, - {{15 } \over { 2097152 }} \, t^6 \, \, -{{ 45} \over {2097152 }} \, t^7 \, \, 
                     -{{2745} \over {67108864}} \, t^8
 \nonumber \\
\hspace{-0.98in}&& \, \,    \quad  \quad   \quad  \quad 
  \, \, -{{1065} \over {16777216}} \, t^9 \, \,  \, 
-{{3011265} \over {34359738368}} \, t^{10} \, \,  \, 
-{{3858885} \over {34359738368}} \, t^{11} \, \,\,  \,  + \, \, \, \cdots  
\end{eqnarray}
Recalling the series expansions (\ref{L3pi3sol}) and  (\ref{solL9pisur3}),
one actually finds that this series (\ref{newSipi3bisexp}) is exactly
\begin{eqnarray}
\label{newSipi3bisactually}
\hspace{-0.98in}&& \, \, \quad  \quad  \quad
-S_3       \, \, = \, \, \, \,
- {{ \theta_2^{(3)}(\pi/3, \, q) } \over {\theta'_2(\pi/3, \, q) }}
   \, \, \, = \, \,\, \, \, {{1} \over {3 }} \cdot \, \mathrm{Sol}(L_3^{(3)})
    \,  \, \,\, +   {{2} \over {3 }} \cdot \, \mathrm{Sol}(L_9^{(3)})
                     \nonumber \\
\hspace{-0.98in}&& \, \,    \quad  \quad  \quad  \quad   \quad  
  \, \, \, = \, \, \,\,\,
     {{1} \over {3 }} \cdot \,
     \Bigl( 4 \cdot \,  (t \, -2) \cdot \, K^2 \, +9 \cdot \, E \, K \Bigr)
   \,  \,\, \, +   {{2} \over {3 }} \cdot \, \mathrm{Sol}(A_3^{(3)}) \cdot \,  K^2.
\end{eqnarray}

{\bf Remark 1:} Let us recall the hypergeometric function
$\, t^{-7/6} \cdot \, (1\, -t)^{-1/6} \cdot \,  _2F_1([5/6,1/6],[1/3],t)$
which is an algebraic function and its order-two linear differential operator
$\, A_2^{(3)}$ (see (\ref{newSipi3Algebr})). Let us also recall
 the order-two  linear differential operator
$\, L_K$ annihilating $\, K \, = \, \, _2F_1([1/2,1/2],[1],t)$. Let us consider the order-three
linear differential operators
corresponding to the symmetric square of these two order-two  linear differential operators,
and let us consider the {\em symmetric product} of these two symmetric squares.  One gets
that way an order-{\em nine} linear differential operator:
\begin{eqnarray}
\label{Z9}
  \hspace{-0.98in}&& \, \, \quad      \quad   \quad         \quad    \quad    \quad    \quad   
    \Omega_9 \,\, = \, \, \, \mathrm{SymProd}\Bigl( \mathrm{Sym}^2(L_K),  \,  \mathrm{Sym}^2(A_2^{(3)})  \Bigr).
\end{eqnarray}
This order-nine linear differential operator $\,  \Omega_9\, $ has a structure of solutions very similar
to the one of the order-nine linear differential operator $\, L_9$.
One finds, in fact, that this order-nine linear differential
operator (\ref{Z9}) is actually non-trivially
homomorphic to the  order-nine linear differential operator $\, L_9$:
\begin{eqnarray}
\label{Z9bis}
  \hspace{-0.98in}&& \, \, \quad \quad     \quad  \quad \quad    \quad    \quad  \quad   
  I_8 \cdot \,  \Bigl(  t^{-7/3}   \cdot \,
  \Omega_9  \cdot \, t^{7/3} \Bigr) \, \,\, = \, \, \, \,  \, L_9^{(3)} \cdot \,  J_8, 
\end{eqnarray}
where   $\, I_8$ and $\, J_8$ are order-eight intertwiners. 
In conclusion the solution of the order-twelve operator corresponding
to $\, S_3$ and thus annihilating (\ref{newSipi3bis}),
is a homomogeneous  (quadratic)  polynomial of $\, E$ and $\, K$
with involved algebraic coefficients.

\vskip .1cm

{\bf Remark 2:} More generally, for $\, u \, = \, \pi/3 \, $ one has:
\begin{eqnarray}
\label{encapsulate5}
\hspace{-0.98in}&& \,  \,  \, \quad\quad  \quad   \quad  \quad   \, \,  \,   \, 
 C_{\rho}(1, \, 1; \, \rho)
 \, \, = \, \, \, 
 - 2 \cdot \, {{ \theta_2'\Bigl({{\pi} \over {3}}, q\Bigr)  } \over {\sqrt{\rho\, +3}
  \cdot \,  \theta_2(0, q) \cdot \, \theta_3(0, q)^2  }} 
\nonumber   \\
\hspace{-0.98in}&& \,  \,  \,
\quad  \quad  \quad \quad  \quad   \quad \quad  \quad   \quad  \quad  \quad    \, \,  \,   \,
 \times \, \sum_{p=0}^{\infty} \, \, 
 \Bigl( \mathrm{\arcsin} \Bigl( {{ \sqrt{\rho\, +3}} \over {2}} \Bigr) \, -{{\pi} \over {3}} \Bigr)^{p} \cdot \,
    {{ S_{(p\, +1) } \over {    p! }}},
\end{eqnarray}
where:
\begin{eqnarray}
\label{encapsulate5Sn}
\hspace{-0.98in}&& \,  \,  \,   \, \,  \,   \,
\quad   \quad  \quad  \quad  \quad  \quad  \quad  \quad   \quad
  S_n \, \, = \, \, \,  {{ \theta_2^{(n)}(\pi/3, \, q) } \over {\theta'_2(\pi/3, \, q) }}.  
\end{eqnarray}

\vskip .1cm

\section{$\lambda$ corresponds to the critical  exponent at $\, t = 1$\\ }
\label{lambdat1}

The lambda extensions $\, C(1,1; \, \lambda) \, $ are a one-parameter family of solutions of the
Okamoto-Painlev\'e VI equation (\ref{jmequation}). It is worth noticing that the parameter lambda
cannot be seen in the non-linear ODE (\ref{jmequation}).  {\em It is not a parameter of the non-linear ODE}
(\ref{jmequation}). 
The parameter lambda {\em actually fixes the critical exponent at} $\, t \, = \, 1$  of the
solution  $\, C(1,1; \, \lambda)$.

Paper~\cite{Connection} gives, in equation (13) and (14), the behaviour of the lambda extensions
$\, C(N,N, \lambda)$ near\footnote[8]{Here $\, \sigma$ is an exponent, which has nothing to do with the
  $\, \sigma$ functions (\ref{sigmam}) or (\ref{sigmap}). Painlev\'e papers
  are famous for their terrible notations.}
$\, t \, = \, 1$:
\begin{eqnarray}
\label{neart1}
  \hspace{-0.98in}&& \, \,\quad \quad 
 C(N,N, \lambda) \,  \,  \simeq  \,  \, \, K(N, \, \sigma) \cdot \, (1\, -t)^{\sigma^2/4}
    \,  \,    \quad \, \,\,   \hbox{where:}  \,  \,  \, \,  \, \,  \,
   \sigma \, = \, \, {{2} \over { \pi}} \cdot \, \mathrm{\arccos}(\lambda),     
\end{eqnarray}
or denoting $\,\,   \lambda \, = \, \mathrm{\cos}(u)$:
\begin{eqnarray}
\label{neart1bis}
  \hspace{-0.98in}&& \, \,\quad \quad \quad \quad \quad \quad \quad \quad \quad
 C(N,N, \lambda) \, \, \, \,     \simeq  \, \,\, \,    \, K(N, \, \sigma) \cdot \, (1\, -t)^{(u/\pi)^2}.    
\end{eqnarray}
One verifies that this power-law formula\footnote[1]{It is very hard to get this result from the
  exact expression (\ref{Saga}) of  $\, C(1,1; \, \lambda) \, $
  in terms of theta functions. } (\ref{neart1bis}) 
is actually valid for all the algebraic expressions displayed in section \ref{lambdaPi4} (see (\ref{actually})),
section \ref{lambdaPi6}  (see (\ref{algM6})),  and   section \ref{lambdaPi3}  (see (\ref{algM3})):

\vskip .1cm

$\bullet$ For $\, \lambda\, = \, 0$, i.e. $\, u \, = \, \pi/2 \, $ one has
a $\, (1\, -t)^{1/4}$ behaviour.

\vskip .1cm

$\bullet$ For $\, \lambda\, = \, 1/\sqrt{2}$, i.e. $\, u \, = \, \pi/4 \, $
one has a $\, (1\, -t)^{1/16} \, $
behaviour (see (\ref{actually}) and (\ref{actually2G0})).

\vskip .1cm

$\bullet$ For $\, \lambda\, = \, \sqrt{3}/2$, i.e. $\, u \, = \, \pi/6 \, $
one has a $\, (1\, -t)^{1/36} \, $ behaviour:
from (\ref{algM6}) one actually gets:
$\, S(t) \, = \,\, 2^{8/9}/3 \cdot \, (1\, -t)^{1/36} \,  \, + \, \cdots $

\vskip .1cm

$\bullet$ For $\, \lambda\, = \, 1/2$, i.e. $\, u \, = \, \pi/3$,
one has a $\, (1\, -t)^{1/9} \, $ behaviour:
from (\ref{algM3}) one actually gets:
$\, \,  S(t) \, = \,\,\,   2^{14/9} \cdot \, 3^{-3/2} \cdot \, (1\, -t)^{1/9} \,  \, + \, \cdots $

\vskip .1cm
\vskip .1cm
\vskip .1cm

\section{Comments and speculations.\\ }
\label{commentspec}

All these calculations, displayed on the low-temperature correlation function $\, C(1, \, 1)$, 
illustrate the extremely rich structures of the lambda extensions
of the two-point square\footnote[2]{One has similar results for the
triangular, honeycomb, ... lattices.
One has similar results for the high-temperature correlation functions. One
has similar results
for the anisotropic correlation functions $\, C(M,\, N)\, $ for $\, \nu = \, -k$.}
Ising correlation functions $\, C(M,\, N)$. For an {\em infinite set} of values of lambda
($\lambda = \, \mathrm{\cos}(\pi \, m/n)$, $m$ and $\, n$ integers)
these lambda extensions become {\em algebraic functions} and for another infinite set of values
of lambda ($\lambda = \,  (m/n)^{1/2}$, $m$ and $\, n$ integers) the series expansions of the
lambda extension are  {\em globally bounded series}~\cite{ChristolDiag} that are
{\em not} D-finite\footnote[5]{Except when
  $\, \lambda \, = 0, \, 1/\sqrt{2}, \, 3/\sqrt{2},\, 1$
where $\, \lambda$ is also of the form $\lambda = \, \mathrm{\cos}(\pi \, m/n)$.}
but {\em only differentially algebraic} (the corresponding
$\, \sigma$ are solutions of a sigma-form of Painlev\'e VI).

Furthermore we have seen, in section \ref{undergraduate}, that 
the ``form-factor-like'' expansions (\ref{morenaturallyM1}) around the (D-finite) two-point
correlation function $\, C(1,1) \, = \, E$, yield new ``form factors'' $\, g_n(t)$'s 
which, at first sight, should be DD-finite
expressions (see section (\ref{Atfirstsight})), 
{\em are, actually, D-finite expressions}. The $\, g_n(t)$'s are, in fact, 
{\em polynomial expressions in} $\, E$ and $\, K$.

The ``form-factor-like''~\cite{Holonomy} expansions  around the
infinite set of algebraic functions at $\lambda = \, \mathrm{\cos}(\pi \, m/n)\, $ 
yield new ``form factors'' $\, G_n(t)$'s (see (\ref{actually2})) which turned out to be
{\em  D-finite expressions}: they are solutions of linear differential 
operators with (quite involved) algebraic functions coefficients. We showed
that  the first $\, G_n(t)$'s are actually {\em polynomial expressions in}
$\, E$ and $\, K$ and, hopefully, one can expect that all these $\, G_n(t)$'s are
polynomial expressions in $\, E$ and $\, K$ (with involved  algebraic functions coefficients).

These results correspond to the (quite puzzling) fact that rational expressions
of the derivatives (at selected values of $\, u$) of {\em Jacobi theta functions}
(like (\ref{Sagag1})) 
can, in fact, be expressed as polynomial expressions in $\, E$ and $\, K$, thus providing
{\em an infinite set of remarkable identities between theta functions and complete
  elliptic integrals of the first and
second kind}\footnote[8]{For identities on products of ratio of Jacobi theta functions see
for instance~\cite{Kare}}. Such calculations provide an infinite set of new D-finite expressions
on the two-dimensional Ising model that will join together with all the previous 
D-finite expressions we have altready encountered  on the two-dimensional Ising model
as $\, n$-fold integrals
that are diagonals of rational functions~\cite{ChristolDiag,2F1,Heun,From}. This
corresponds to the kind of holonomic (i.e. D-finite) studies 
we are used to perform on the  two-dimensional Ising model~\cite{Holonomy,Khi3} in the variable
$\, t\, = \, k^2$. These D-finite expressions emerge from form factor-like
perturbation theory (the kind of perturbation theory physicists are used
to with Feynman diagrams, Periods of algebraic varieties, ...). However, we also see
that the lambda extension $\, C(1,1; \, \lambda) \, $
which is differentially algebraic (solution of a {\em non-linear} ODE (\ref{jmequation})  
with the Painlev\'e property of fixed critical points~\cite{LastJPA}),
 can be understood ``holistically'', globally,
and not using the bread and butter perturbative
physicist's approach, if one switches to a  description
{\em in terms of the nome} $\, q$ (or the ratio $\, \tau$ of the two periods
of the elliptic function) requiring to introduce intensively Jacobi
theta functions~\cite{Holonomy,Saga,LastJPA}. With that alternative holistic
description one has a rather simple exact closed formula for the lambda extension
(see (\ref{Saga})). The ``price to pay'' is that
this exact and elegant holistic expression of the lambda extension (like (\ref{Saga})) is
solution of a {\em non-linear} ODE (\ref{jmequation})  and, for instance, the emergence of
all the D-finite expressions, displayed in this paper, is not obvious
from that non-linear differential equations or
Jacobi theta functions viewpoint~\cite{Schwarzian}.

\vskip .1cm

\subsection{Painlev\'e VI transcendentals as deformations of elliptic functions and
  why theta functions are well-suited:  Jacobi forms\\ }
\label{deformManinsub}

\vskip .1cm

The occurrence of {\em Jacobi theta functions}~\cite{Nova,Jacobi}
for the exact closed expression  (\ref{Sagag1}) of
the lambda extension solution of sigma-form of Painlev\'e VI is, in fact,  highly relevant
as far as all the symmetries of the model are concerned.

Let us first recall that Painlev\'e VI transcendents should be seen as
{\em deformations of elliptic functions}~\cite{Manin}. Along this
line it is worth recalling
Manin's idea~\cite{Manin} that the Painlev\'e VI equation
for a particular choice of the four
Okamoto parameters, can be written extremely simply in terms of
 the ratio of periods $\, \tau$.
Let us denote $\,  {\cal P}(z, \, \tau)$  the  $\,  {\cal P}$-Weierstrass function
and  $\, {\cal P}_z(z, \, \tau) \, = \, \,
{{ \partial {\cal P}(z, \, \tau) } \over { \partial z}}$.
Manin's result means that the Painlev\'e VI equation can
be written in a form (see equation (1.16) in~\cite{Manin}): 
\begin{eqnarray}
  \label{Manin_sum}
 &&\hspace{-.58in} \quad \quad \quad \quad 
        {{ d^2 z(\tau)} \over { d\tau^2}} \, \,\, = \, \, \, \,
        \Bigl({{1} \over {2\, \pi \, i}}\Bigr)^2 \cdot \, \sum_{i=0}^{3}
\alpha_i  \cdot \, 
 {\cal P}_z\Bigl(z\, + {{T_i} \over {2}}, \, \tau\Bigr).
\end{eqnarray}
In previous studies of the $\, C(M, \, N)$ correlation functions
and their non-linear Painlev\'e ODEs, we have
underlined the fundamental role of Landen transformations~\cite{LastJPA}. 
The crucial role of {\em Landen transformations} is underlined in~\cite{Heegner,LastJPA,Manin}.  
It is also worth  recalling that the Weierstrass ${\cal P}$-function is simply
related to theta functions.
The Weierstrass ${\cal P}$-function is
related\footnote[1]{The constant $\, c$ is defined so that the Laurent expansion
of  $\, {\cal P}(u, \, \tau)$
at $\, u \, = \, 0$ has zero constant term ($\theta_1'(0, \, q)$
is the derivative with respect to $\, u$),
see (B.7), (B.8) in~\cite{Zabrodin}.
  See for instance https://handwiki.org/wiki/Theta$\_$function
in the paragraph Relation to the Weierstrass elliptic function. See also~\cite{Ohyama}.} to the
second log derivative of $\, \theta_1(u, \, q)$:
\begin{eqnarray}
 \label{Weierstrass}
 &&\hspace{-.98in} \quad 
    {\cal P}(u, \, \tau) \, \, = \, \, \, \,
    - \, { {\partial^2 \ln(\theta_1(u, \, \tau))} \over { \partial u^2}} \, \, \,  +   c
    \, \, = \, \, \, \,
    - \, { {\partial^2 \ln(\theta_1(u, \, \tau))} \over { \partial u^2}} \, \,
   + \, {{1} \over {3}} \, {{\theta_1'''(0, \, q)} \over { \theta_1'(0, \, q)}},
\end{eqnarray} 
The  closed expressions  (\ref{Sagag1}) for the lambda-extension $\, C(1,1; \, \lambda)$
underlines the occurrence of the partial derivative with respect to the $\, u$-deformation parameter
(or equivalently the lambda parameter). Along this line one can recall another 
interesting property of the theta functions. They are solutions of the heat equation:
\begin{eqnarray}
 \label{theta}
 &&\hspace{-.58in} \quad  \quad \quad  \quad 
 { {\partial \theta(u, \, \tau)} \over { \partial \tau}}
 \, \, = \, \, \, q \cdot \, { {\partial \theta(u, \, q)} \over { \partial q}} \, \, = \, \, \,
 { {\partial^2 \theta(u, \, q)} \over { \partial u^2}}.  
\end{eqnarray}
Consequently, and to some extent,
the partial derivatives in $\, u$ can be replaced by partial derivatives in $\, \tau$.

It is also worth  mentioning the {\em modular group relations}
on the Weirstrass ${\cal P}$-functions
as well as the similar  ``modular group transformations'' on the theta functions~\cite{Farkas,Kloosterman}:
\begin{eqnarray}
\label{WeierstrassPmodgroup}
\hspace{-0.98in}&& \quad \quad \quad  \quad \quad  \quad 
{\cal P}\Bigl( {{z} \over {c\, \tau \, +d }}, \, \, {{a\, \tau \, + \, b} \over {c \, \tau \, +d }} \Bigr)
\, \,\, = \, \, \,
(c\, \tau \, +d)^2 \cdot \, {\cal P}(z, \, \tau),
  \\
\hspace{-0.98in}&& \quad \quad \quad  \quad \quad  \quad 
{\cal P}_z\Bigl( {{z} \over {c\, \tau \, +d }},
       \, \, {{a\, \tau \, + \, b} \over {c \, \tau \, +d }} \Bigr)
\, \, \,= \, \, \,
(c\, \tau \, +d)^3 \cdot \, {\cal P}(z, \, \tau),     
\end{eqnarray}
and\footnote[5]{See equation (2.16) in~\cite{Farkas}.  }
\begin{eqnarray}
\label{Thetamodgroup}
\hspace{-0.98in}&&  \, \, \,\,\,
\kappa \cdot \, (c\, \tau \, +d)^{1/2} \cdot \, \theta_{\alpha}(u, \, \tau)
\, \,\, = \, \, \,\,\,
\exp\Bigl(- \, i\, \pi  {{ c u^2} \over { c\, \tau \, + d }}\Bigr) \cdot \, 
\theta_{\beta}\Bigl({{u} \over {c\, \tau \, +d }},
\, \, {{a\, \tau \, + \, b} \over {c \, \tau \, +d }}   \Bigr),
\end{eqnarray}
where $\, \kappa \, \, $ is a constant, and where the integers $\, a, \, b, \, c, \, d$
are such that $\, a \, d \, -\, b\, c \, = \, 1$.
For $\, u \, = \, \, 0 \, $ the previous modular group transformations
(\ref{WeierstrassPmodgroup}), (\ref{Thetamodgroup})
is reminiscent of the modular forms of weight $\, k$:
\begin{eqnarray}
\label{modularform}
\hspace{-0.98in}&& \quad \quad \quad  \quad \quad  \quad \quad  \quad \quad
(c\, \tau \, +d)^{k} \cdot \, f(\tau) \, \, = \, \, \,
 \,
f\Bigl( {{a\, \tau \, + \, b} \over {c \, \tau \, +d }}   \Bigr).
\end{eqnarray}
With some abuse of language we could say that the theta functions are ``some kind'' of
``modular forms of weight $\, 1/2$''.

Recalling the relation (\ref{Saga}) between $\, \lambda$ and $\, u$,   
{\em the theta functions thus provide, because of} (\ref{Thetamodgroup}),
{\em   some natural $\, u$-extension, and thus
lambda-extension, of the modular forms} (Jacobi forms).
From the  closed expression  (\ref{Sagag1}) it is clear that the lambda-extension will naturally
inheritate from (\ref{Thetamodgroup}), some symmetry properties
{\em  with respect to the modular group}.
This kind of global (holistic) symmetry is almost impossible to see in the holonomic (D-finite)
world of the linear differential operators in the variable $\, t$. Conversely all the D-finite
results, we have displayed in this paper, are not an obvious consequence of the emergence of
$\, \theta'_2(u, \, q)$ in  (\ref{Sagag1}). All these D-finite results are ``hidden''
in the theta functions (considered at selected values of $\, u$). This is similar to the
situation one encounters with
{\em modular forms}~\cite{Heun,Modular,Schwarz} where the fact that they
are D-finite in the variable $\, t$ is not totally
straightforward\footnote[1]{See in particular Proposition 21 page 61 in~\cite{Zagier2}. 
  One can find in~\cite{Lester} why automorphic forms are solutions
  of linear differential equations}.

\vskip .1cm
\vskip .1cm
\vskip .1cm

\section{Conclusion}
\label{Conclusion}

The lambda-extensions of the two-point correlation functions $\, C(M,N)$
of the square Ising model are
a good illustration of the mirror-map $ \, t \, \leftrightarrow \, q \, $ duality 
in mirror symmetries~\cite{Doran2,LianYau,Straten}, where all the
holonomic (D-finite) structures
are well seen in the $\, t$ variable but are hard to see
in the nome~\cite{Saga,IsingCalabi} $\, q$
(or in the ratio of periods $\, \tau$),
and conversely the modular group, modular forms structures are easily seen in the nome $\, q$
variable  (or in the ratio of periods $\, \tau$) but are very  hard to see
in the original $\, t$ variable.
In the $\, t$ variables the perturbative
approach provides a large set of D-finite expressions 
which are $\, n$-fold integrals (and in fact diagonals of rational
functions~\cite{ChristolDiag}),
when the description in the nome variable (or the $\, \tau$ variable)
provides a holistic understanding (see (\ref{Saga})) which makes crystal clear modular
group symmetries and the emergence of  Landen transformations~\cite{Heegner,LastJPA},
and of modular forms~\cite{Heun,Ohyama,Modular}, but requires
to consider non-linear ODEs~\cite{Modular,Schwarz,LastJPA}.
Both descriptions are complementary and necessary to describe efficiently
these lambda-extensions.

Focusing, for pedagogical reasons, on a very simple example of lambda-extension,
namely $\, C(1,1; \, \lambda)$,
we have considered the series expansion in $\, t$ as different form-factor-like
expansions around
the D-finite subcase $\, C(1,1) \, = \, E$ or a large set of algebraic functions subcases
(see (\ref{recalling}),  (\ref{actually}),  (\ref{algM6}),  (\ref{algM3})). For the first
form-factor-like expansion (\ref{morenaturallyM1}), 
the corresponding form-factors $\, g_n(t)$, which should, at first sight,
be DD-finite, turn out to be D-finite and simple polynomials of the
complete elliptic integrals of the first
and second kind $\, K$ and $\, E$. On the other hand, the form-factors $\, G_n(t)$,
corresponding to a deformation around
the algebraic functions subcases of the lambda-extension, have been seen to be  D-finite,
and, either, shown to be polynomials of  $\, K$ and $\, E$, or can be very
reasonably conjectured to be
polynomials of  $\, K$ and $\, E$.  These results can be seen as remarkable, non-trivial
(and rather unexpected ...), identities between ratio of Jacobi theta functions and
the complete elliptic
integrals of the first and second kind $\, K$ and $\, E$.

These identities are a nice illustration of this complementary description of
the D-finite $\, t$-variable
(elliptic integrals) viewpoint and the non-linear (modular group,
Jacobi theta functions~\cite{Holonomy,Saga,Ohyama})
nome viewpoint.  

\vskip .1cm
\vskip .1cm

\vskip .3cm

{\bf Acknowledgments.} We wish to thank Pr B. M. McCoy for support, interesting suggestions and
for so many lambda-extension exchanges during the achievement of this work.
J-M. Maillard would like to thank I. Dornic and R. Conte for many fruitful discussions on Painlev\'e equations.
 JMM also wants to thank A. Bostan and J-A. Weil for many fruitful diff. algebra discussions.

\vskip .1cm
\vskip .1cm
\vskip .1cm

\vspace{.1in}

\appendix

\section{Calculation of the coefficient $\, g_3(t)$}
\label{g3}

The series $\, g_3(t)$
can also be seen to be D-finite, being
solution of an order-{\em twelve} linear differential operator which turns out to be
the direct-sum (LCLM) of the
previous order-two linear differential operator $\, L_E$, of the previous
order-four $\, L_4$, of the previous order-six linear differential
operator homomorphic to the symmetric fifth power
of $\, L_E$, and of an order-{\em eight} linear differential
operator homomorphic to the {\em symmetric seventh power}
of $\, L_E$, with the following order-seven intertwiner
\begin{eqnarray}
\label{R3}
\hspace{-0.99in}&& \quad 
{{256} \over {315}} \cdot \, R_3 \, \,  \, = \, \, \, \,
      {{1} \over {8}}  \cdot \, (t-1) \cdot \, (8\,t^2-33\,t+33) \cdot \, t^7  \cdot \, D_t^7
\nonumber \\
\hspace{-0.99in}&&    \quad  \quad 
+ {{7} \over {16}} \,  \cdot \, (t-1) \cdot \, (40\,t^2 -173\,t +181) \cdot \, t^6  \cdot \, D_t^6
\nonumber \\
\hspace{-0.99in}&&     \quad  \quad 
+  {{7} \over {32}} \cdot \, (360\, t^3-2077\, t^2+3795\, t-2166)) \cdot  \, t^5 \cdot \, D_t^5
 \\
\hspace{-0.99in}&&     \quad  \quad 
+{{35} \over {64}} \,  \cdot \, {{120\, t^4-975\, t^3+2968\, t^2-3933\, t+1900} \over {t-1}}
\cdot\, t^4\, D_t^4 \, \,
- {{7} \over {128}}\,  \cdot \, {{ q_5 } \over { (t-1)^2}} \cdot \,t^3\, D_t^3
 \nonumber 
\end{eqnarray}
\begin{eqnarray}
\label{R3suite}
\hspace{-0.99in}&&   \quad  \quad    \quad    \quad  
+{{7} \over {256}}\,  \cdot \,
{{  q_6  } \over { (t-1)^3}} \cdot \, t^2 \cdot  \,D_t^2
\, \, \, \, \,    -{{1} \over {512}}  \cdot  \,  {{ q_7  } \over {(t-1)^4}} \cdot \, t \cdot \,  D_t
\, \, \, \,  \,  +{{7 } \over {1024}}  \cdot  \,  {{ q_8  } \over {(t-1)^5}},
  \nonumber 
\end{eqnarray}
where the $\, q_n$ polynomials read:
\begin{eqnarray}
\label{R3suitesuitesuiteqn}
\hspace{-0.98in}&& \quad  \quad 
q_5 \,\, = \, \,\,
600\, t^5 -5379\, t^4 +16550\, t^3 -15061\, t^2-8708\, t \, +13854,
\nonumber \\
\hspace{-0.98in}&& \quad  \quad 
q_6 \,\, = \, \,\,
1080\, t^6 -10287\, t^5 +30197\, t^4 -9695\, t^3 -59739\, t^2+51338\, t \, +4402, 
\nonumber \\
\hspace{-0.98in}&& \quad  \quad 
q_7 \,\, = \, \,\,
12600\, t^7-125991\, t^6 +346295\, t^5 +108127\, t^4 \, -1210745\, t^3
\nonumber \\
\hspace{-0.98in}&& \quad \quad \quad \quad  \quad   \quad 
  +868060\, t^2 +142022\, t \, +4016, 
\nonumber \\
\hspace{-0.98in}&&   \quad  \quad 
q_8 \,\, = \, \,\,
1800\, t^8 -18801\, t^7 +47986\, t^6 +43466\, t^5 -233350\, t^4 +147125\, t^3
 \nonumber \\
  \hspace{-0.98in}&& \quad \quad \quad \quad \quad  \quad 
 +40936\, t^2 \,  +1378\, t \, +180.  
\end{eqnarray}
One finally finds that the series expansion for $\, g_3(t)$ is exactly the
linear combination of  $\, E$, of the order-three linear differential  operator (\ref{R1})
acting on $\, E^3$, of an order-five linear differential  operator (\ref{R2}) acting on $\, E^5$
and the order-{\em seven} linear differential  operator (\ref{R3}) acting on $\, E^5$:
\begin{eqnarray}
  \label{gg2}
  \hspace{-0.98in}&& \,
g_3(t) \,\, = \, \,\,\,
{{5} \over {7168}} \cdot \, E \,\,\,
+  {{37} \over {46080}} \cdot \,  R_1(E^3) \,\,\,
-  {{1} \over {9216}} \cdot \,  R_2(E^5)\,\,\,
+  {{1} \over {322560}} \cdot \,  R_3(E^7)
\nonumber \\   
  \hspace{-0.98in}&& \quad  \quad 
     \,\, = \, \,\, \,   {{5} \over {7168}} \,\cdot  \, E \, \, \, \, 
     -{{37} \over {15360}} \,\cdot  \, K\, \, E^2 \,  \,  \,  \,
     -{{37} \over {23040}} \,\cdot  \, (t-1)\,\cdot  \, K^3\,
\nonumber \\   
  \hspace{-0.98in}&& \quad \quad \, \,\quad  \, 
  +{{5} \over {3072}} \,\cdot  \, K^2\, \, E^3\, \,  \,
  +{{5} \over {1536}} \,\cdot  \, (t-1)\,\cdot  \, K^4\, \, E \,  \, \,
  +{{1} \over {1152}} \,\cdot  \, (t-1)\, \cdot \, (t-2)\,\cdot  \, K^5
\nonumber \\   
\hspace{-0.98in}&& \quad \quad \,  \,   \,  \,  \, \,
-{{1} \over {3072}}  \,\cdot  \, K^3\, \, E^4 \, \,
-{{1} \over {768}} \,\cdot  \, (t-1)\, \cdot  \, K^5\,  \, E^2 \, \, \,
-{{1} \over {1440}} \,\cdot  \, (t-1)\,\cdot  \, (t-2)\,\cdot  \, K^6\, \, E
\nonumber \\   
\hspace{-0.98in}&& \quad \quad \quad \quad \quad \quad \quad \,  \,
-{{1} \over {80640}} \,\cdot  \,
(t-1)\,\cdot  \, (8\, \, t^2-33\, \, t+33)\, \cdot  \, K^7.
\end{eqnarray}

\section{Low temperature lambda extension $\, C_{-}(0,0, \, \lambda)$}
\label{Appendixchin}

Similarly to the Taylor expansion (\ref{encapsulate2}), we can write a
similar identity for  the lambda extension $\, C(0,0, \, \lambda)$.
Introducing
\begin{eqnarray}
\label{encapsulate71}
\hspace{-0.98in}&& \,  \,  \,   \, \,  \,   \,
\quad  \quad \quad  \quad  \quad  \quad  \quad  \quad  \quad   \quad
  S_n \, \, = \, \, \,  {{ \theta_4^{(n)}(0, \, q) } \over {\theta_4(0, \, q) }},  
\end{eqnarray}
the lambda extension $\, C_{-}(0,0, \, \lambda) \, $ can be written
\begin{eqnarray}
\label{encapsulate7}
\hspace{-0.98in}&& \,  \,  \,   \quad  
C_{-}(0, \, 0; \, \lambda) 
\, \, = \, \, \,\,  \, {{ \theta_3(\mathrm{\arccos} \, \lambda, \, q)} \over {\theta_3(0, q)}}
\nonumber \\
  \hspace{-0.98in}&& \,  \,  \,    \quad  \quad
\, \, = \, \, \,\,  \, (1\, -t)^{1/4}  \cdot  \,
\sum_{p=0}^{\infty} \, \, \Bigl(\mathrm{\arcsin} \lambda\Bigr)^{(2\, p)} \cdot \,                   
     {{ S_{2\, p} } \over {(2\, p)!}}
\nonumber \\
  \hspace{-0.98in}&& \,  \,  \,    \quad  \quad
 = \, \, \,  (1\, -t)^{1/4} \cdot \Bigl( 1 \,\, \, + {{S_2} \over {2}}  \cdot \,  \lambda^2
  \,   \, +  \Bigl(  {{S_2} \over {6}}  \, + {{S_4} \over {24}}   \Bigr) \cdot \,  \lambda^4
  \, \, + \Bigl(  {{ 4 \, S_2} \over {45}}  \, + {{S_4} \over {36}} \, + {{S_6} \over {720}}     \Bigr)
  \cdot \,  \lambda^6
 \nonumber \\
 \hspace{-0.98in}&& \,  \,  \,
 \quad \quad    \quad  \quad   \quad   \quad  \quad   \quad \quad     \quad  \quad  \quad \quad            
  \, +  \Bigl(  {{ 2 \, S_2} \over {35}}  \,
  + {{7 \, S_4} \over {360}} \, + {{S_6} \over {720}}   \, + {{S_8} \over {40320}}     \Bigr)
  \cdot \,  \lambda^8
 \,\,\,\,\,\, + \, \,\, \cdots \Bigr) 
\nonumber \\
  \hspace{-0.98in}&& \,  \,  \,    \quad \quad  \quad  \quad 
  \, \, = \, \, \,\,  \,
  (1\, -t)^{1/4}  \cdot  \,  \sum_{p=0}^{\infty} \, \kappa^{(2\, p)} \cdot \,     \lambda^{(2\, p)},                
\end{eqnarray}
where $\, \kappa^{(2\, p)} = \, f_{0,0}^{(2\, p)}$.
From (\ref{encapsulate7}) one can deduce the expression of the
$\, S_{2\, p}$'s in terms of the $\, \kappa_{(2\, p)}$'s:
\begin{eqnarray}
\label{encapsulate72}
\hspace{-0.98in}&& \,  \,  \,    \, \quad  \quad  \quad
 {{S_2} \over { 2!}}  \,  \,  =  \, \,  \,  \kappa^{(2)},
\nonumber \\
\hspace{-0.98in}&& \,  \,  \,   \quad \quad      \quad            
 {{S_4} \over { 4!}}  \,  \,  =  \, \,  \, \kappa^{(4)} \, -{{ 1} \over {3}} \, \kappa^{(2)},
  \nonumber \\
\hspace{-0.98in}&& \,  \,  \,   \quad \quad    \quad              
 {{S_6} \over { 6!}}  \,  \,  =  \, \,  \,
   \kappa^{(6)} \,  - \, {{ 2 } \over {3}}  \, \kappa^{(4)}\,  +{{ 2 } \over {45}}\, \kappa^{(2)},
\nonumber \\
\hspace{-0.98in}&& \,  \,  \,   \quad \quad   \quad               
  {{S_8} \over { 8!}}  \,  \,  =  \, \,  \,
  \kappa^{(8)} \,  -\kappa^{(6)} \,
  + \, {{ 1 } \over {5}}\, \kappa^{(4)} \, -{{ 1} \over {315}} \,  \kappa^{(2)},
\nonumber \\
\hspace{-0.98in}&& \,  \,  \,   \quad \quad    \quad              
 {{S_{10}} \over { 10!}}  \,  \,  =  \, \,  \,
  \kappa^{(10)} \, - {{ 4 } \over {3}} \, \kappa^{(8)} \,  +{{ 7 } \over {15}}\, \kappa^{(6)}
  \,  - \, {{ 34 } \over {945}} \, \kappa^{(4)} \, +{{ 2 } \over {14175}}\,  \kappa^{(2)},
  \,   \quad \quad      \cdots            
\end{eqnarray}

\vskip .1cm

\section{Exterior squares and absolute factorisation.}
\label{AppendixA}

\subsection{Absolute factorisation.}
\label{AppendixAabsolu}

Let us recall a simple example of an {\em absolute factorisation} of an
order-four linear differential operator given in~\cite{CompointWeil}:
\begin{eqnarray}
\label{exampleabsolute}
  \hspace{-0.98in}&& \, \,\, \,  \quad \quad \quad \quad  \quad \quad  
A_4 \,\,  = \, \, \,
D_t^4  \,   \,  -{{1} \over { t}}\cdot \, D_t^3 \,  +{{3} \over {4\, t^2}} \cdot  \, D_t^2 \, -t
\nonumber \\
\hspace{-0.98in}&& \, \,\, \,  \quad \quad  \quad \quad   \quad \quad \quad  \quad  
\,\,  = \, \, \,
\Bigl( D_t^2 \, -{{1} \over { t}} \cdot \, D_t + {{3} \over {4\, t^2}} \, +\sqrt{t}  \Bigr)
 \cdot \, \Bigl( D_t^2 \, - \sqrt{t}  \Bigr).
\end{eqnarray}
The fact that such  a {\em factorisation over an algebraic extension} of $\, \mathbb{C}(t)$
exists can be deduced~\cite{CompointWeil} from the fact that one has a direct-sum (LCLM)
decomposition of the (order-five) exterior square
of the order-four linear differential operator $\, A_4$:
\begin{eqnarray}
\label{exampleabsolutedirect}
  \hspace{-0.98in}&& \, \,\, \,  \quad  \quad 
\mathrm{Ext}^2(A_4) \, \, = \,\, \,\,
    D_t \oplus \,
    \Bigl( D_t^4  \,   -{{3} \over {2 \, t}} \cdot \, D_t^3
    \, + {{9} \over {4\, t^2}} \cdot  \, D_t^2
 \,  -{{15} \over {8 \, t^3}} \cdot  \, D_t \, +4 \, t \Bigr).
\end{eqnarray}

\vskip .1cm

\subsection{Exterior square of $\, M_4$ and absolute factorisation of $\, M_4$ .}
\label{AppendixAM4}

Let us now study here the order-four linear differential operator
$\, M_4$ occurring in section \ref{otherpisur6}
for the  deformations of $\,  u \, = \pi/6$.

The order-four linear differential operator $\, M_4$ 
is slightly more difficult to analyse than the first order-four
linear differential operator $\, L_4$
in (\ref{otherpisur6}).
We seem to have  a solution of this order-four linear differential operator
$\, M_4$ of the form
$\, \, \, \alpha(t) \cdot \, E \, + \,  \beta(t) \cdot \, K$,
$\, \alpha(t)$ and  $\,  \beta(t)$ being (very) involved algebraic functions,
however finding a symmetric product form,
like in the previous order-four linear differential operator $\, L_4$,
is difficult.  Let us show, in a quite indirect way, that this is probably the case.
Let us consider the exterior square of this order-four linear differential operator $\, M_4$.
This is an order-six linear differential operator $\, M_6$, which is actually the direct-sum
(LCLM) of two order-three linear differential operators $\, A_3$ and $\, B_3$
\begin{eqnarray}
\label{LCLM}
\hspace{-0.98in}&& \, \,\, \,  \quad \quad \quad \quad
   M_6 \, \, = \, \, \, \mathrm{Ext}^2(M_4) \, \, = \, \, \,
            \mathrm{\ LCLM}(A_3, \, B_3) \, \, = \, \, \,  A_3 \, \oplus \, B_3, 
\end{eqnarray}
where one finds easily that the first  order-three linear differential operator $\, A_3$
corresponds to {\em algebraic solutions} associated with the polynomial equation:
\begin{eqnarray}
\label{LCLM}
  \hspace{-0.99in}&& 
  (16\, t^{17} \, -184\, t^{16} \, -135149\, t^{15} \, +1128329\, t^{14} \, -6708683\, t^{13} \, +26956928\, t^{12}
\nonumber \\
\hspace{-0.99in}&& \, \,\, \,  \quad 
 -65809991\, t^{11} +96341783\, t^{10}-88006226\, t^9 +63929329\, t^8-60215242\, t^7
\nonumber \\
  \hspace{-0.99in}&& \, \,\, \,  \quad
 +59165527\, t^6-37633087\, t^5+12783832\, t^4-1787515\, t^3 -7679\, t^2-1957\, t-32)
\nonumber \\
\hspace{-0.99in}&& \, \,\, \,  \quad 
+4\, (t-1) \, (t^2-t+1)\, (20\, t^{15}-186\, t^{14}-20481\, t^{13}+138367\, t^{12} -473685\, t^{11}
 \nonumber \\
  \hspace{-0.99in}&& \, \,\, \,  \quad \quad 
 +1069635\, t^{10}-1516399\, t^9 +1115037\, t^8-53199\, t^7-617857\, t^6+547761\, t^5
\nonumber \\
  \hspace{-0.99in}&& \, \,\, \,  \quad \quad \quad 
 -255237\, t^4+78967\, t^3-12885\, t^2+156\, t-16) \cdot \, y(t)
 \\
\hspace{-0.99in}&& \,  \, 
\, +18 \cdot  \, (8\, t^6-33\, t^5-447\, t^4+943\, t^3-447\, t^2-33\, t+8)
\, (t-1)^2\, (t^2-t+1)^5 \, t \cdot \, y(t)^2
  \nonumber \\
\hspace{-0.99in}&& \, \,\, \,  \quad 
\, +108  \cdot \, (t-1)^4 \, (t^2-t+1)^7 \cdot \, y(t)^3
\, \, +27\, t^3\, (t-1)^4\, (t^2-t+1)^7 \, t^2 \cdot \, y(t)^4 \, \, \, = \,\,\,  \, 0.
\nonumber 
\end{eqnarray}
The second order-three linear differential operator $\, B_3$ is {\em homomorphic}\footnote[1]{With
order-two intertwiners $\, I_2$ and $\, J_2$.} to
the symmetric square of an order-two linear differential operator
$\, L_2$ which is simply conjugated to the
order-two linear differential operator $\, L_K$ annihilating the complete elliptic integral
of the first kind $\, K \, = \, _2F_1([1/2,1/2],[1],t)$
\begin{eqnarray}
\label{B_3}
\hspace{-0.98in}&& \quad \quad
B_3 \, \cdot \,  I_2 \, \, = \, \, \,   J_2 \cdot \,  \mathrm{Sym}^2(L_2),
 \quad  \quad  \quad  \quad  \quad \hbox{where:} 
 \\
 \hspace{-0.98in}&& \, \,\, \,  \, \, 
\label{B_3L2}
L_2   \, \, = \, \,  \,  {{1} \over {\rho(t)}} \cdot \, L_K \cdot \, \rho(t)
\, \, \, = \, \, \,\,\,  
 D_t^2 \, \, + {{4} \over {3}} \cdot \,  {{2\,t \, -1 } \over { t \, (t-1)}} \cdot \, D_t
    \, \, + {{ 25\, t^2 \, -25\,t \, +1 } \over { 36 \,  t^2 \, (t-1)^2}}, 
\end{eqnarray}
where $\,\,  \rho(t) \, = \, \, t^{1/6} \cdot \, (1\, -t)^{1/6}$.
It is worth comparing these results with similar calculations (see \ref{Appendix} for a
general identity on exterior square of symmetric products and direct sum of
symmetric square) for the first order-four
linear differential operator $\, L_4$ in section \ref{otherpisur6} which was the direct-sum of two 
linear differential operators (\ref{twoOper}). In that case the  
exterior square of $\, L_4$ is an order-six linear differential operator 
\begin{eqnarray}
\label{LCLM2}
\hspace{-0.98in}&& \, \,\, \,  \quad \quad \quad \quad \quad
  L_6 \, \, = \, \, \, \mathrm{Ext}^2(L_4) \, \, = \, \, \,
 \mathrm{LCLM}({\tilde A}_3, \, {\tilde B}_3)
     \, \, = \, \, \,  \tilde{A}_3 \, \oplus \, \tilde{B}_3, 
\end{eqnarray}
where the two order-three linear differential operators $\, \tilde{A}_3$ and  $\, \tilde{B}_3 \, $
are both {\em symmetric squares} of order-two linear differential operators having respectively the solutions
\begin{eqnarray}
\label{LCLMsol}
\hspace{-0.98in}&& \, \,\, \,  \quad
t^{5/6} \,\cdot \, (1\, -t)^{5/6} \,\cdot \, (t^2-t+1)^{-1/2} \cdot \,
  _2F_1\Bigl([{{7} \over {6}}, \, {{5} \over {2}}], \, [{{7} \over {3}}], \, \,  t \Bigr),
 \\
\hspace{-0.98in}&& \, \,\, \,  \quad 
 t^{1/6} \, \cdot \,(1\, -t)^{1/6} \, \cdot \, (t^2-t+1)^{-3/4} \cdot \,
_2F_1\Bigl([ -{{1} \over {12}}, \, {{7} \over {12}}], \, [1], \, \,
  {{27} \over {4}}\,{{ t^2 \cdot \, (1\, -t)^2 } \over { (1 \, -t \, +t^2)^{3} }} \Bigr),
\nonumber
\end{eqnarray}
totally reminiscent of the two solutions (\ref{twoOpersol}) and  (\ref{twoOpersol2}).

According to~\cite{CompointWeil} the direct-sum decomposition (\ref{LCLM2}) means that
the order-four operator $\, M_4$  is {\em absolutely reducible}, i.e. it admits a {\em factorization
over an algebraic extension} of $\, \mathbb{C}(t)$. This is confirmed by relation (\ref{LCLML4M4})
in section \ref{otherpisur6}
\begin{eqnarray}
\label{LCLMapp}
\hspace{-0.98in}&& \, \,\, \,  \quad \quad \quad
L_4 \cdot \,  I_3  \, \, \, = \, \, \,   \,
 J_3 \cdot \, \Bigl( {{1} \over {\rho}} \cdot \, M_4  \cdot \,  \rho\Bigr)
\quad \quad \hbox{with:}  \quad \quad
\rho \, = \, \, t^{2/3} \cdot \, (1\, -t)^{2/3}, 
\end{eqnarray}
where $\, I_3$ and $\, J_3$ are order-three intertwiners
and where the order-four operator $\, L_4$ is a symmetric product of two
order-two linear differential operators (\ref{twoOper}).   

\vskip .1cm

\subsection{Exterior square of symmetric products and direct sum of symmetric squares.}
\label{Appendix}

Let us consider two order-two linear differential operators
\begin{eqnarray}
\label{L2M2}
  &&\hspace{-.58in} \quad \quad \quad \quad
L_2 \, \, = \, \, \,  D_t^2 \,  \,\,
- {{1} \over { w_L(t)}} \cdot \, {{ d  w_L(t)} \over { dt}} \cdot \, D_t  \,  \,  \,  + l(t),
     \nonumber \\ 
  &&\hspace{-.58in} \quad \quad  \quad \quad
M_2 \, \, = \, \, \,
 D_t^2 \, \, \,  - {{1} \over { w_M(t)}} \cdot \, {{ d  w_M(t)} \over { dt}} \cdot \, D_t  \,   \, \,  + m(t),
\end{eqnarray}
where $\,  w_L(t)$ is the wronskian of $\, L_2$ and  $\,  w_M(t)$ is the wronskian of $\, M_2$.
We have the following identity between the exterior square of
symmetric product of these two linear differential operators
and the LCLM (i.e. direct sum) of the symmetric squares of these two linear differential
operators\footnote[1]{In Maple, with DEtools, the identity reads:
  $\, exterior$\_$power( symmetric$\_$product(L_2, \, M_2))$
$\, = \, LCLM(  symmetric$\_$power(L_2, \, 2), \,   symmetric$\_$power(M_2, \, 2)  ) $.}:
\begin{eqnarray}
 \label{L2M2identity}
 &&\hspace{-.98in} \quad \quad
 Ext^2\Bigl( \mathrm{SymProd}(L_2, \, M_2)  \Bigr)   \, \, = \, \, \,
 \nonumber \\ 
  &&\hspace{-.98in} \quad \quad \quad
  \, \, = \, \, \, 
  \Bigl(   w_M(t) \cdot \, \mathrm{Sym}^2(L_2)  \cdot \,   {{1} \over { w_M(t)}}  \Bigr)
  \oplus  \Bigl( w_L(t)  \cdot \,  \mathrm{Sym}^2(M_2)    \cdot \,  {{1} \over { w_L(t)}} \Bigr).
\end{eqnarray}
In a more general framework, like in (\ref{B_3}),  we do not have an
identity but an {\em equivalence} (homomorphisms)
between the LHS and the RHS: see for instance Lemma 8 in~\cite{CompointWeil}.

\end{document}